\documentclass[10pt,aps,prd,twocolumn,showpacs,superscriptaddress,eqsecnum,longbibliography,nofootinbib]{revtex4-2}
\usepackage{mathrsfs,amsmath,amsthm,latexsym,amssymb,amsfonts,epsfig,cancel,enumerate,graphicx,subcaption,txfonts,overpic,makecell,array,booktabs}
\usepackage{multirow}
\usepackage{xcolor}
\definecolor{navy}{RGB}{0,0,150}
\usepackage[colorlinks, linkcolor=blue, citecolor=blue, urlcolor=blue, plainpages=false, pdfstartview=FitH]{hyperref}
\usepackage{appendix}
\allowdisplaybreaks
\newcommand{\GZU}{College of Physics, Guizhou University, Guiyang 550025, China}
\newcommand{\BJC}{Beijing Capital Air Environmental Science and Technology Co., Ltd., Beijing 100176, China}
\newcommand{\SU}{School of Physics and Electronic Engineering, Shanxi Normal University, Taiyuan 030031, China}

\begin{document}
	
	\title{Shadow and thin accretion disk around Ay\'{o}n-Beato-Garc\'{i}a black hole coupled with cloud of strings}
	
	\author{Ziqiang Cai}
	\email{gs.zqcai24@gzu.edu.cn}
	\affiliation{\GZU}
	
	\author{Zhenglong Ban}
	\email{gs.zlban22@gzu.edu.cn}
	\affiliation{\GZU}
	
	\author{Lu Wang}
	\email{beautifulmind@163.com}
	\affiliation{\BJC}
	
	\author{Haiyuan Feng}
	\thanks{Corresponding author}
	\email{fenghaiyuanphysics@gmail.com}
	\affiliation{\SU}
	
	\author{Zheng-Wen Long}
	\thanks{Corresponding author}
	\email{zwlong@gzu.edu.cn}
	\affiliation{\GZU}
	
\begin{abstract}
In this paper, we investigate the shadow and thin accretion disk around Ay\'{o}n-Beato-Garc\'{i}a (ABG) black hole (BH) coupled with a cloud of strings (CS), characterized by the nonlinear electrodynamics (NLED) parameter $g$, and the CS parameter $a$. By comparing shadow diameters with Event Horizon Telescope (EHT) observations of M87$^{*}$ and Sgr A$^*$, we have established constraints on the BH parameters $g$ and $a$. Additionally, we analyze the BH shadow, lensing ring, and photon ring features for the ABG BH coupled with CS. Our results indicate that the shadow radius increases monotonically with the CS parameter $a$, while it decreases with increasing $g$. Finally, the study explores the physical properties and observational signatures of thin accretion disks around ABG BH with CS. The results show that an increase in parameter $g$ leads to a hotter and more luminous disk, while an increase in parameter $a$ results in a cooler and less luminous disk.
\end{abstract}

\maketitle
\section{Introduction}
General Relativity (GR) gives rise to one of the most fascinating predictions in modern physics — the existence of BHs, often described as the most mysterious objects in the universe. The gravitational wave detections of colliding BHs by the Laser Interferometer Gravitational-Wave
Observatory (LIGO) represent a landmark achievement that both confirms fundamental aspects of GR and provides researchers with an unprecedented tool for probing BH properties \cite{LIGOScientific:2016aoc}. The first direct visual evidence of a BH was obtained in 2019 by the EHT Collaboration \cite{EventHorizonTelescope:2019dse,EventHorizonTelescope:2019uob,EventHorizonTelescope:2019jan,EventHorizonTelescope:2019ths,EventHorizonTelescope:2019pgp,EventHorizonTelescope:2019ggy}, imaging the supermassive BH at the center of M87$^{*}$. The EHT collaboration also succeeded in imaging the supermassive BH in the Galactic Center, Sgr A$^*$ \cite{EventHorizonTelescope:2022wkp,EventHorizonTelescope:2022apq,EventHorizonTelescope:2022wok,EventHorizonTelescope:2022exc,EventHorizonTelescope:2022urf,EventHorizonTelescope:2022xqj}. These groundbreaking observations not only validate theoretical models of BH spacetime but also carry imprints of the dynamic processes governing the surrounding material \cite{EventHorizonTelescope:2021bee}.

In light of these remarkable observational advances, theoretical efforts have increasingly focused on developing more realistic BH models that go beyond the classical solutions of GR. One promising approach involves incorporating nonlinear electrodynamics (NLED) into gravitational theories. This idea was originally proposed by Born and Infeld, who initially proposed NLED to resolve the problematic singularities and energy divergences inherent in point charges \cite{Born:1934gh}. Indeed, NLED has demonstrated its versatility and relevance by finding applications in both cosmological contexts \cite{DeLorenci:2002mi,Novello:2003kh,Mignani:2016fwz} and string theory \cite{Seiberg:1999vs,Fradkin:1985qd}. These diverse applications highlight the theory's adaptability and its crucial role in addressing a wide range of problems in modern theoretical physics. The first BH solution incorporating a NLED field that satisfies the weak energy condition was proposed by ABG \cite{Ayon-Beato:1998hmi}. Letelier \cite{Letelier:1983vka} modeled a CS as a perfect fluid without pressure. This model is crucial for understanding the basic constituents of the universe and has important applications in both astrophysics and cosmology. The CS has been employed as a potential material source in the Einstein field equations, leading to the discovery of generalized BH solutions \cite{Letelier:1983du,Letelier:1979ej,Singh:2020nwo}. A regular charged BH constructed within the framework of ABG geometry, incorporating both NLED and a CS, was recently proposed in \cite{Kumar:2024cnh}. This work highlights the impact of including CS as a matter source, demonstrating its influence on the BH's structural and dynamical characteristics.

Moreover, the success of the EHT in imaging BH shadows has opened a new frontier in observational relativity, prompting extensive research into how such images can be used to infer the fundamental properties of BHs. In particular, recent studies have focused on BH shadows in the context of modified theories of gravity and non-standard spacetime geometries near the horizon \cite{Okyay:2021nnh,Roy:2021uye,Khodadi:2020jij,Vagnozzi:2022moj,Wang:2018prk,Cunha:2019hzj,Pantig:2020uhp}. These studies are particularly valuable for constraining deviations from general relativity in the strong-field regime.

Given the increasing observational capabilities in BH imaging and electromagnetic spectroscopy, it becomes essential to investigate how modifications to the spacetime geometry — such as those arising from nonlinear electrodynamics and exotic matter like a cloud of strings — influence key observables. One of the most promising avenues for this exploration lies in the study of accretion disks around BHs, which provide a powerful probe of the underlying gravitational theory. Over the years, the physical properties of thin accretion disks have been widely investigated across different spacetime contexts (Ref. \cite{Harko:2009rp,Chen:2011wb,Liu:2021yev,Heydari-Fard:2021ljh,Karimov:2018whx,Chen:2011rx,Kazempour:2022asl,Guzman:2005bs,Gyulchev:2021dvt,Heydari-Fard:2020ugv,Liu:2020vkh,Liu:2019mls,Pun:2008ua,Gyulchev:2020cvo,Gyulchev:2019tvk,Zhang:2021hit,Wu:2024sng,Liu:2024brf,Feng:2024iqj}). These disks' energy flux and spectral characteristics are crucial for understanding BH properties and testing modified gravity theories \cite{Bambi:2015kza}. Since the 1970s, images of accretion disks surrounding BHs have captivated observational astronomers, prompting significant advancements in theoretical models. Early work by Shakura et al. introduced a standard model for geometrically thin and optically thick accretion disks, which has since served as a fundamental framework for interpreting these systems \cite{Shakura:1972te}. This model was later extended into a relativistic context by Novikov and Thorne, leading to the widely recognized Novikov-Thorne model \cite{Page:1974he}. In astrophysical observations, BHs are often accompanied by accretion disks, making such models indispensable for analyzing data and enhancing our understanding of BH environments. Imaging techniques for thin accretion disks typically employ two main approaches: semi-analytic methods and ray-tracing combined with radiative transfer. For instance, Luminet utilized the semi-analytic approach to produce direct and secondary images of a thin accretion disk around a Schwarzschild BH, calculating its brightness using an analytical formula for radiation flux derived in Ref.\cite{Luminet:1979nyg}. To simulate the accretion structures around BHs, researchers have developed numerous numerical ray-tracing codes (e.g., \cite{37,38,39,Broderick:2005my,Dexter:2016cdk,Vincent:2011wz,43,44,Cunha:2016bjh}). Furthermore, the optical properties and physical characteristics of thin accretion disks in various background spacetimes have been extensively studied \cite{Hou:2022eev,Zhang:2024lsf,Gyulchev:2019tvk,Shaikh:2019hbm,Bambi:2019tjh,Johannsen:2016uoh,Gates:2020sdh,Okyay:2021nnh}. Concurrently, investigations such as \cite{Huang:2023ilm,Guo:2023grt,Liu:2021lvk,Guo:2022rql} have explored the imaging of BHs and naked singularities within modified gravity frameworks.

The structure of this paper is as follows. In Sec. \ref{section2}, we analyze the behavior of photon geodesics in the equatorial plane of the ABG BH coupled with a CS. In Sec. \ref{section3}, we investigate how the parameters of an ABG BH coupled with a CS influence its shadow characteristics. In Sec. \ref{section4}, we analyze the radiant energy flux, radiation temperature, and observed luminosity of thin accretion disks around the ABG BH coupled with a CS. Additionally, we employ the deflection angle diagram $(\varphi(b))$ to study the imaging features of BH accretion disks. Finally, in Sec. \ref{section5}, we summarize our findings. 

\section{Null Geodesics of the Ay\'{o}n-Beato-Garc\'{i}a Black Hole Coupled with a Cloud of Strings}
\label{section2}
We consider the gravitational theory coupled to ABG NLED and a CS term, with the action given by:
		\begin{equation}
			S=\int d^{4}x\sqrt{-\tilde{g}}(R+\mathcal{L}(F))+S_{CS},\label{S}
		\end{equation}
where $\tilde{g}$ is the determinant of the metric, $R$ is the scalar curvature, and $\mathcal{L}(F)$ is the Lagrangian density of a NLED theory, with $F=\frac{1}{4}F_{\mu\nu}F^{\mu\nu}$. Here, $F_{\mu\nu}=2\nabla_{[\mu} A_{\nu]}$ is the electromagnetic field-strength tensor. The function $\mathcal{L}(F)$ takes the form:
		\begin{equation}
			\mathcal{L}(F)=\frac{F(1-3\sqrt{2g^{2}F})}{(1+\sqrt{2g^{2}F})^{3}}-\frac{3M}{g^{3}}\left(\frac{(2g^{2}F)^{5/4}}{(1+\sqrt{2g^{2}F})^{5/2}}\right),\label{LF}
		\end{equation}
where $M$ and $g$ denote the mass and the magnetic monopole (MM) charge of the BH, respectively. The electromagnetic field-strength tensor $F_{\mu\nu}$ has the following form:
		\begin{equation}
			F_{\mu\nu}=2\delta_{[\mu}^{\theta}\delta_{\nu]}^{\phi}B(r,\theta)=2\delta_{[\mu}^{\theta}\delta_{\nu]}^{\phi}f(r)\sin\theta,\label{Fmu}
		\end{equation}
resulting in $f(r)=g$, with $g$ representing the MM charge.
		\begin{equation}
			\frac{1}{4\pi}\int_{s^{\infty}}Fds=\frac{g}{4\pi}\int_{0}^{\pi}\int_{0}^{2\pi}\sin\theta d\theta d\phi=g,\label{gint}
		\end{equation}
where $s^{\infty}$ denotes the spherical surface at spatial infinity. The cloud of strings source, governed by the Nambu-Goto action \cite{Letelier:1979ej}, is described by
		\begin{equation}
			S_{CS}=\int\sqrt{-\gamma}\mathcal{M}d\lambda^{0}d\lambda^{1}=\int\mathcal{M}(-\frac{1}{2}\Sigma^{\mu\nu}\Sigma_{\mu\nu})^{\frac{1}{2}}d\lambda^{0}d\lambda^{1},\label{SCS}
		\end{equation}
where $\mathcal{M}$ denotes a dimensionless constant associated with the string, and $(\lambda^{0}, \lambda^{1})$ are the timelike and spacelike worldsheet coordinates, respectively. The determinant $\gamma$ corresponds to the induced metric $\gamma_{ab}=g_{\mu\nu}\frac{\partial x^{\mu}}{\partial\lambda^{a}}\frac{\partial x^{\nu}}{\partial\lambda^{b}}$, on the string worldsheet. The bivector $\Sigma^{\mu\nu}$, defined as $\Sigma^{\mu\nu}=\epsilon^{ab}\frac{\partial x^{\mu}}{\partial\lambda^{a}}\frac{\partial x^{\nu}}{\partial\lambda^{b}}$, encodes the worldsheet's orientation, with $\epsilon^{ab}$ being the two-dimensional Levi-Civita tensor satisfying $\epsilon^{01}=-\epsilon^{10}=1$.
		
\begin{figure}[htbp]
	\centering
	\begin{subfigure}{0.45\textwidth}
		\includegraphics[width=3.2in, height=5.5in, keepaspectratio]{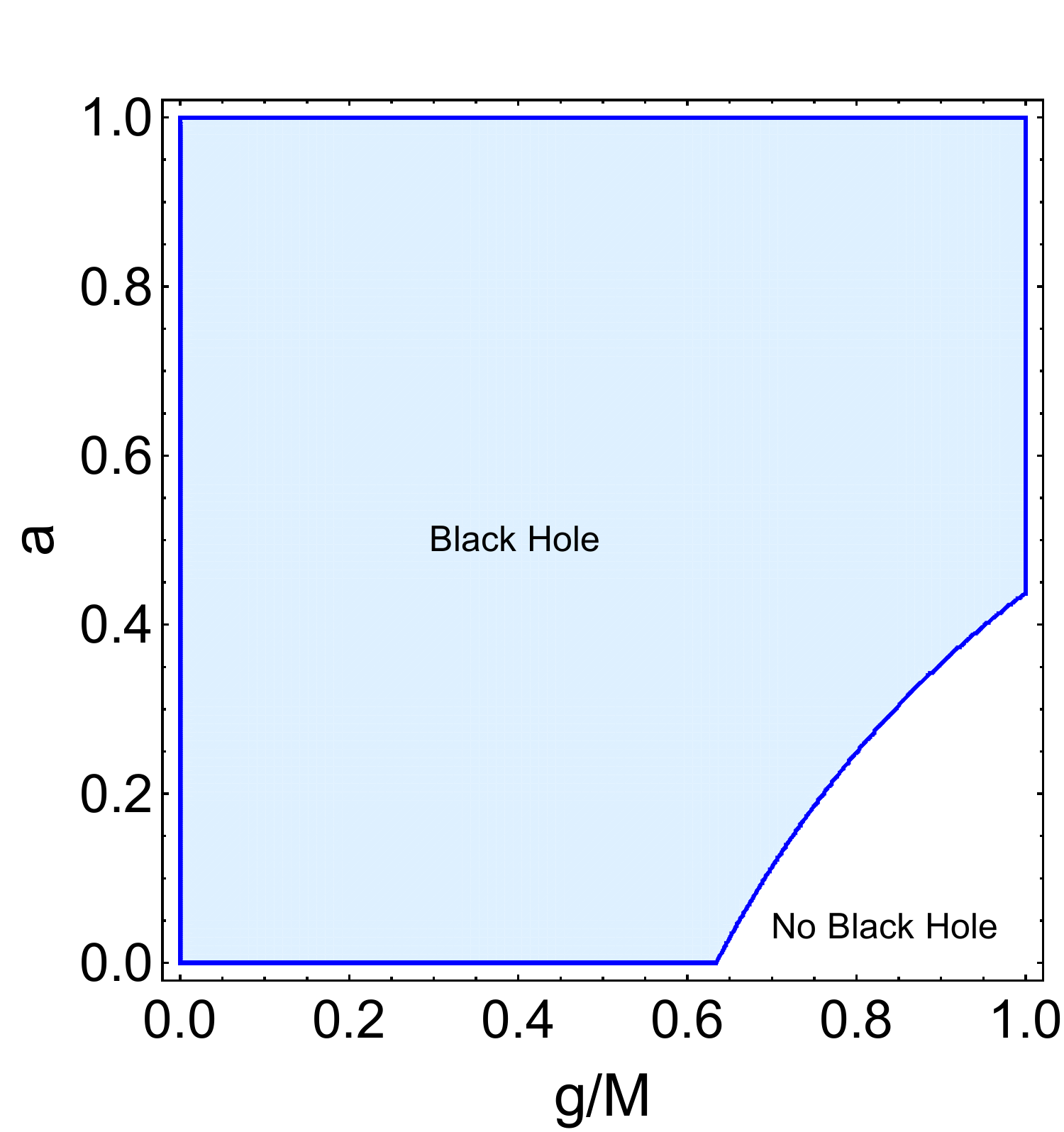}
	\end{subfigure}
	\caption{Regions of black hole horizon existence in the $(a, g/M)$ parameter space.}
	\label{horizon}
\end{figure}

The equations of motion are derived by varying the action (\ref{S}) with respect to $g_{\mu\nu}$ and $A_{\mu}$:
		\begin{equation}
			R_{\mu\nu}-\frac{1}{2}g_{\mu\nu}R=T_{\mu\nu},\label{eq}
		\end{equation}
		\begin{equation}
			\nabla_{\mu}\left(\frac{\partial\mathcal{L}(F)}{\partial F}F^{\mu\nu}\right)=0 \quad \text{and} \quad \nabla_{\mu}(\ast F^{\mu\nu})=0,\label{eq2}
		\end{equation}
where the matter energy-momentum tensor $T_{\mu\nu}$ is expressed as
		\begin{equation}
			T_{\mu\nu}=2\left[\frac{\partial\mathcal{L}(F)}{\partial F}F_{\mu\sigma}F_{\nu}^{\sigma}-g_{\mu\nu}\mathcal{L}(F)\right]+\frac{\rho\Sigma_{\mu\sigma}\Sigma{^{\sigma}}_{\!\!\nu}}{\sqrt{-\gamma}}, \label{Tmu}
		\end{equation}
where $\rho$ is the proper density of the CS. We consider the static, spherically symmetric Ay\'{o}n-Beato-Garc\'{i}a black hole solution coupled to nonlinear electrodynamics and a cloud of strings.
The corresponding spacetime geometry for this configuration is described by the following line element \cite{Kumar:2024cnh}:
\begin{equation}
	\label{1}
	ds^{2}=-f(r)dt^{2}+\frac{dr^{2}}{f(r)}+r^{2}\left(d\theta^{2}+\sin^{2}\theta d \phi^{2}\right),
\end{equation}
where 
\begin{equation}
	\label{2}
	f(r)=1-a-\frac{2Mr^{2}}{(r^{2}+g^{2})^{3/2}}+\frac{g^{2}r^{2}}{(r^{2}+g^{2})^{2}}.
\end{equation}
The BH solution given in Eq. (\ref{2}) is characterized by three key parameters: the BH mass 
$M$, the magnetic monopole (MM) charge $g$, and the CS parameter $a$. This exact solution provides a novel BH spacetime coupled to both NLED and a CS. In the limiting case where both $a = 0$ and $g = 0$, the solution reduces to the standard Schwarzschild BH spacetime. Figure.~\ref{horizon} illustrates the parameter space $(a, g/M)$ for ABG BH coupled with CS, indicating the region in which physical BH solutions exist. The blue solid line marks the extremal limit of BH configurations, and the associated light blue region delineates the existence domain of BH solutions within this parameter space.

\begin{figure*}[htbp]
	\centering
	\begin{tabular}{ccc}
		\begin{minipage}[t]{0.3\textwidth}
			\centering
			\begin{overpic}[width=0.75\textwidth]{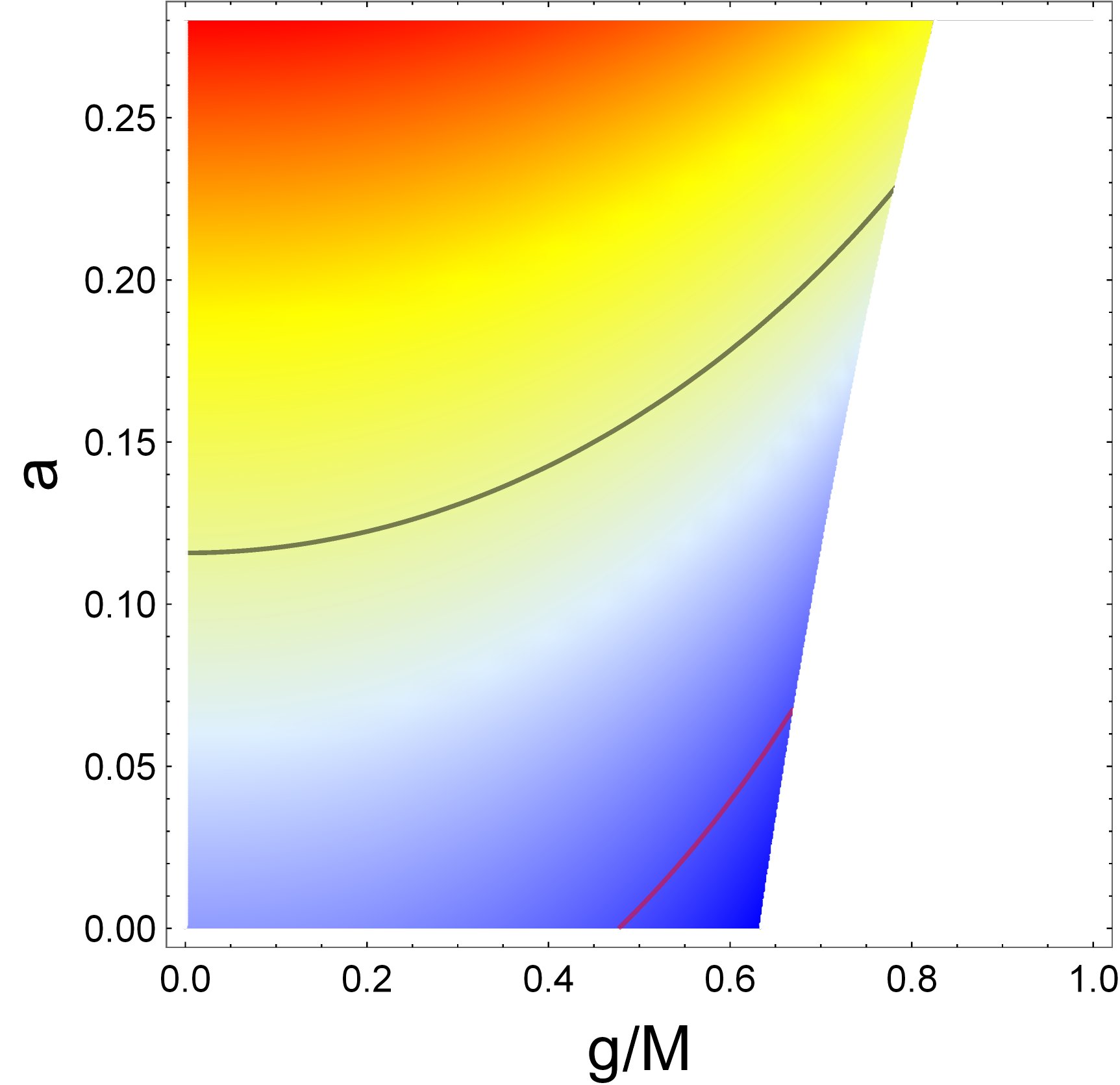}
			\end{overpic}
			\raisebox{0.06\height}{ 
				\begin{overpic}[width=0.13\textwidth]{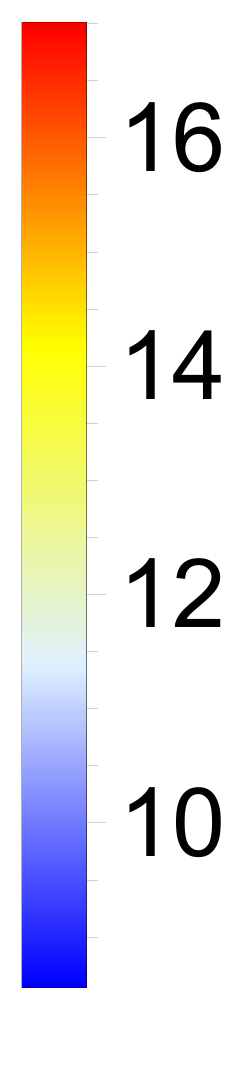}
					\put(0,103){\color{black}\large $d_{sh}$} 
				\end{overpic}
			}
		\end{minipage}
		&
		\begin{minipage}[t]{0.3\textwidth}
			\centering
			\begin{overpic}[width=0.75\textwidth]{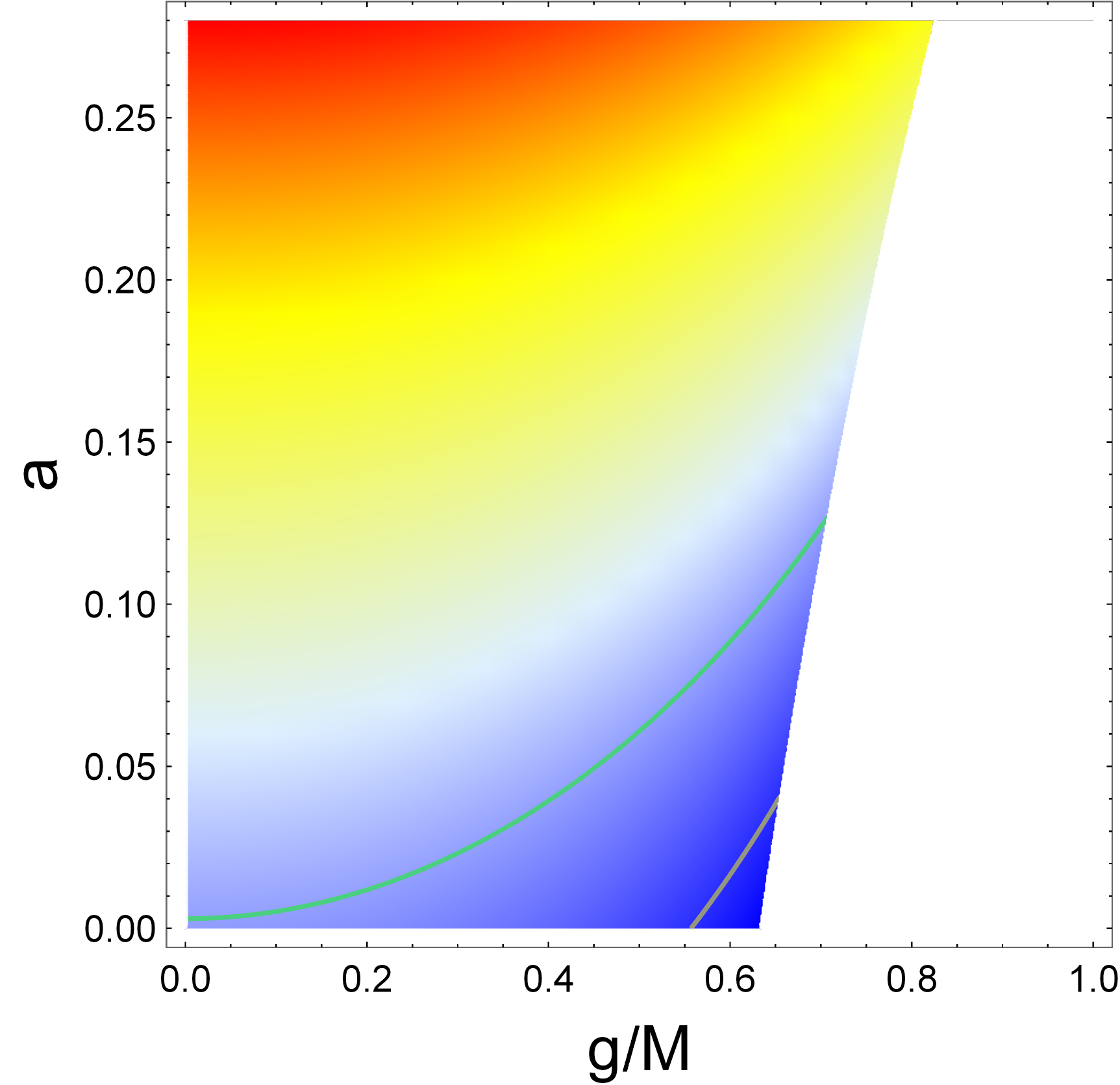}
			\end{overpic}
			\raisebox{0.06\height}{ 
				\begin{overpic}[width=0.13\textwidth]{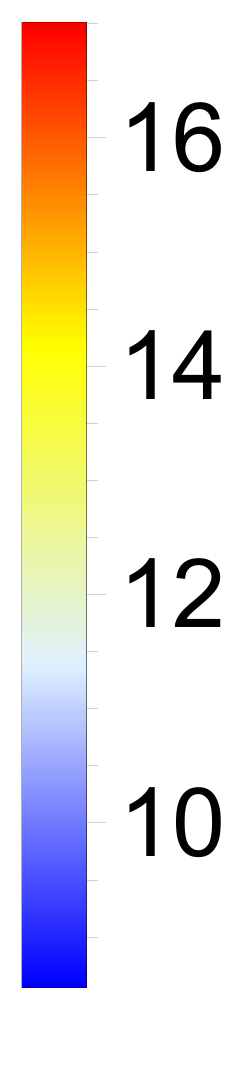} 
					\put(0,103){\color{black}\large $d_{sh}$} 
				\end{overpic}
			}			
		\end{minipage}
	\end{tabular}
	\caption{Shadow diameter $d_{sh} = 2b_{c}$ as a function of the parameters $(a, g/M)$. The red and black curves represent the M87$^*$ shadow diameter at $d_{sh} = 9.5$ and $d_{sh} = 12.5$, respectively. The region between these curves corresponds to the 1$\sigma$ bound for the M87$^*$ shadow measurement, highlighting the parameter space in which the theoretical model is consistent with observational data (left panel). For Sgr A$^*$, the yellow line denotes $d_{sh} = 9.1$, and the green curve represents $d_{sh} = 10.44$. The shaded region indicates the 1$\sigma$ confidence interval of the observed shadow diameter, highlighting the range of parameters that are consistent with the measured Sgr A$^{*}$ shadow (right panel).}
	\label{xianzhi}
\end{figure*}

Next, we consider the equations of motion for photons near the BH. The Lagrangian $\mathcal{L}$ describing the motion of a point particle in the spacetime (\ref{1}) is expressed as
\begin{equation}
	\label{3}
	\mathcal{L}=\frac{1}{2}g_{\mu\nu}\frac{dx^{\mu}}{d\lambda}\frac{dx^{\nu}}{d\lambda}.
\end{equation}
Here, $\lambda$ denotes the affine parameter. Without loss of generality, we can assume that the particle moves in the equatorial plane $\theta=\pi/2$. In this case, two conserved quantities naturally emerge:
\begin{equation}
	\label{4}
	E=-g_{tt}\frac{dt}{d\lambda}=f(r)\frac{dt}{d\lambda},
\end{equation}
\begin{equation}
	\label{5}
	L=g_{\phi\phi}\frac{d\phi}{d\lambda}=r^{2}\frac{d\phi}{d\lambda}.
\end{equation}
Here, $E$ and $L$ denote the energy and angular momentum of the photon, respectively. We focus on the null geodesics with vanishing Lagrangian ($\mathcal{L}=0$), for which the corresponding orbit equation can be derived
\begin{equation}
	\label{6}
	\left(\frac{dr}{d\phi}\right)^{2}=r^{4}\left(\frac{1}{b^{2}}-\frac{f(r)}{r^{2}}\right)\equiv V_{eff},
\end{equation}
where $b=L/E$ is the impact parameter of the light ray, $V_{eff}$ is the effective potential, and $r_{ph}$ denotes the radius of the photon sphere formed by a bounded light orbit. This radius is determined by
\begin{equation}
	\label{7}
	V_{eff}|_{r=r_{ph}}=0,
\end{equation}
\begin{equation}
	\label{8}
	\frac{dV_{eff}}{dr}|_{r=r_{ph}}=0.
\end{equation}
The critical impact parameter $b_{c}$, associated with the photon sphere, can be obtained by solving Eqs. (\ref{7}) and (\ref{8}), yielding
\begin{equation}
	\label{9}
	b_{c}=\frac{r_{ph}}{\sqrt{f(r_{ph})}}.
\end{equation}
This relationship defines a direct mapping between the theoretically predicted photon sphere radius and the observed BH shadow radius, offering a systematic framework for the quantitative analysis of BH parameters. From an observational standpoint, the BH shadow diameter $d_{sh}$ can be extracted from EHT data. Based on these observations, the shadow diameters of M87$^*$ and Sgr A$^*$ have been measured as $d_{\text{M87}^*} \simeq 11.0 \pm 1.5$ \cite{Bambi:2019tjh} and $d_{\text{SgrA}^*} \simeq 9.77 \pm 0.67$ \cite{Vagnozzi:2022moj}, respectively. Using the aforementioned observational data in conjunction with Eq. (\ref{9}), we compute the shadow diameter $d_{sh}=2b_{c}$ as a function of the parameters $\alpha$ and $g$, and present the results in Fig.~\ref{xianzhi} for the ABG BH coupled with a CS. This visualization demonstrates the dependence of shadow size on parameters $\alpha$ and $g$. We now leverage high-resolution EHT observations of M87$^{*}$ and Sgr A$^{*}$ to quantitatively constrain these parameters.

Constraints from M87$^{*}$: The diameter of the shadow $(d_{sh})$ for the M87$^{*}$ BH, measured to lie within the range of $9.5$ to $12.5$ microarcseconds with a $1\sigma$ interval according to EHT observations, imposes meaningful constraints on the parameters of the ABG BH model coupled with a CS, namely $g$ and $a$. As illustrated in Fig.~\ref{xianzhi} (left), $d_{sh}$ is depicted as a function of these two parameters, with the red and black contours representing constant values of $d_{sh} = 9.5$ and $d_{sh} = 12.5$, respectively. The $1\sigma$ observational interval effectively restricts the permissible values of $\alpha$, with the bounds depending on the chosen value of $g$. For instance, when $g$ is set to $0.5M$, $a$ is constrained to lie between $0.006$ and $0.158$. In contrast, for $g = 0.6M$, the corresponding range for $a$ is narrowed to $0.039$ to $0.178$ (see Table \ref{tab} for further details). Additionally, the values of the BH parameters selected for the figures in the following sections are based on the constrained values provided in Table \ref{tab}. These findings highlight the power of high-resolution EHT observations in refining theoretical models of BHs by precisely constraining their key physical parameters.

Constraints from Sgr A$^{*}$: As shown in Fig.~\ref{xianzhi} (right), the diameter of the BH shadow $(d_{sh})$ is presented as a function of the parameters $g$ and $a$, illustrating how variations in these parameters influence the shadow size for the ABG BH model coupled with a CS. The yellow line represents $d_{sh} = 9.1$, whereas the green solid line corresponds to $d_{sh} = 10.44$, matching the observed shadow diameter of Sgr A$^{*}$. By analyzing the dependence of the shadow diameter on these theoretical parameters, we compare the predictions of the ABG BH model coupled with a CS model with the EHT observations of Sgr A$^{*}$ within the $1\sigma$ confidence interval. This comparison yields meaningful constraints on the parameter $a$, narrowing its permissible range to maintain consistency with observational data. For example, when $g$ is fixed at $0.5M$, the value of $a$ is constrained from above, yielding an upper limit of $a\leq0.061M$. This constraint ensures that the theoretically predicted shadow size remains compatible with the observed diameter of Sgr A$^{*}$.

The seminal work by Ay\'{o}n-Beato and Garc\'{i}a \cite{Ayon-Beato:1998hmi}, which introduced the regular ABG black hole solution, established a theoretical bound on the NLED charge parameter from the requirement of the existence of an event horizon, yielding $|q|\leq 2s_{c}m\approx0.6m$ (where $q$ is equivalent to our parameter $g$). Similarly, Letelier's original paper on clouds of strings \cite{Letelier:1979ej} demonstrated that for a spherically symmetric string cloud surrounding a mass, the parameter $a$ must satisfy $0<a<1$ to preserve the static nature of the spacetime outside the cloud. These are fundamental, theory-driven constraints. However, to the best of our knowledge, our study presents the first observational and joint constraints on the parameters $a$ and $g$ for an ABG black hole coupled with a Cloud of Strings. We utilize the high-precision shadow diameter measurements from the Event Horizon Telescope (EHT) for both M87$^*$ and Sgr A$^*$ to simultaneously restrict the allowed region in the $a-g/M$ parameter space (as summarized in our Table~\ref{tab}). Comparison and significance: For instance, for $g=0.5M$, our analysis constrained the string parameter to $0.006<a<0.158$ based on M87$^*$ data. This observational bound is not only consistent with Letelier's theoretical upper limit $(a<1)$ but provides a much tighter, astrophysically relevant constraint. Similarly, our constraints on $g$ for given values of $a$ are complementary to the theoretical bound $g\leq0.6M$. This demonstrates the power of EHT observations to serve as a new, independent probe for testing fundamental physics in the strong-gravity regime. We believe this addition effectively contextualizes our findings against the existing literature and underscores the advancement our work represents.

\begin{table}[h]
	\centering
	\caption{Maximum and minimum values of the CS parameter $a$ for different values of $g$, constrained by EHT shadow observations of M87$^*$ and Sgr A$^*$.}
	\label{tab}
	\begin{tabular}{c c c c c}
		\hline \hline
		$g/M$ & \multicolumn{2}{c}{$d^{\mathrm{M87}^{*}}_{\mathrm{sh}}$} & \multicolumn{2}{c}{$d^{\mathrm{Sgr\,A}^{*}}_{\mathrm{sh}}$} \\
		& $a_{\mathrm{min}}$ & $a_{\mathrm{max}}$ & $a_{\mathrm{min}}$ & $a_{\mathrm{max}}$ \\ \hline
		0.0 & 0 & 0.115 & 0 & 0.003 \\
		0.1 & 0 & 0.117 & 0 & 0.005 \\
		0.2 & 0 & 0.122 & 0 & 0.011 \\
		0.3 & 0 & 0.131 & 0 & 0.023 \\
		0.4 & 0 & 0.143 & 0 & 0.038 \\
		0.5 & 0.006 & 0.158 & 0 & 0.061 \\
		0.6 & 0.039 & 0.178 & 0.016 & 0.088 \\ \hline \hline
	\end{tabular}
\end{table}
\section{shadow, photon rings, and lensing rings}
\label{section3}
In this section, the BH shadow, photon rings, and lensing rings are examined for an ABG BH coupled with a CS, in the presence of a highly luminous accretion disk. To lay the groundwork for this analysis, we begin by studying the trajectories of light rays propagating in the vicinity of the BH. For convenience, we introduce the transformation $u=1/r$, which allows us to reformulate the orbit equation in a more analytically tractable form
\begin{equation}
	\label{10}
	\left(\frac{du}{d\phi}\right)^{2}=\frac{1}{b^{2}}-u^{2}f(\frac{1}{u})\equiv G(u).
\end{equation}
For impact parameters satisfying $b<b_{c}$, the trajectory of a photon remains external to the event horizon. In this regime, the total change in the azimuthal angle $\phi$ along the trajectory can be computed by integrating the corresponding orbit equation
\begin{equation}
	\label{11}
	\phi = \int_{0}^{u_{h}} \frac{1}{\sqrt{G(u)}} \, du, \quad b < b_{c}.
\end{equation}
Here, $u_{h}=1/r_{h}$, where $r_{h}$ denotes the radius of the outermost event horizon. In the case where the impact parameter satisfies $b>b_{c}$, the total variation in the azimuthal angle $\phi$ along a light trajectory can be computed by integrating the corresponding orbit equation
\begin{equation}
	\label{12}
	\phi=2\int_{0}^{u_{min}}\frac{1}{\sqrt{G(u)}} \, du, \quad b > b_{c},
\end{equation}
where $u_{min}$ is the minimum positive root of $G(u)=0$ when $b>b_{c}$.

\begin{figure*}[htbp]
	\centering
	
	\begin{subfigure}{0.3\textwidth}
		\centering
		\begin{overpic}[width=\textwidth]{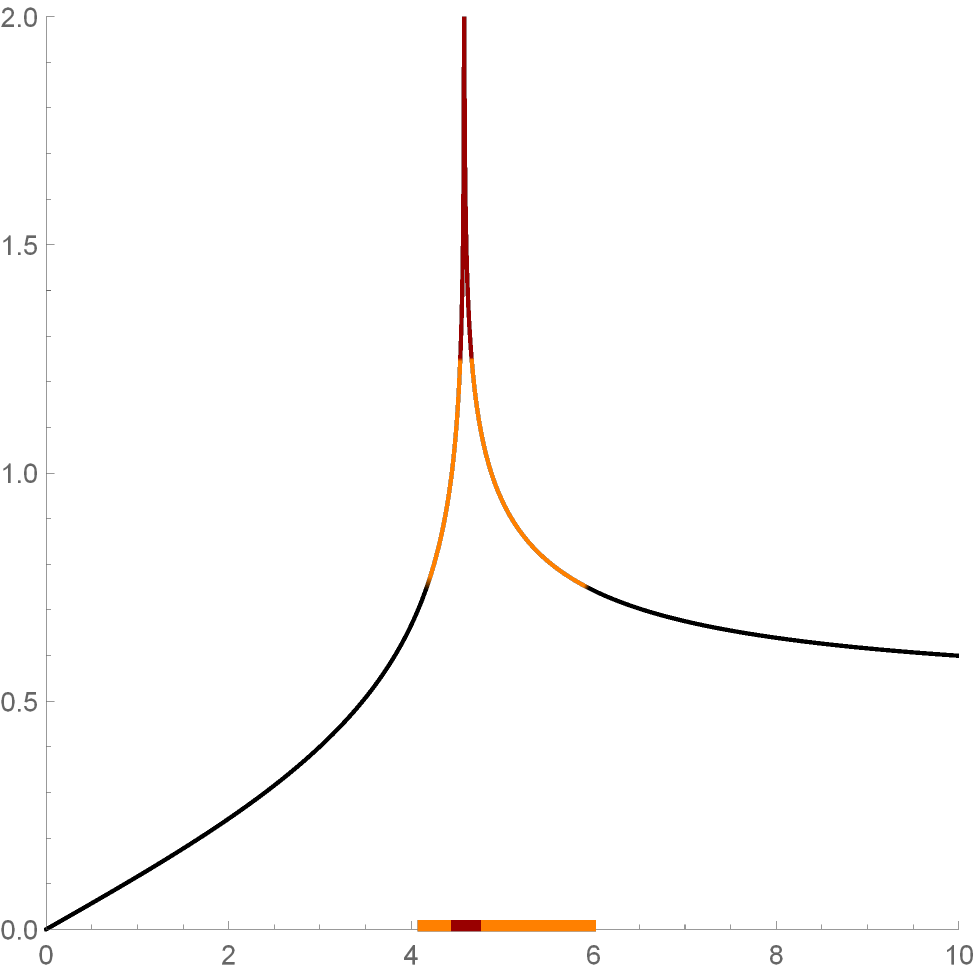}
			\put(103,2){\color{black} $b$}
			\put(0,100){\color{black} $n=\phi/2\pi$}
		\end{overpic}
		\includegraphics[width=\textwidth]{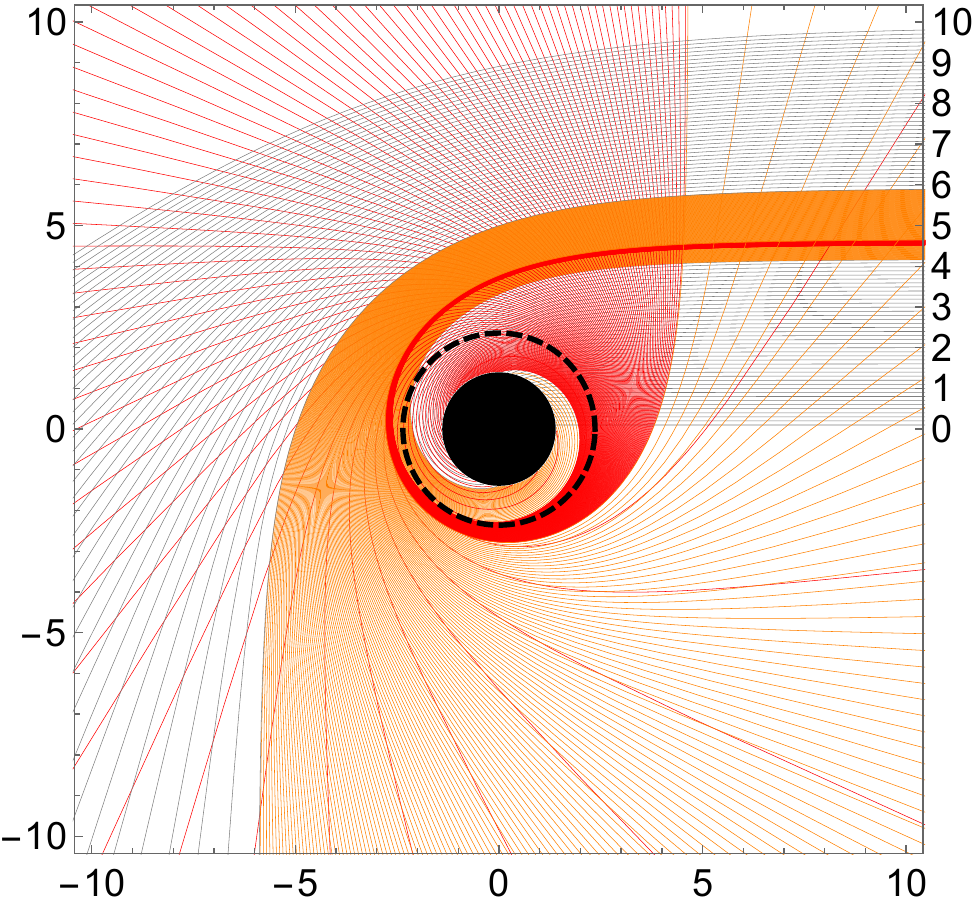}
	\end{subfigure}
	\hfill
	\begin{subfigure}{0.3\textwidth}
		\centering
		\begin{overpic}[width=\textwidth]{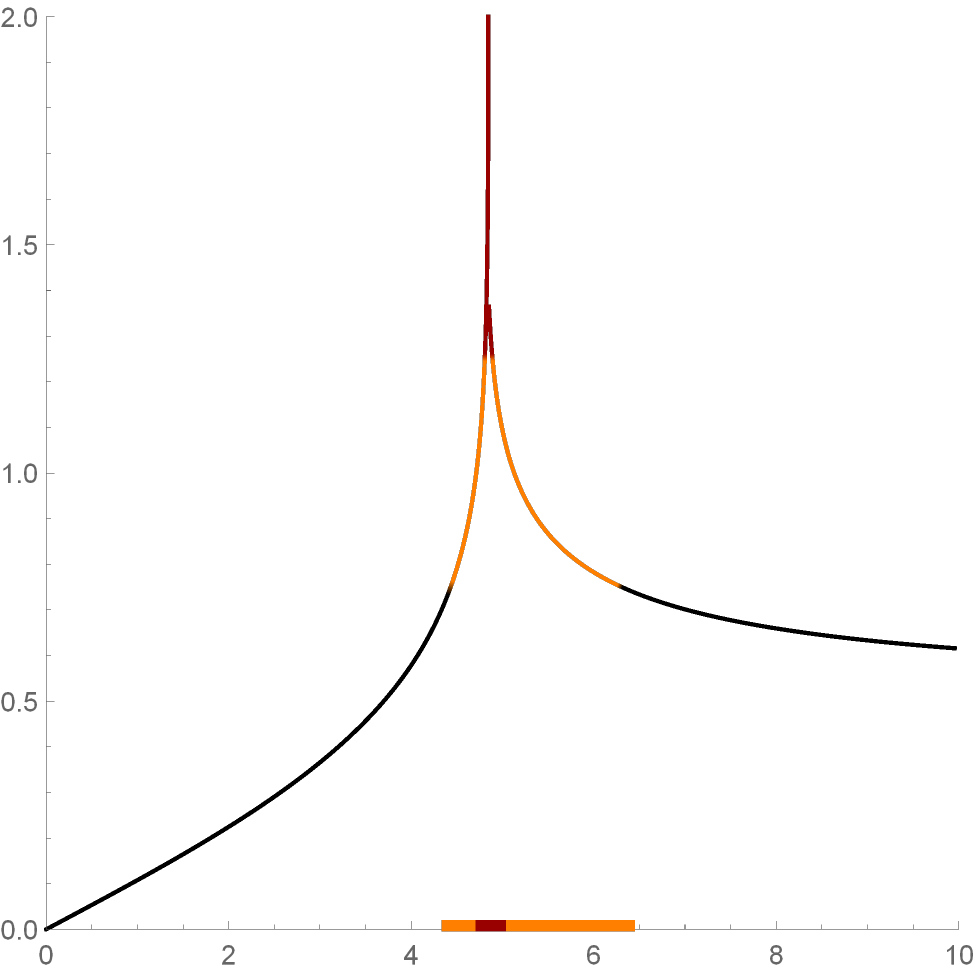}
			\put(103,2){\color{black} $b$}
			\put(0,100){\color{black} $n=\phi/2\pi$}
		\end{overpic}
		\includegraphics[width=\textwidth]{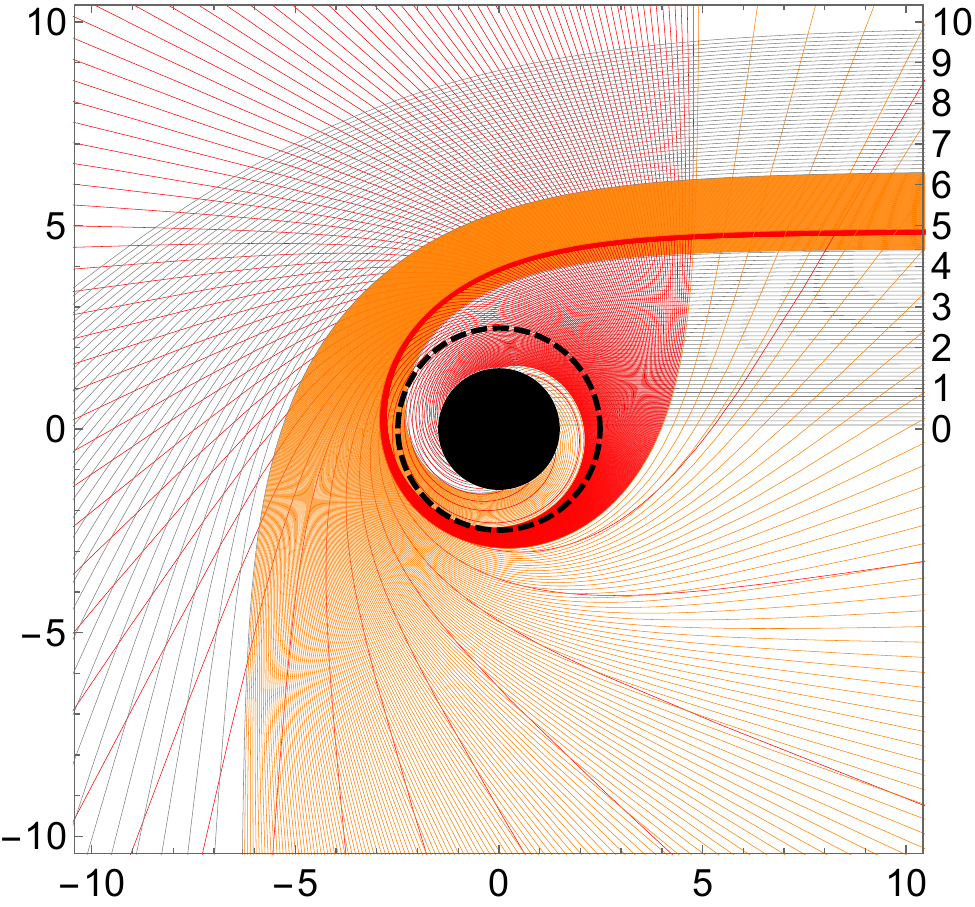}
	\end{subfigure}
	\hfill
	\begin{subfigure}{0.3\textwidth}
		\centering
		\begin{overpic}[width=\textwidth]{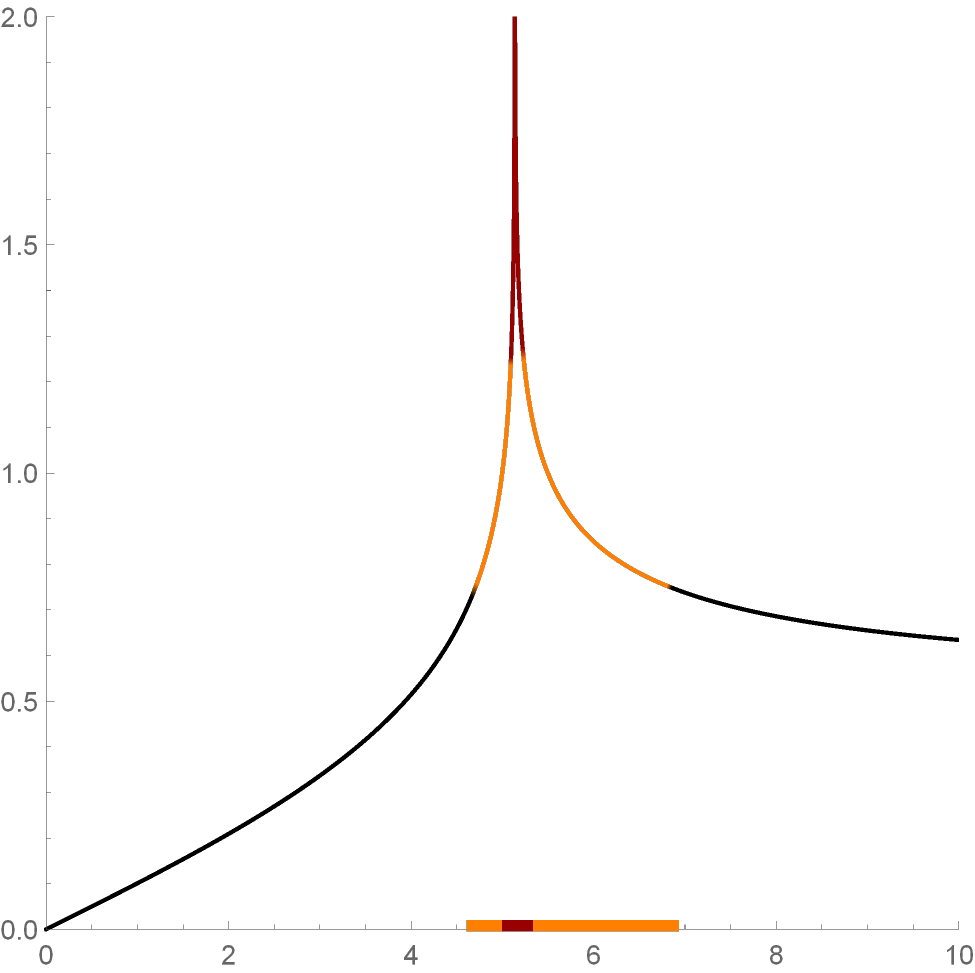}
			\put(103,2){\color{black} $b$}
			\put(0,100){\color{black} $n=\phi/2\pi$}
		\end{overpic}
		\includegraphics[width=\textwidth]{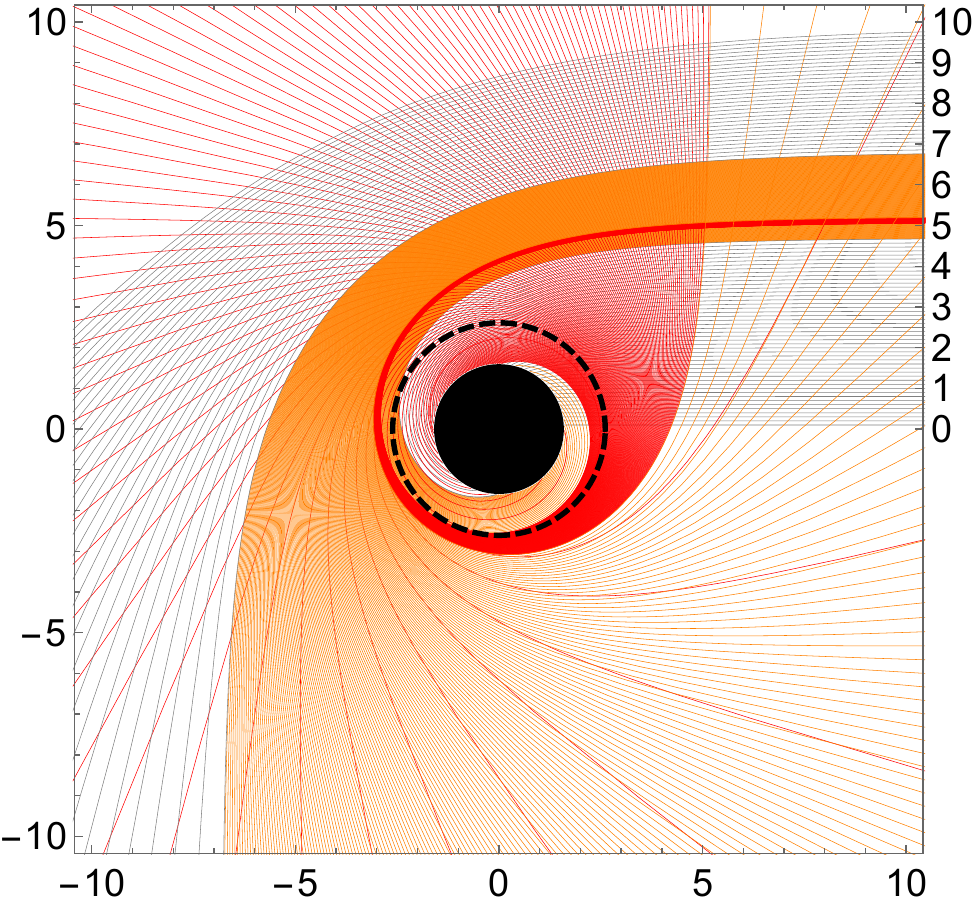}
	\end{subfigure}
	
	\caption{The behavior of photon trajectories around an ABG BH coupled with a CS as a function of the impact parameter $b$. The upper panel displays the total number of orbits $(n = \phi/2\pi)$, classifying trajectories into three categories based on $n$: direct emission $(n<3/4)$ depicted in black, lensed trajectories $(3/4 < n < 5/4)$ shown in orange, and photon ring trajectories $(n > 5/4)$ colored in red. In the lower panel, selected photon paths are visualized using Euclidean polar coordinates $(r, \phi)$. The impact parameter spacing is adjusted to $1/10$, $1/100$, and $1/1000$ for direct emissions, lensed paths, and photon rings, respectively. Three scenarios are analyzed: setting $g=0.6,a=0.02$ in the first column; $g=0.6,a=0.05$ in the second column; and $g=0.6,a=0.08$ in the third column.}
	\label{wanqu}
\end{figure*}

In order to interpret the observed features of radiation emitted near a BH, Ref. \cite{Gralla:2019xty} introduced a classification of photon trajectories into direct, lensed, and photon ring contributions. Let $n=\frac{\phi}{2\pi}$ denote the total number of orbits completed by a photon trajectory. This quantity is generally a function of the impact parameter $b$, and we denote its solution as
\begin{equation}
	\label{13}
	n(b) = \frac{2m - 1}{4}, \quad m = 1, 2, 3, \dotsc
\end{equation}
by $b_{m}^{\pm}$. It is important to note that $b_{m}^{-}<b_{c}$ and $b_{m}^{+}>b_{c}$. Based on this distinction, we can classify all photon trajectories as follows:
\begin{itemize}
	\item direct: $\dfrac{1}{4} < n < \dfrac{3}{4} \Rightarrow b \in (b_{1}^{-}, b_{2}^{-}) \cup (b_{2}^{+}, \infty)$
	\item lensed: $\dfrac{3}{4} < n < \dfrac{5}{4} \Rightarrow b \in (b_{2}^{-}, b_{3}^{-}) \cup (b_{3}^{+}, b_{2}^{+})$
	\item photon ring: $n > \dfrac{5}{4} \Rightarrow b \in (b_{3}^{-}, b_{3}^{+})$
\end{itemize}
The physical significance of this classification can be understood from the trajectory plots in Fig.~\ref{wanqu}. Assuming emission from the north pole (on the far right), trajectories with $1/4 < n < 3/4$ intersect the equatorial plane once, those with $3/4 < n < 5/4$ intersect it twice, and trajectories with $n > 5/4$ intersect it at least three times.
Tables~\ref{tab2} presents the impact parameter ranges corresponding to the lensing and photon rings. It is evident that these ranges progressively expand as either $a$ or $g$ increases.
\begin{table}[htbp]
	\centering
	\caption{The region of lensing ring and photon ring for different parameters $a$ and $g$.}
	\label{tab2}
	\begin{tabular}{|c|c|c|}
		\hline
		& Lensing Ring & Photon Ring \\  
		\hline
		$a=0.02$, $g=0.6$ & 
		\makecell[l]{$4.17073<b<4.53744$ \\ $4.66202<b<5.92482$} & 
		$4.53744<b<4.66202$ \\
		\hline
		$a=0.05$, $g=0.6$  & 
		\makecell[l]{$4.43217<b<4.80598$ \\ $4.93636<b<6.35403$} & 
		$4.80598<b<4.93636$ \\  
		\hline
		$a=0.08$, $g=0.6$  & 
		\makecell[l]{$4.70497<b<5.09280$ \\ $5.23217<b<6.83637$} & 
		$5.09280<b<5.23217$ \\  
		\hline
		$a=0.05$, $g=0.45$  & 
		\makecell[l]{$4.94270<b<5.20637$ \\ $5.28424<b<6.57639$} & 
		$5.20637<b<5.28424$ \\  
		\hline
		$a=0.05$, $g=0.55$  & 
		\makecell[l]{$4.64992<b<4.96816$ \\ $5.07183<b<6.43724$} & 
		$4.96816<b<5.07183$ \\  
		\hline
		$a=0.05$, $g=0.6$  & 
		\makecell[l]{$4.43217<b<4.80598$ \\ $4.93636<b<6.35403$} & 
		$4.80598<b<4.93636$ \\ 
		\hline
	\end{tabular}
\end{table}
The orbit equation corresponding to the time-like geodesic with $\mathcal{L}=-1/2$ is given by
\begin{equation}
	\label{14}
	\left(\frac{dr}{d\phi}\right)^{2}=r^{4}\left(\frac{1}{b^{2}}-\frac{f(r)}{r^{2}}-\frac{f(r)}{L^{2}}\right)\equiv \tilde{V}_{eff}.
\end{equation}
The radius $r_{isco}$ of the innermost stable circular orbit is given by 
\begin{equation}
	\label{15}
	\tilde{V}_{eff}|_{r=r_{isco}}=\frac{d\tilde{V}_{eff}}{dr}|_{r=r_{isco}}=\frac{d^{2}\tilde{V}_{eff}}{dr^{2}}|_{r=r_{isco}}=0.
\end{equation}

\begin{table}[htbp]
	\centering
	\renewcommand{\arraystretch}{1.8}  
	\caption{Variation of $r_{h}$ (horizon radius), $r_{\mathrm{ph}}$ (photon sphere radius), $b_{c}$ (critical impact parameter), and $r_{\mathrm{isco}}$ (innermost stable circular orbit radius) with the parameter $a$ for $g=0.6$.}
	\label{tab3}
	\begin{tabular}{cccc}
		\toprule  
		$a$ & 0.02 & 0.05 & 0.08 \\  
		\midrule  
		$r_{h}$       & 1.3939 & 1.4944 & 1.5934 \\  
		$r_{\mathrm{ph}}$ & 2.3592 & 2.4826 & 2.6100 \\  
		$b_{c}$       & 4.5796 & 4.8474 & 5.1349 \\ 
		$r_{\mathrm{isco}}$ & 4.7857 & 5.0178 & 5.2605 \\  
		\bottomrule  
	\end{tabular}
\end{table}

\begin{table}[htbp]
	\centering
	\renewcommand{\arraystretch}{1.8}
	\caption{Variation of $r_{h}$, $r_{\mathrm{ph}}$, $b_{c}$, and $r_{\mathrm{isco}}$ with the parameter $g$ for $a=0.05$.}
	\label{tab4}
	\begin{tabular}{cccc}
		\toprule
		$g$ & 0.45 & 0.55 & 0.60 \\  
		\midrule
		$r_{h}$       & 1.8225 & 1.6356 & 1.4944 \\  
		$r_{\mathrm{ph}}$ & 2.8264 & 2.6222 & 2.4826 \\  
		$b_{c}$       & 5.2254 & 4.9972 & 4.8474 \\  
		$r_{\mathrm{isco}}$ & 5.6517 & 5.2665 & 5.0178 \\  
		\bottomrule
	\end{tabular}
\end{table}
Tables~\ref{tab3} and \ref{tab4} summarize the dependence of $r_{h}$, $r_{ph}$, $b_{c}$, and $r_{isco}$ on the parameters $a$ and $g$. These quantities increase with $a$ and decrease with $g$.

We now proceed to investigate the shadow of the ABG BH coupled with a CS, assuming the presence of a geometrically and optically thin accretion disk located on its equatorial plane. The observer is positioned along the north polar direction, enabling a clear distinction between direct and lensed photon trajectories. Additionally, we assume that the emitted specific intensity depends solely on the radial coordinate and is denoted by $I_{\nu}^{em}$, where $\nu$ represents the emission frequency measured in the static frame. The observed frequency is determined by
\begin{equation}
	\label{16}
	I^{obs}(b)=\sum_{m}f^{2}I^{em}|_{r=r_{m}(b)}.
\end{equation}
Here, $I^{em}=\int I_{\nu}^{em}d\nu$ denotes the total emitted intensity from the accretion disk at radius $r$. The function $r_{m}(b)$, commonly referred to as the transfer function, represents the radial coordinate of the $m$-th intersection between the photon trajectory and the emitting disk.

\begin{figure*}[htbp]
	\centering
	\begin{subfigure}{0.3\textwidth}
		\includegraphics[width=\textwidth, keepaspectratio]{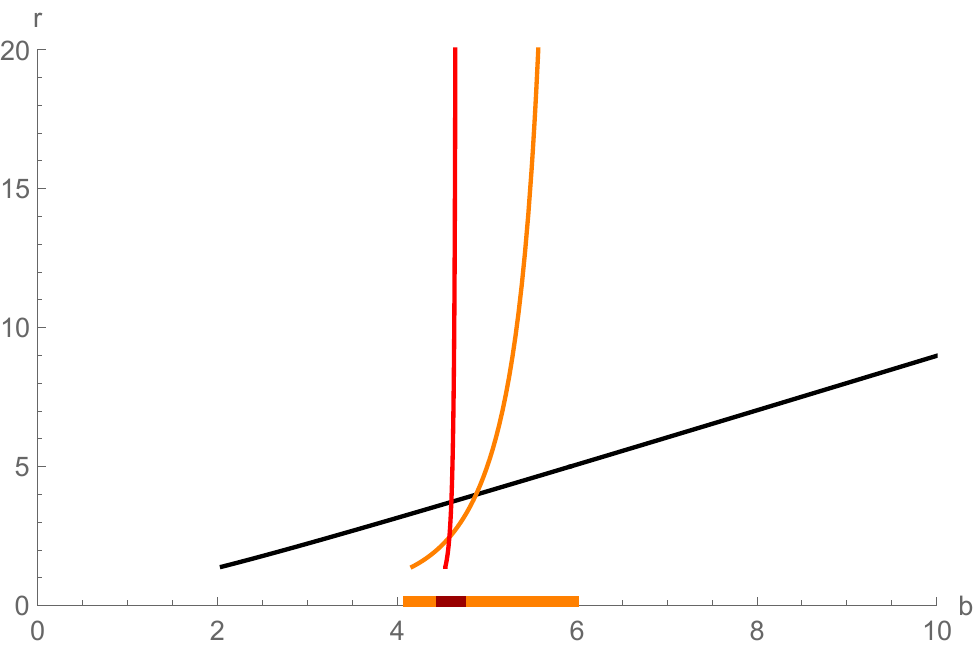}
	\end{subfigure}
	\hfill
	\begin{subfigure}{0.3\textwidth}
		\includegraphics[width=\textwidth, keepaspectratio]{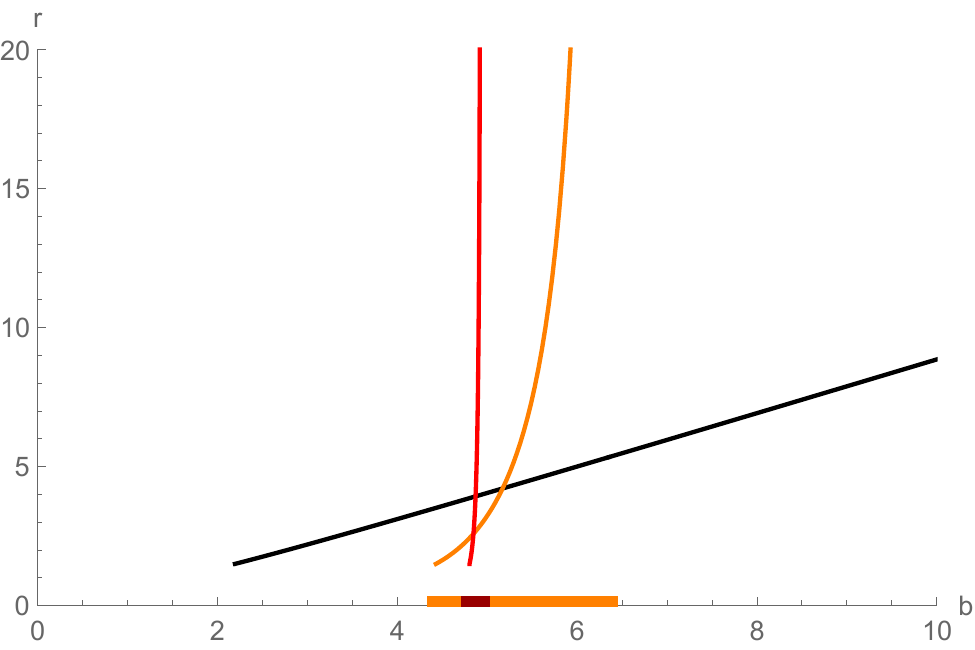}
	\end{subfigure}
	\hfill
	\begin{subfigure}{0.3\textwidth}
		\includegraphics[width=\textwidth, keepaspectratio]{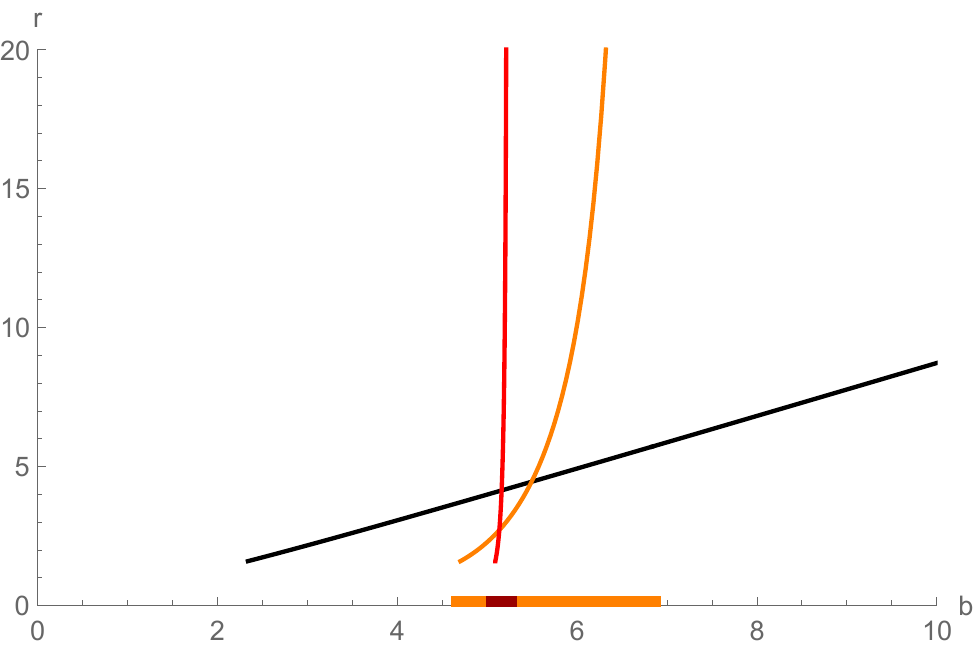}
	\end{subfigure}
	
	\caption{The first three transfer functions for BHs corresponding to different values of $\alpha$ and $g$. From left to right, the panels represent the cases with $(g=0.6,a=0.02)$, $(g=0.6,a=0.05)$, and $(g=0.6,a=0.08)$, respectively. These curves correspond to the radial positions of the first (black), second (orange), and third (red) intersections between the light rays and the emission disk.}
	\label{zhuanyi}
\end{figure*}

\begin{figure*}[htbp]
	\centering
	
	\begin{subfigure}[b]{0.31\textwidth}
		\centering
		\begin{overpic}[width=\linewidth]{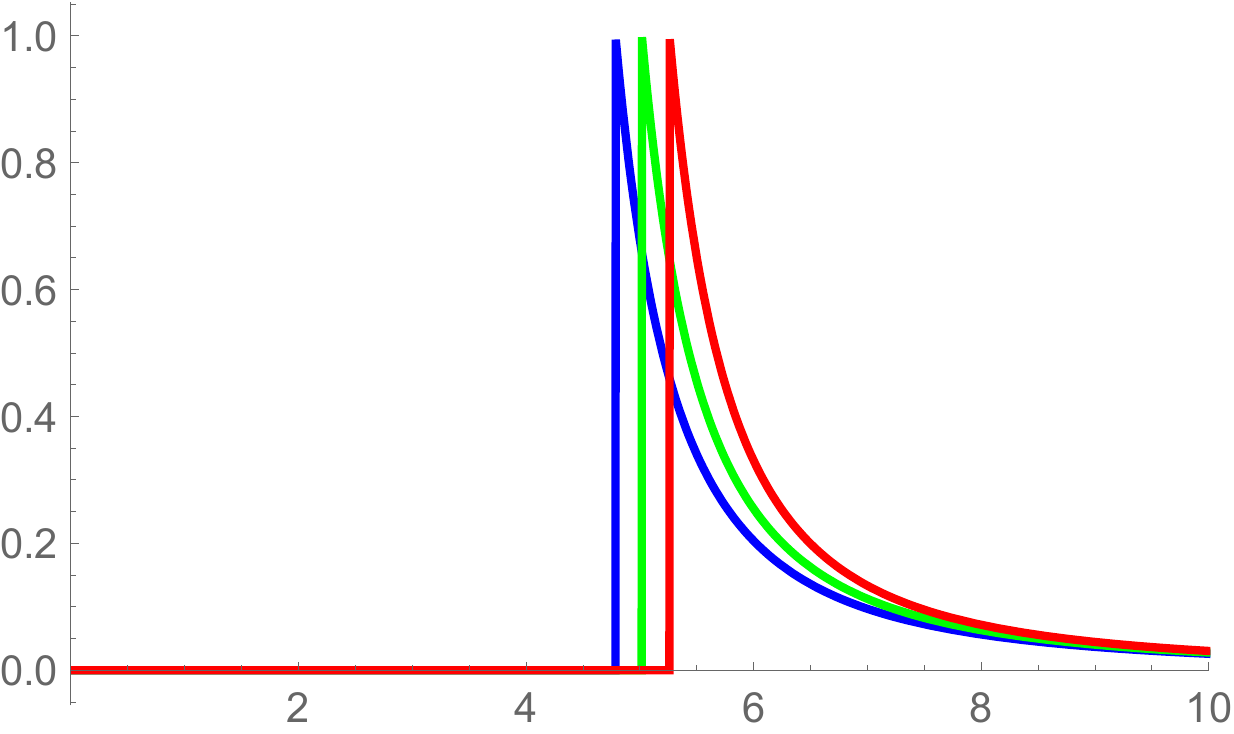}
			\put(103,2){\color{black} $b$}
			\put(0,60){\color{black} $I^{\text{em}}/I^{0}$}
		\end{overpic}
	\end{subfigure}
	\hfill
	\begin{subfigure}[b]{0.31\textwidth}
		\centering
		\begin{overpic}[width=\linewidth]{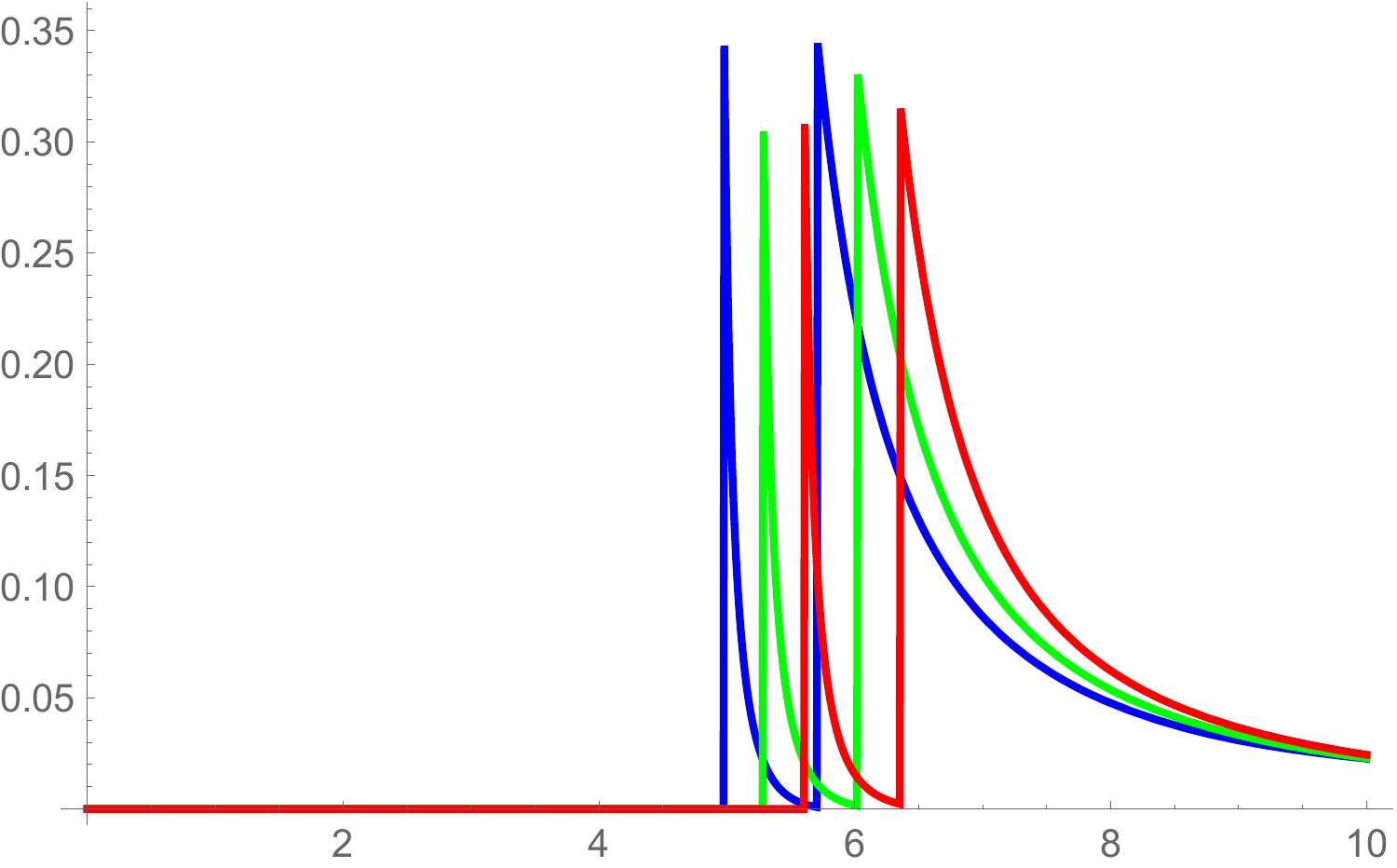}
			\put(103,2){\color{black} $b$}
			\put(0,60){\color{black} $I^{\text{obs}}/I^{0}$}
		\end{overpic}
	\end{subfigure}
	\hfill
	\begin{subfigure}[b]{0.31\textwidth}
		\centering
		\includegraphics[width=0.7\linewidth]{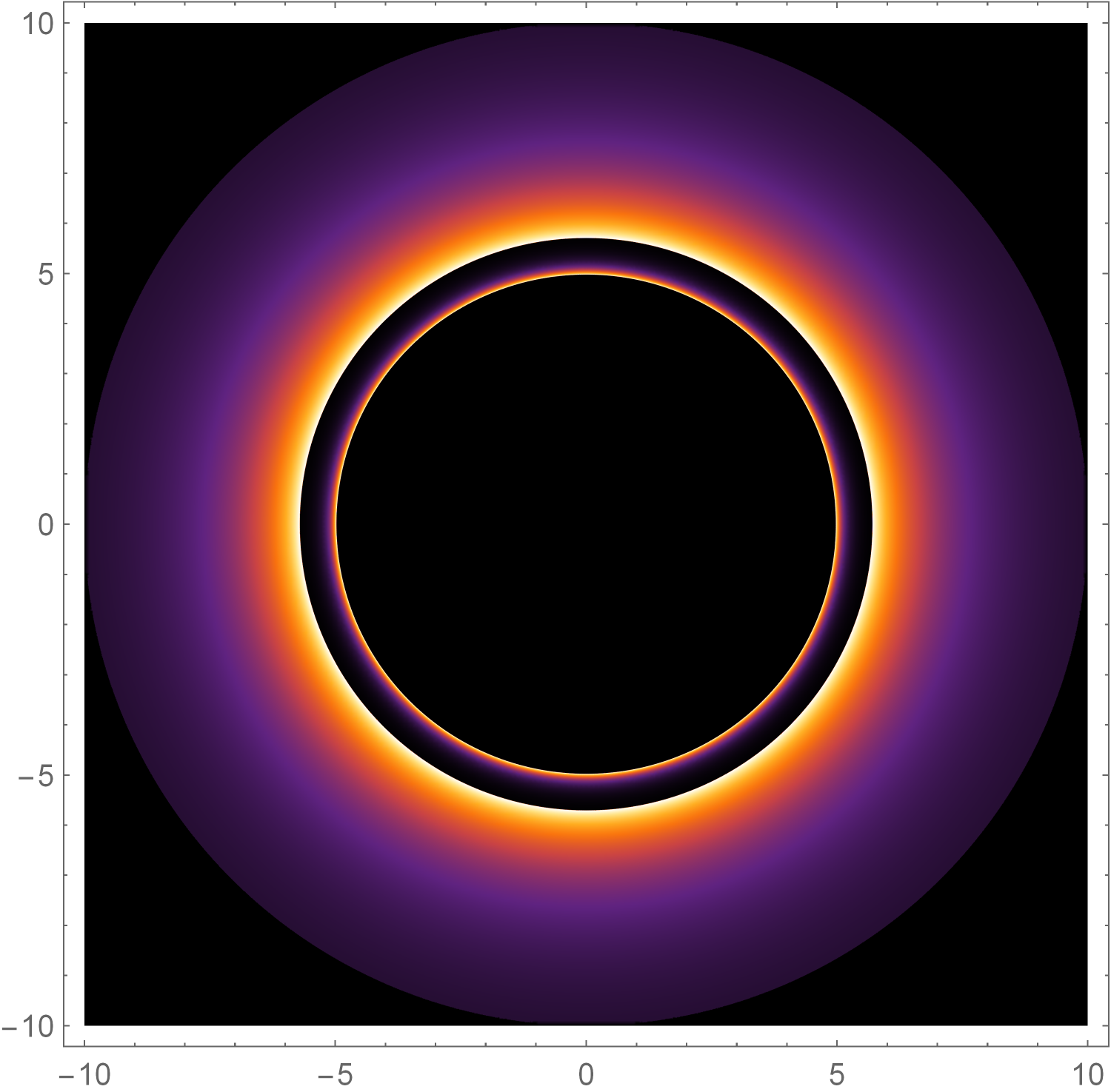}
		\hspace{0.1cm}
		\includegraphics[width=0.075\linewidth]{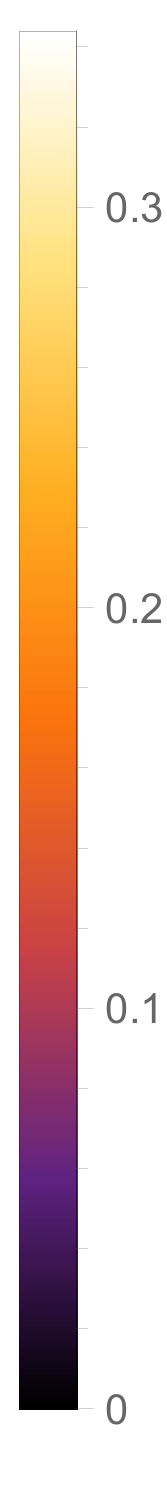}
	\end{subfigure}
	
	\vspace{0.05cm} 
	
	\begin{subfigure}[b]{0.31\textwidth}
		\centering
		\begin{overpic}[width=\linewidth]{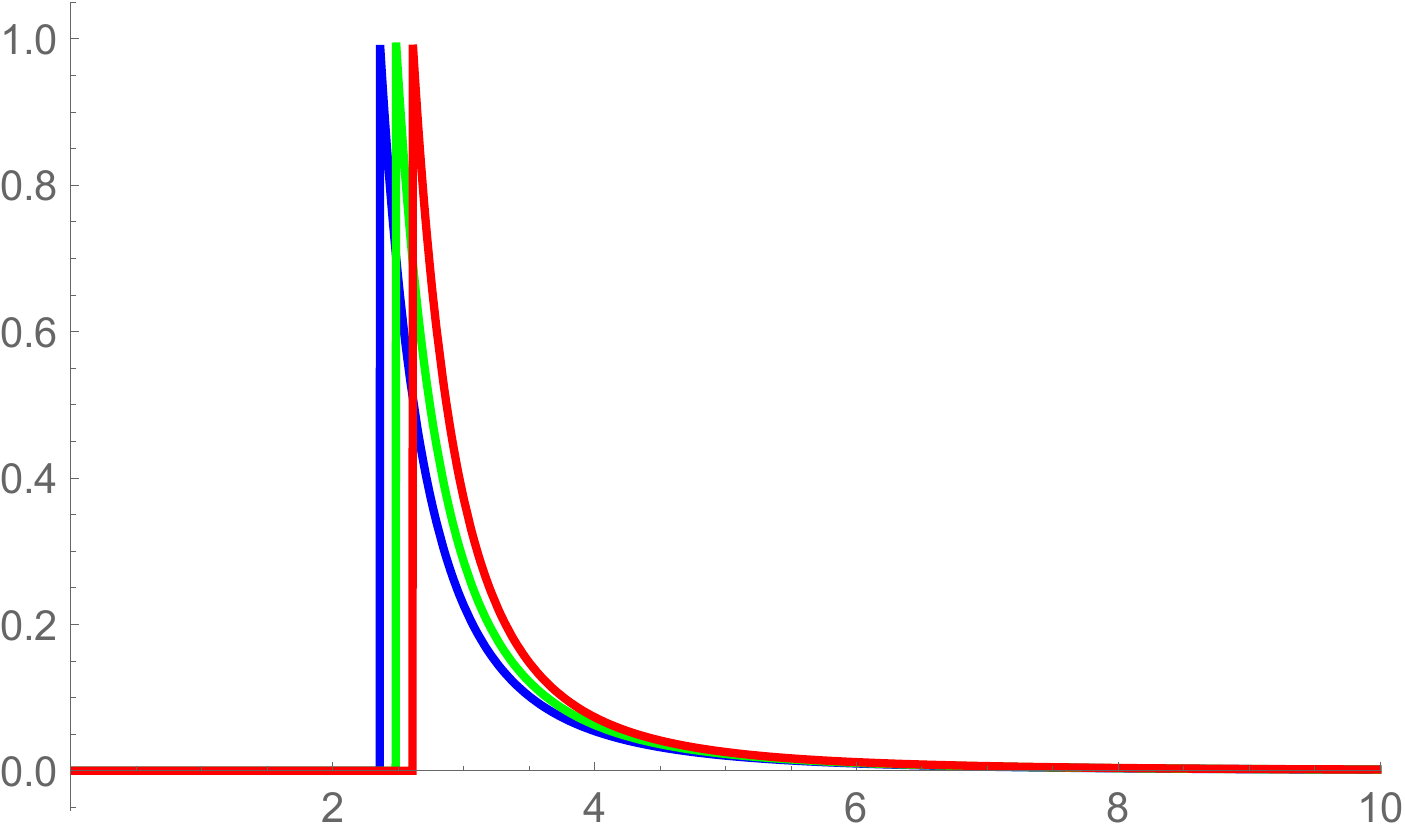}
			\put(103,2){\color{black} $b$}
			\put(0,60){\color{black} $I^{\text{em}}/I^{0}$}
		\end{overpic}
	\end{subfigure}
	\hfill
	\begin{subfigure}[b]{0.31\textwidth}
		\centering
		\begin{overpic}[width=\linewidth]{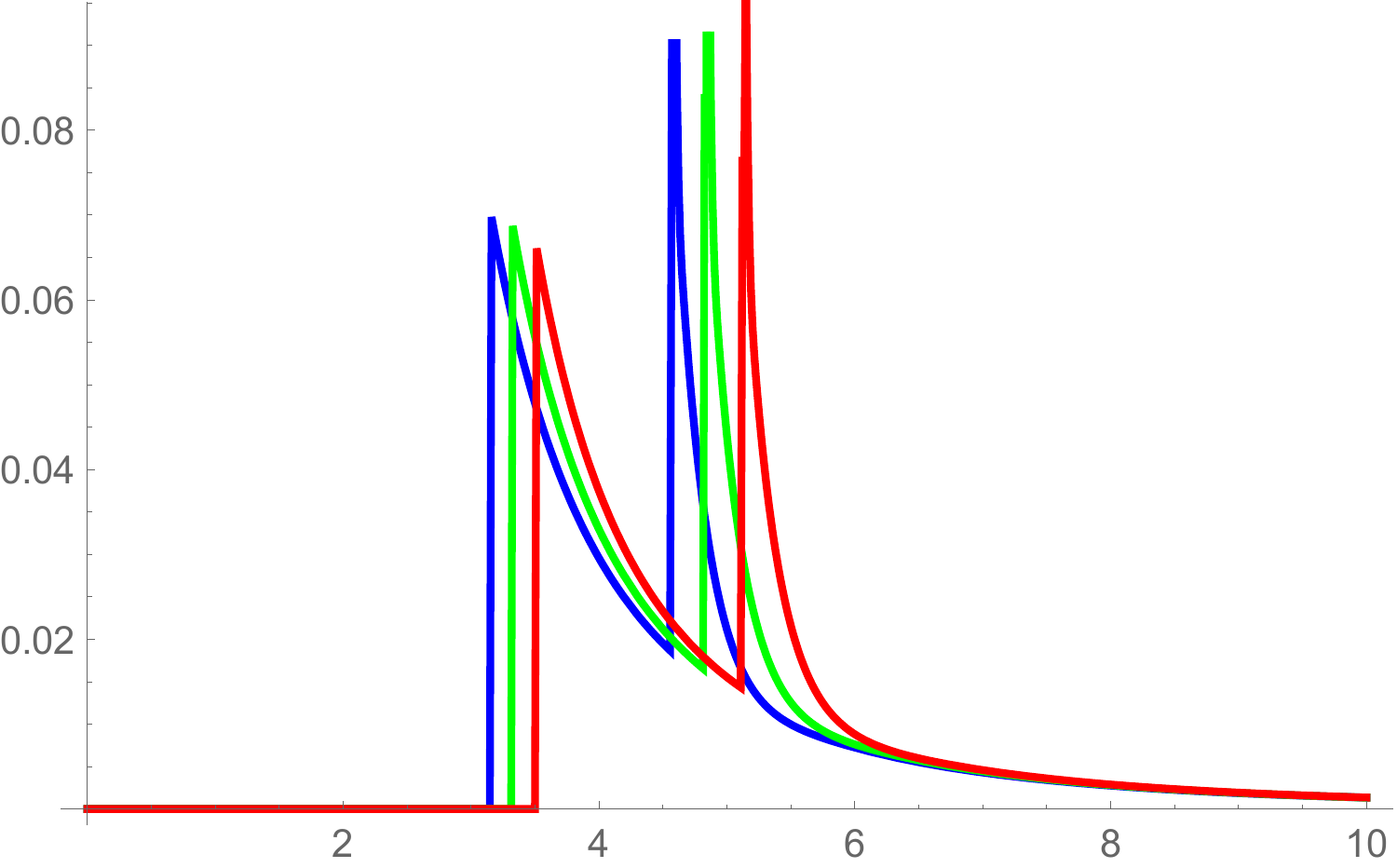}
			\put(103,2){\color{black} $b$}
			\put(0,63){\color{black} $I^{\text{obs}}/I^{0}$}
		\end{overpic}
	\end{subfigure}
	\hfill
	\begin{subfigure}[b]{0.31\textwidth}
		\centering
		\includegraphics[width=0.7\linewidth]{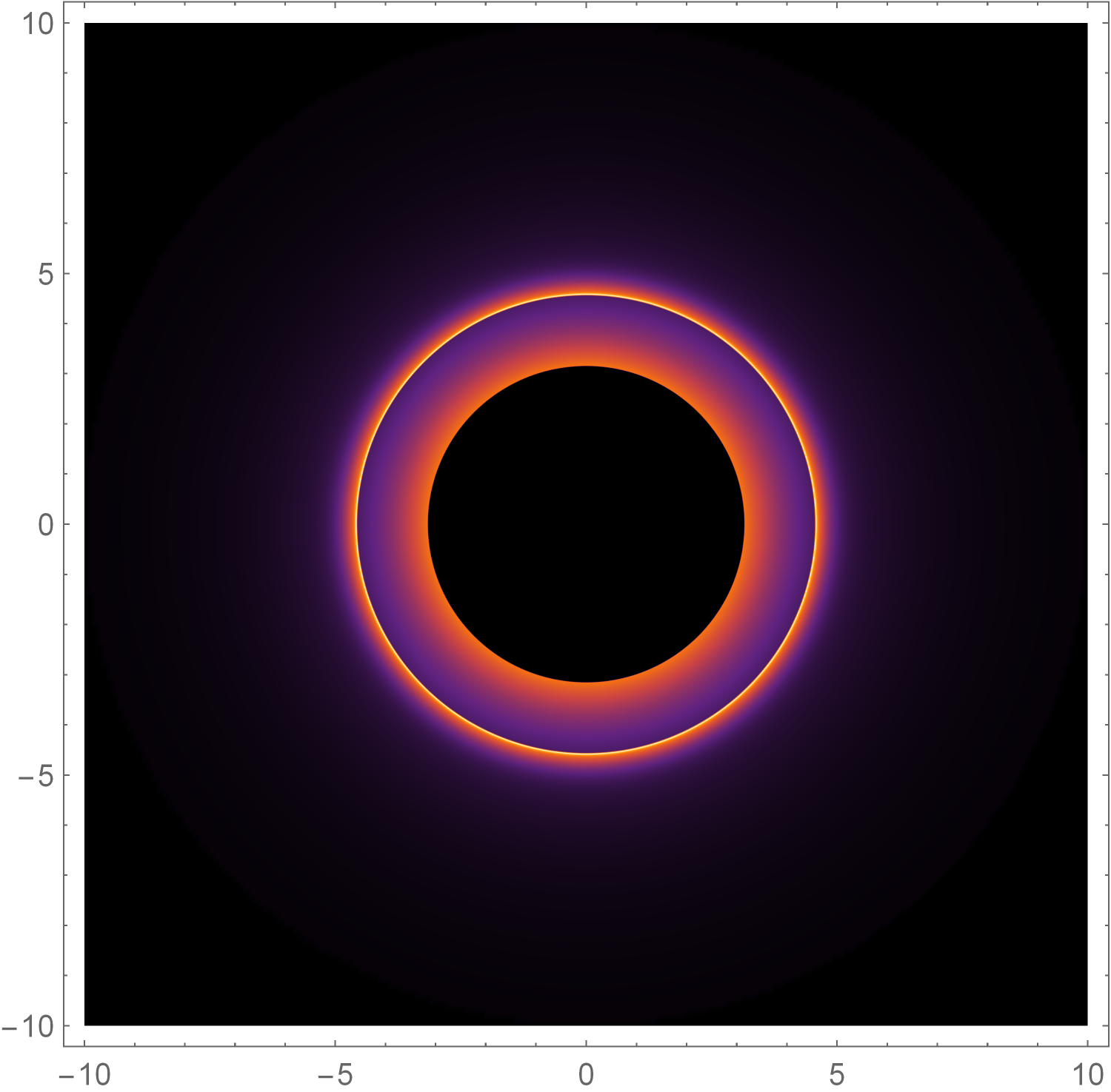}
		\hspace{0.1cm}
		\includegraphics[width=0.095\linewidth]{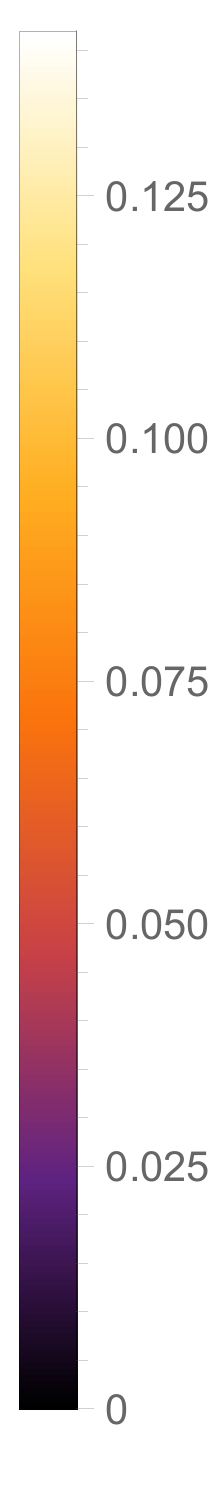}
	\end{subfigure}
	
	\vspace{0.05cm}
	
	\begin{subfigure}[b]{0.31\textwidth}
		\centering
		\begin{overpic}[width=\linewidth]{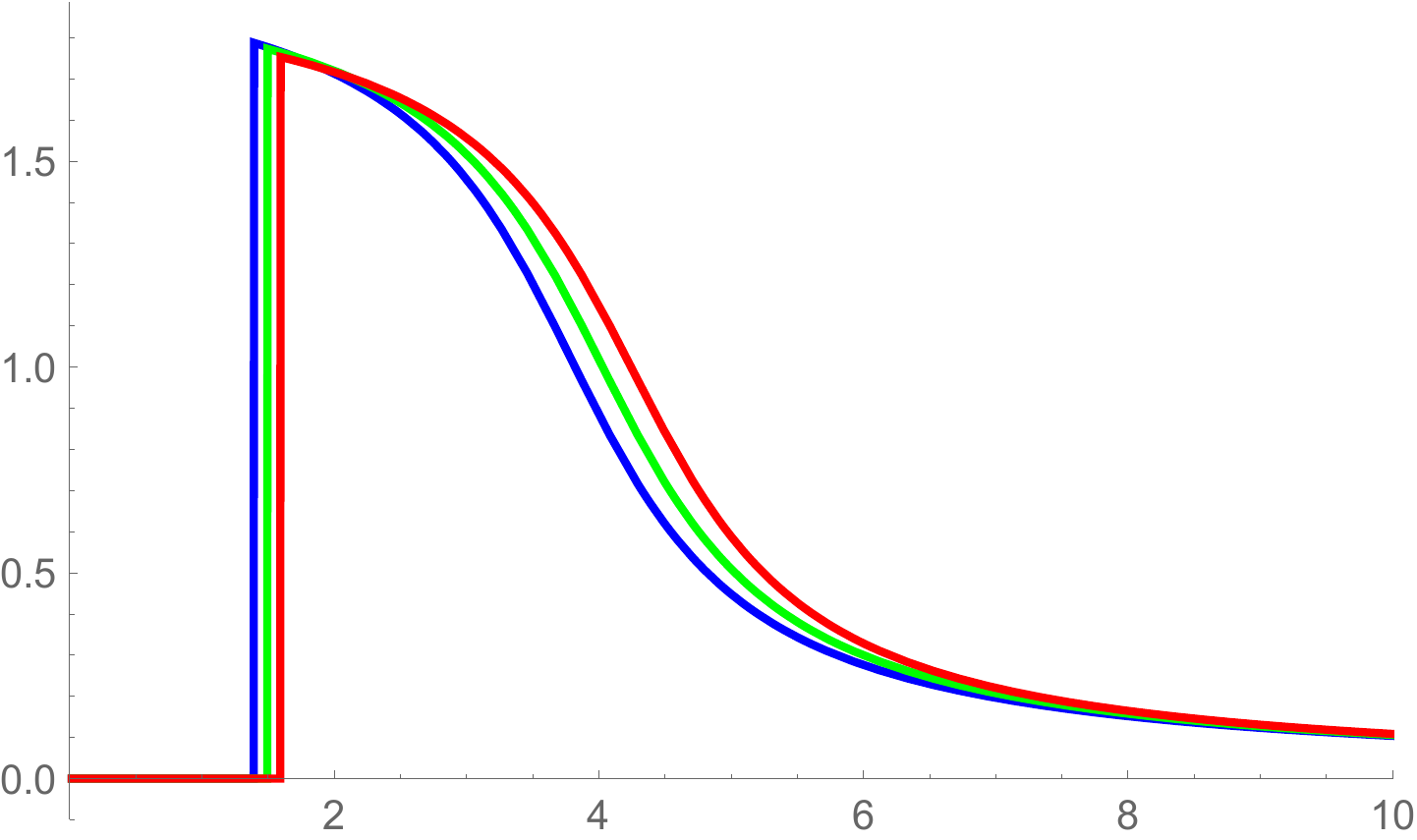}
			\put(103,2){\color{black} $b$}
			\put(0,60){\color{black} $I^{\text{em}}/I^{0}$}
		\end{overpic}
	\end{subfigure}
	\hfill
	\begin{subfigure}[b]{0.31\textwidth}
		\centering
		\begin{overpic}[width=\linewidth]{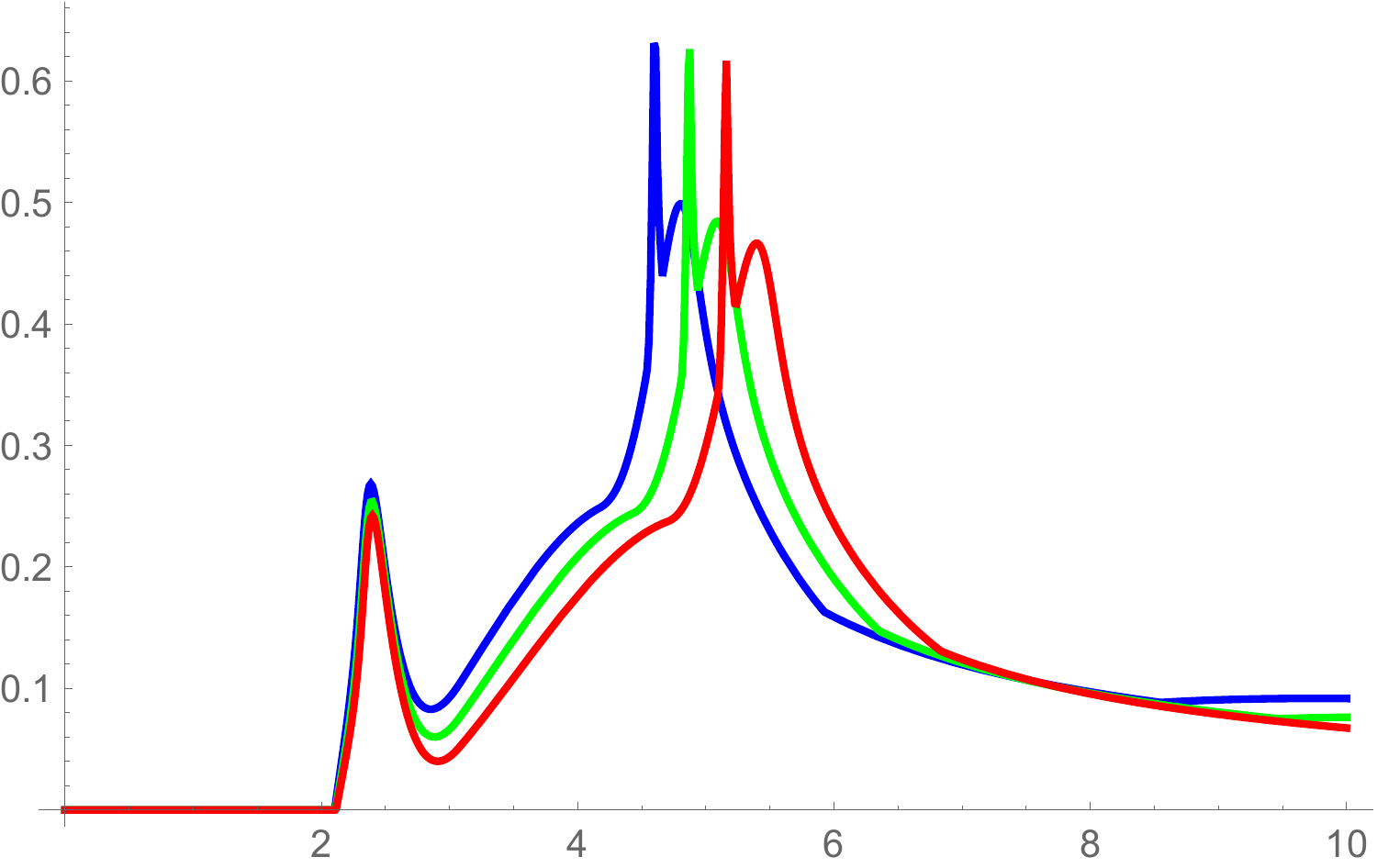}
			\put(103,2){\color{black} $b$}
			\put(0,64){\color{black} $I^{\text{obs}}/I^{0}$}
		\end{overpic}
	\end{subfigure}
	\hfill
	\begin{subfigure}[b]{0.31\textwidth}
		\centering
		\includegraphics[width=0.7\linewidth]{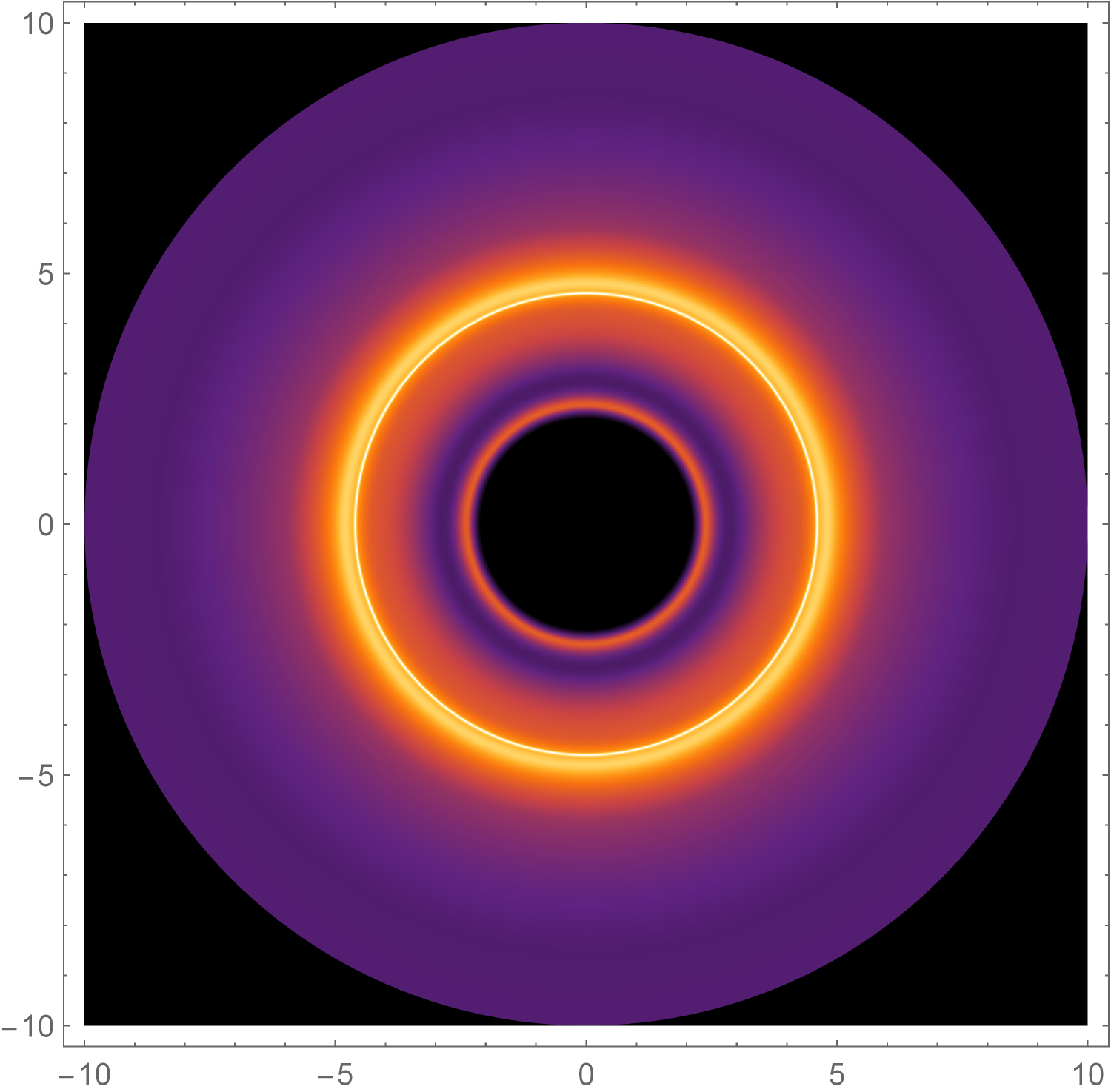}
		\hspace{0.1cm}
		\includegraphics[width=0.075\linewidth]{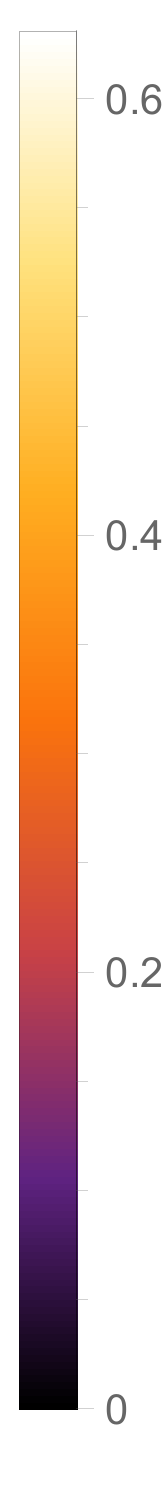}
	\end{subfigure}
	
	\caption{Observational appearance of a geometrically and optically thin disk with different emission profiles near an ABG BH coupled with a CS. The left column shows the emission profiles $I^{\text{em}}(r)$ for various cases. The middle column displays the corresponding observed intensities $I^{\text{obs}}$ as functions of the impact parameter $b$, for the ABG BH coupled with a CS, with results shown for $a = 0.02$ (blue), $a = 0.05$ (green), and $a = 0.08$ (red). The right column presents the density plots of the observed intensity $I^{\text{obs}}(b)$ for the case $a = 0.02$. The parameter $g$ is fixed at $0.6$.}
	\label{shadow1}
\end{figure*}

\begin{figure*}[htbp]
	\centering
	
	\begin{subfigure}[b]{0.31\textwidth}
		\centering
		\begin{overpic}[width=\linewidth]{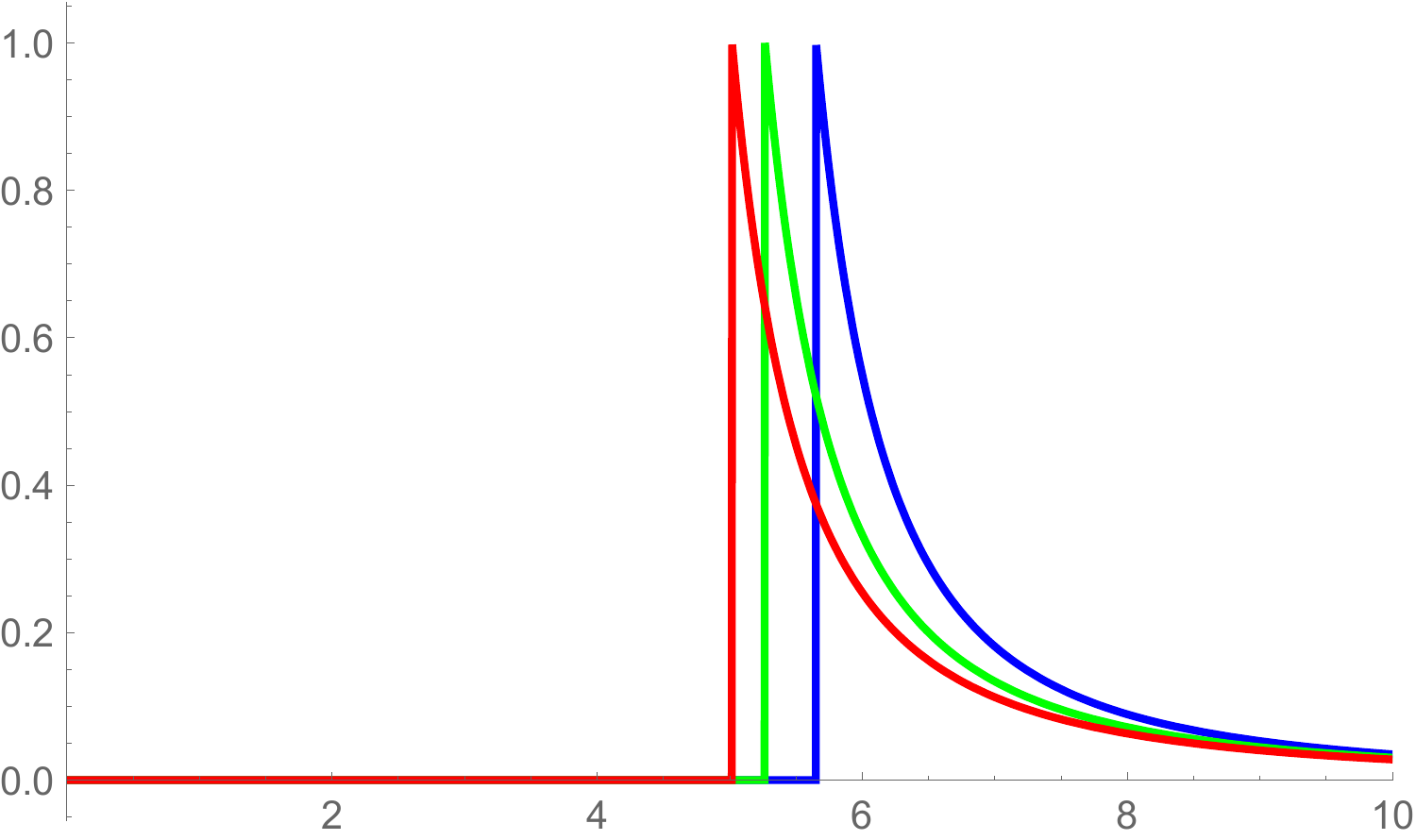}
			\put(103,2){\color{black} $b$}
			\put(0,60){\color{black} $I^{\text{em}}/I^{0}$}
		\end{overpic}
	\end{subfigure}
	\hfill
	\begin{subfigure}[b]{0.31\textwidth}
		\centering
		\begin{overpic}[width=\linewidth]{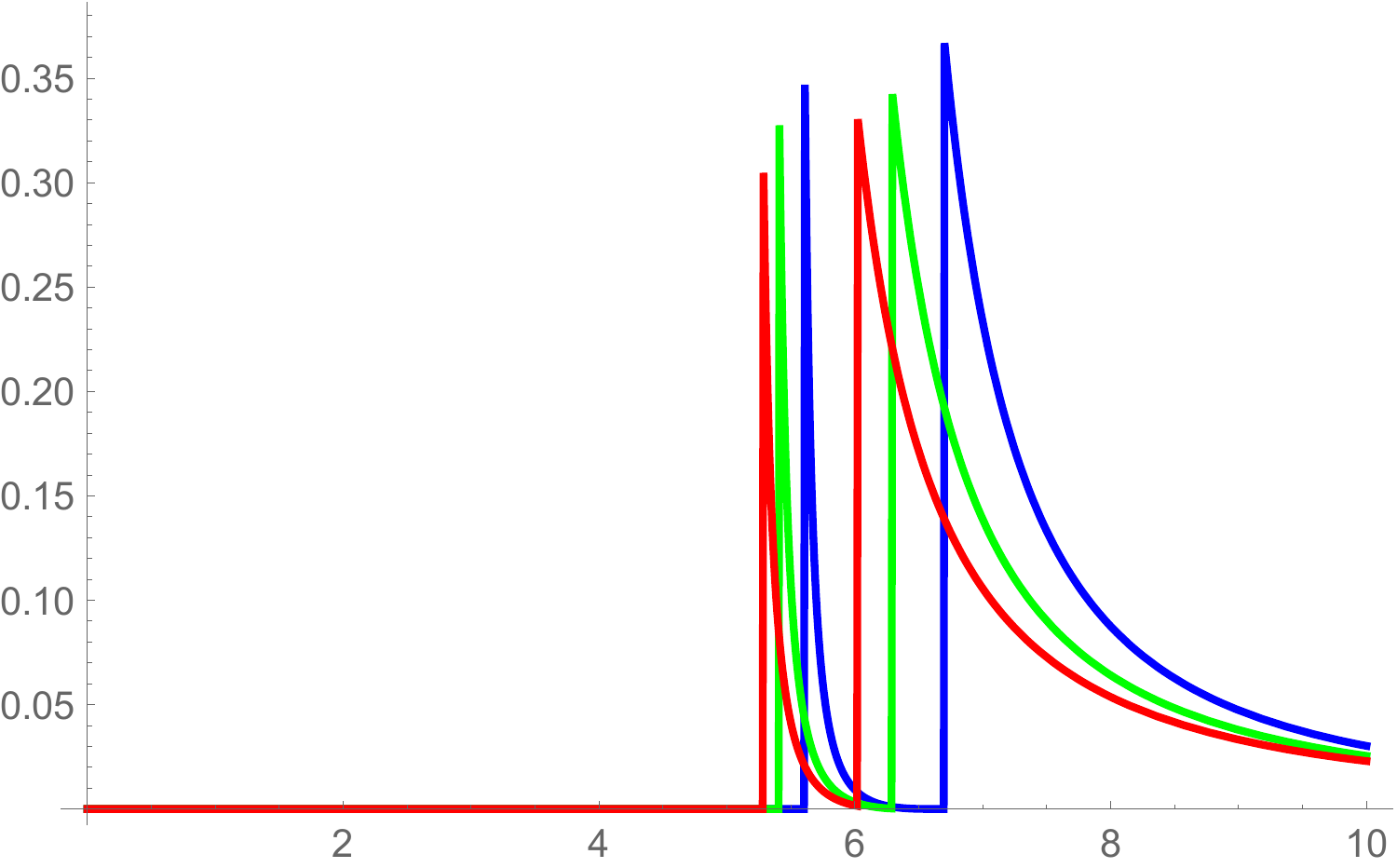}
			\put(103,2){\color{black} $b$}
			\put(0,60){\color{black} $I^{\text{obs}}/I^{0}$}
		\end{overpic}
	\end{subfigure}
	\hfill
	\begin{subfigure}[b]{0.31\textwidth}
		\centering
		\includegraphics[width=0.7\linewidth]{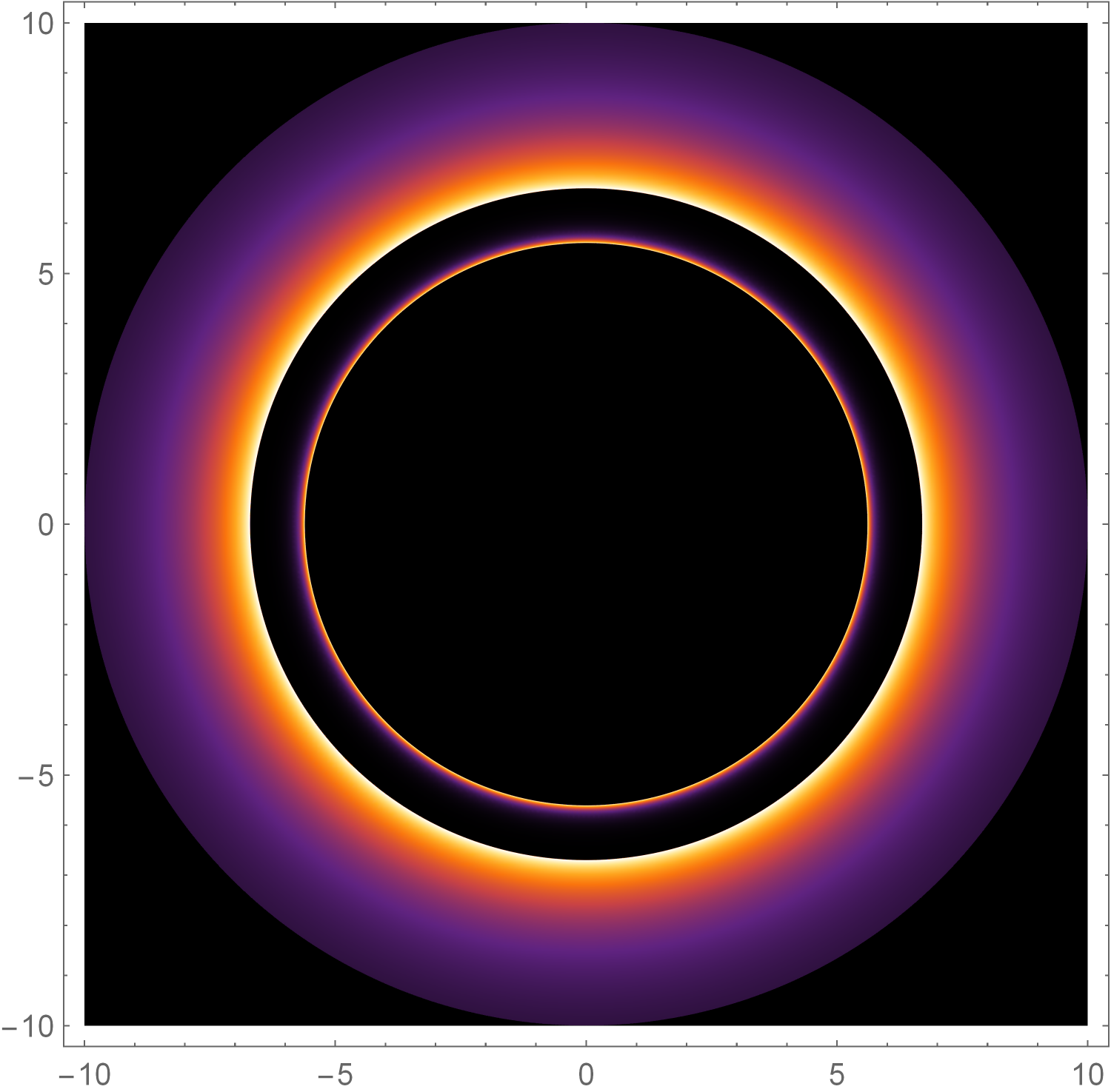}
		\hspace{0.1cm}
		\includegraphics[width=0.075\linewidth]{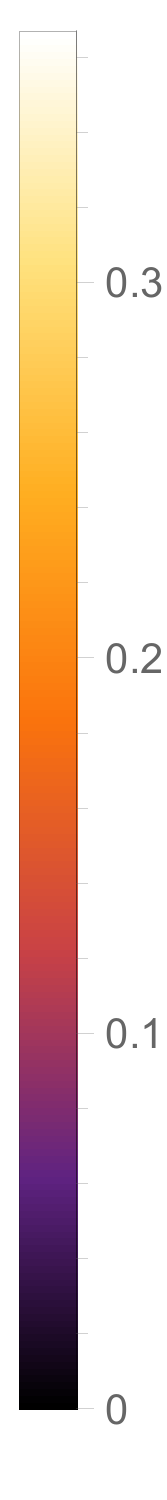}
	\end{subfigure}
	
	\vspace{0.05cm} 
	
	\begin{subfigure}[b]{0.31\textwidth}
		\centering
		\begin{overpic}[width=\linewidth]{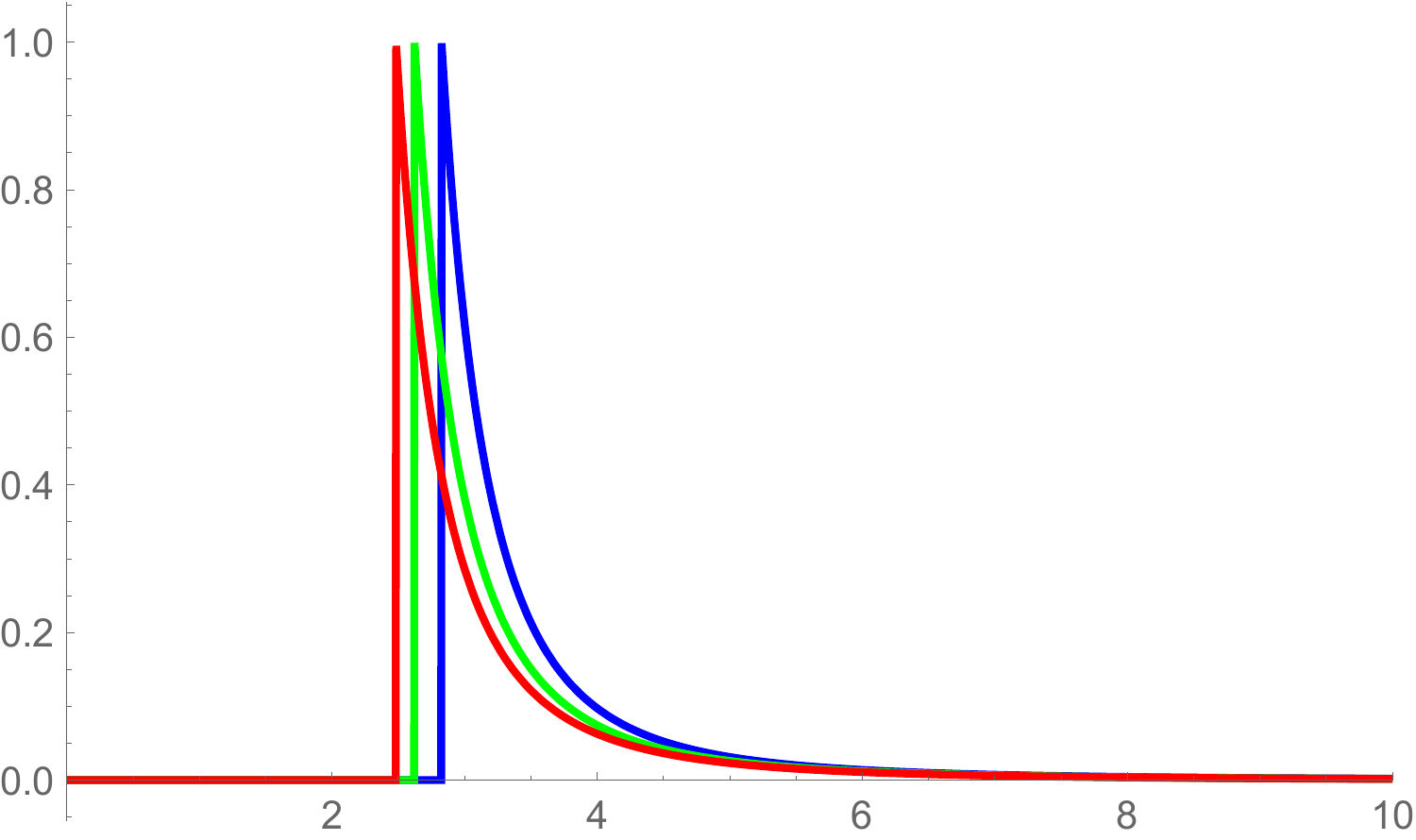}
			\put(103,2){\color{black} $b$}
			\put(0,60){\color{black} $I^{\text{em}}/I^{0}$}
		\end{overpic}
	\end{subfigure}
	\hfill
	\begin{subfigure}[b]{0.31\textwidth}
		\centering
		\begin{overpic}[width=\linewidth]{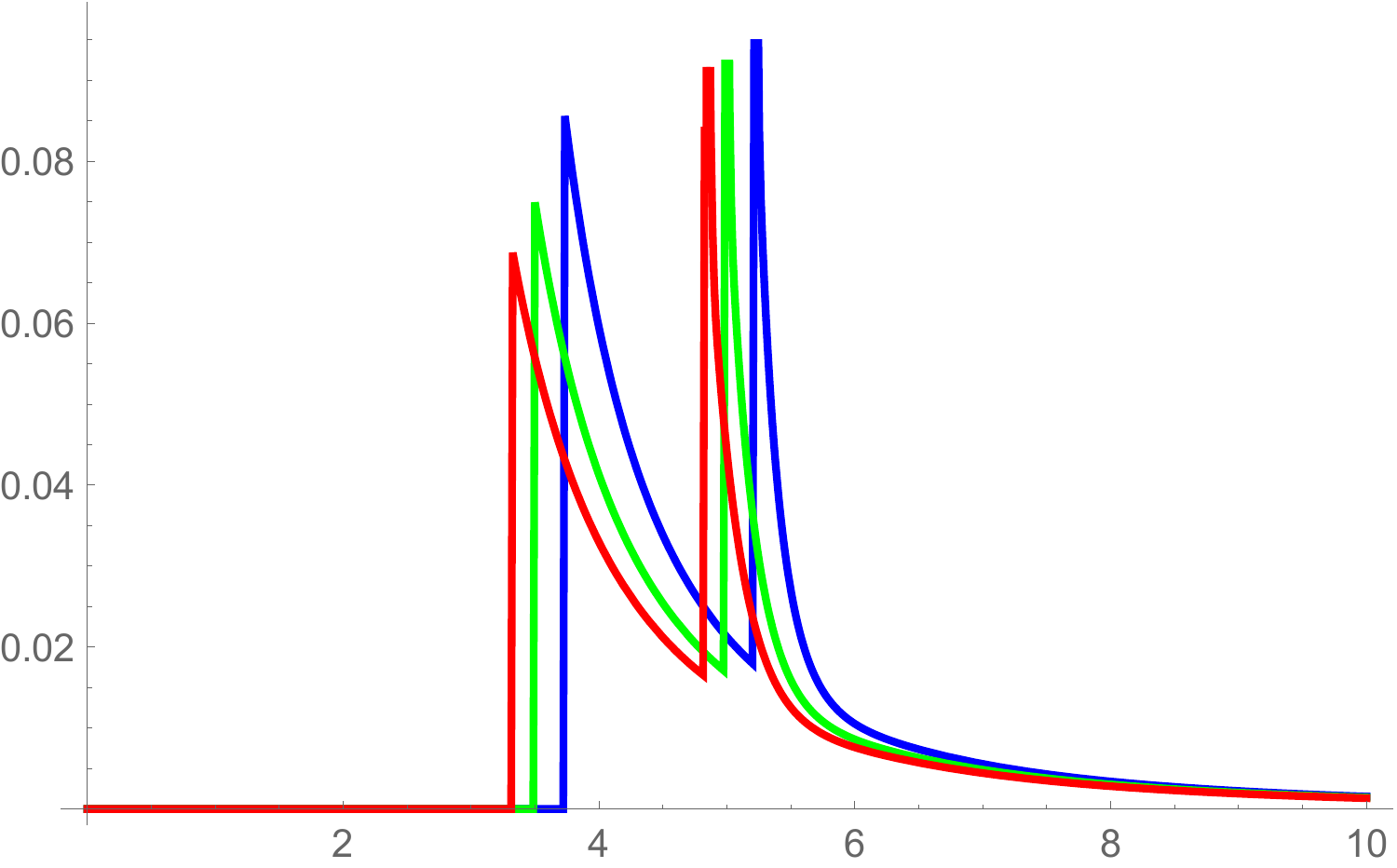}
			\put(103,2){\color{black} $b$}
			\put(0,63){\color{black} $I^{\text{obs}}/I^{0}$}
		\end{overpic}
	\end{subfigure}
	\hfill
	\begin{subfigure}[b]{0.31\textwidth}
		\centering
		\includegraphics[width=0.7\linewidth]{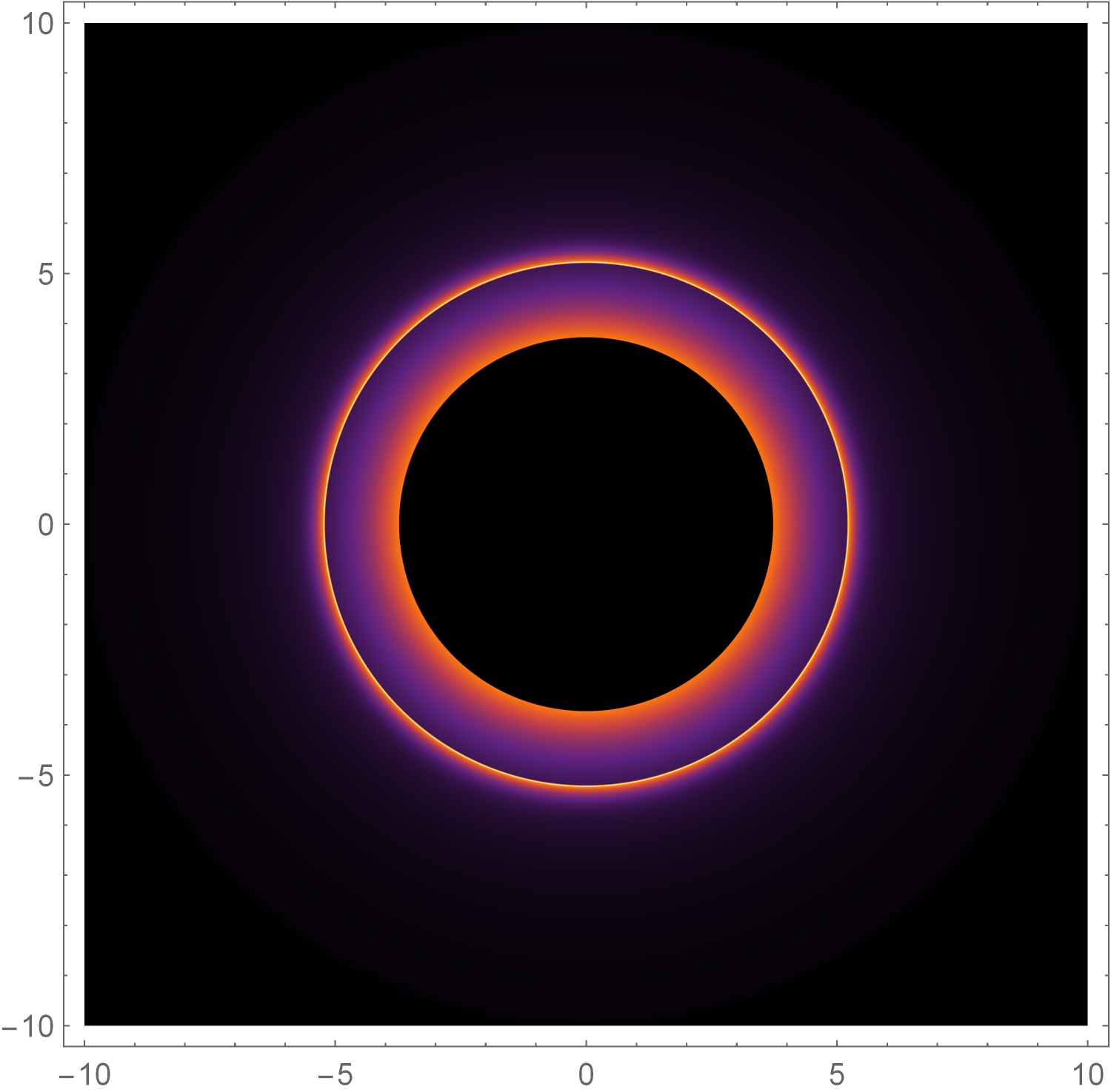}
		\hspace{0.1cm}
		\includegraphics[width=0.085\linewidth]{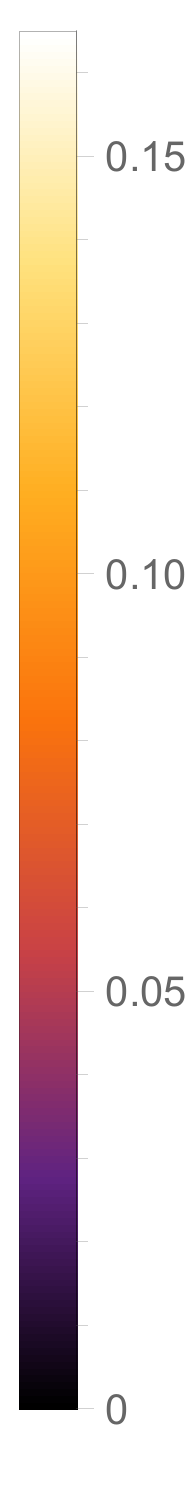}
	\end{subfigure}
	
	\vspace{0.05cm}
	
	\begin{subfigure}[b]{0.31\textwidth}
		\centering
		\begin{overpic}[width=\linewidth]{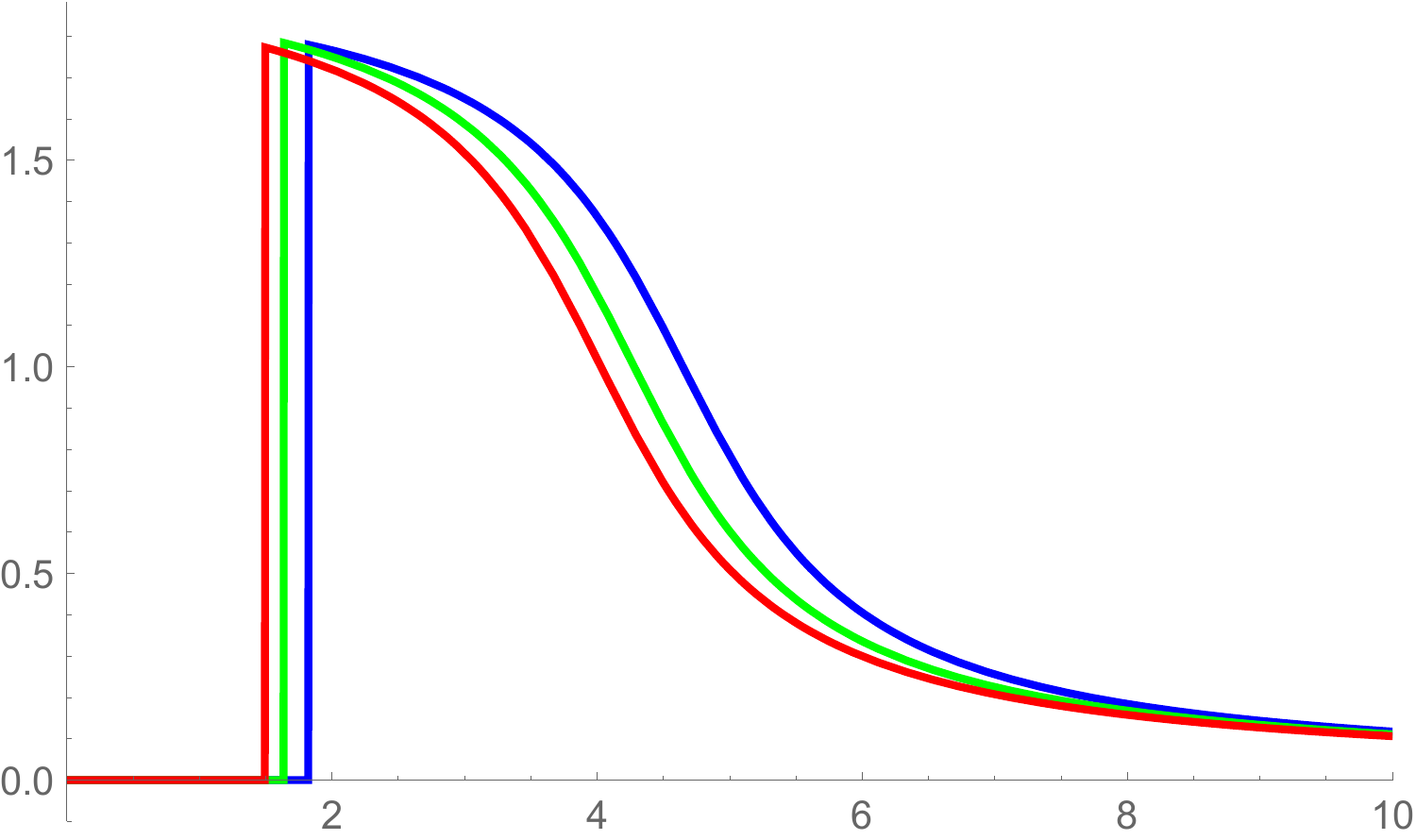}
			\put(103,2){\color{black} $b$}
			\put(0,60){\color{black} $I^{\text{em}}/I^{0}$}
		\end{overpic}
	\end{subfigure}
	\hfill
	\begin{subfigure}[b]{0.31\textwidth}
		\centering
		\begin{overpic}[width=\linewidth]{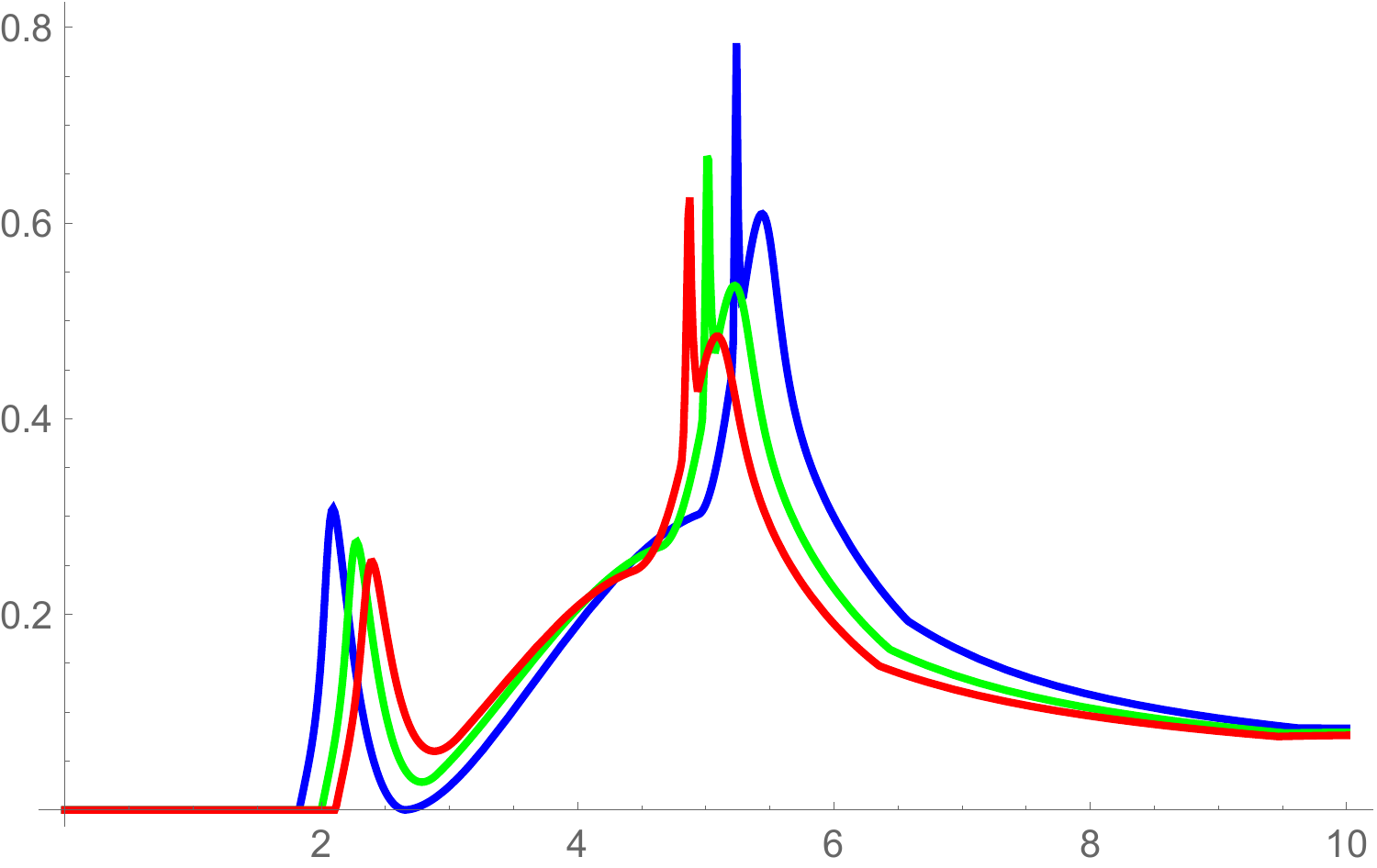}
			\put(103,2){\color{black} $b$}
			\put(0,64){\color{black} $I^{\text{obs}}/I^{0}$}
		\end{overpic}
	\end{subfigure}
	\hfill
	\begin{subfigure}[b]{0.31\textwidth}
		\centering
		\includegraphics[width=0.7\linewidth]{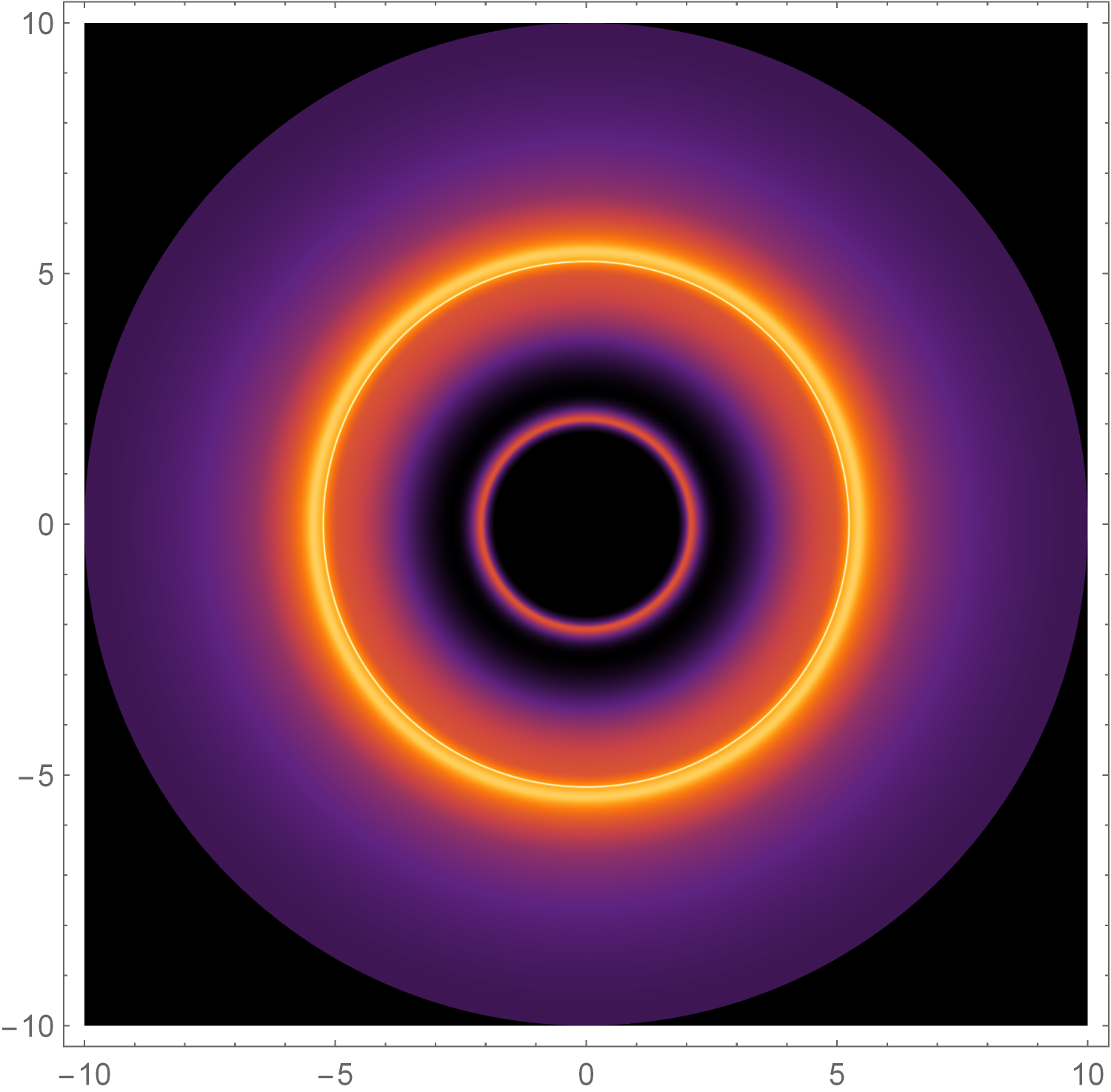}
		\hspace{0.1cm}
		\includegraphics[width=0.075\linewidth]{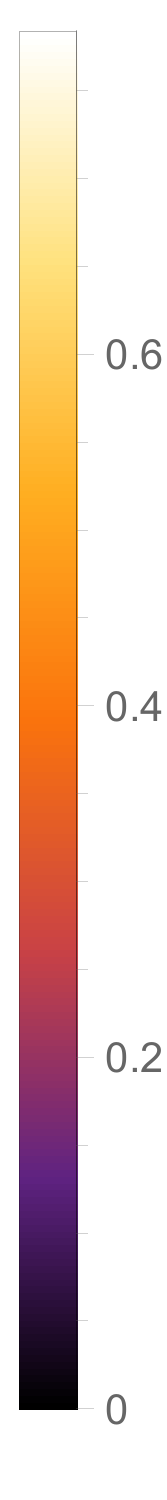}
	\end{subfigure}
	
	\caption{Observational appearance of a geometrically and optically thin disk with different emission profiles near an ABG BH coupled with a CS. The left column shows the emission profiles $I^{\text{em}}(r)$ for various cases. The middle column displays the corresponding observed intensities $I^{\text{obs}}$ as functions of the impact parameter $b$, for the ABG BH coupled with a CS, with results shown for $g = 0.45$ (blue), $g = 0.55$ (green), and $g = 0.6$ (red). The right column presents the density plots of the observed intensity $I^{\text{obs}}(b)$ for the case $g = 0.45$. The parameter $a$ is fixed at $0.05$.}
	\label{shadow2}
\end{figure*}

We denote the solution of the orbit equation by $ u(\phi, b) $ and focus on the first three transfer functions, which can be derived from the trajectory intersections with the emission plane:
\begin{equation}
	\label{17}
	\left\{
	\begin{aligned}
		r_{1}(b) &= \frac{1}{u\left( \frac{\pi}{2}, b \right)}, & b &\in (b_{1}^{-}, \infty), \\
		r_{2}(b) &= \frac{1}{u\left( \frac{3\pi}{2}, b \right)}, & b &\in (b_{2}^{-}, b_{2}^{+}), \\
		r_{3}(b) &= \frac{1}{u\left( \frac{5\pi}{2}, b \right)}, & b &\in (b_{3}^{-}, b_{3}^{+}).
	\end{aligned}
	\right.
\end{equation}
The first three transfer functions are illustrated in Fig.~\ref{zhuanyi}. The first transfer function corresponds to the ``direct image'' of the disk, which essentially reflects the redshifted emission from the source. The second represents a highly demagnified view of the disk's far side, known as the ``lensing ring''. The third captures an extremely demagnified image of the near side of the disk, referred to as the ``photon ring''.
To investigate the observational appearance of emission, it is necessary to define the emissivity 
$I_{em}$. We now consider the following three specific forms of emissivity:
\begin{align}
	I_{1}^{\text{em}}(r) &:= 
	\begin{cases}
		I^{0} \left[ \frac{1}{r - (r_{\text{isco}} - 1)} \right]^2, & r > r_{\text{isco}} \\
		0, & r \leq r_{\text{isco}}
	\end{cases} \\
	I_{2}^{\text{em}}(r) &:= 
	\begin{cases}
		I^{0} \left[ \frac{1}{r - (r_{\text{ph}} - 1)} \right]^3, & r > r_{\text{ph}} \\
		0, & r \leq r_{\text{ph}}
	\end{cases} \\
	I_{3}^{\text{em}}(r) &:= 
	\begin{cases}
		I^{0}\frac{\frac{\pi}{2}-\arctan[r-(r_{isco}-1)]}{\frac{\pi}{2}-\arctan[r_{h}-(r_{isco}-1)]}, & r > r_{\text{h}} \\
		0. & r \leq r_{\text{h}}
	\end{cases} 
\end{align}
Here, $I^{0}$ denotes the maximum value of the emitted intensity. Figures~\ref{shadow1} and~\ref{shadow2} display the observational appearances of the thin disk near black holes with three distinct matter distributions. The numerically computed shadow radii $(r_{sh})$ are summarized in Table~\ref{tab5}. As shown in Table~\ref{tab5} and Fig.~\ref{shadow1}, within the thin disk accretion model, an increase in the parameter 
$a$ leads to a larger black hole shadow. In contrast, Fig.~\ref{shadow2} indicates that a higher value of the parameter $g$ reduces the shadow size.

\begin{table}[htbp]
	\centering
	\caption{The numerical values of shadow radius $r_{sh}$ for different parameter combinations.}
	\label{tab5}
	\resizebox{\linewidth}{!}{
		\begin{tabular}{ccccccccc}
			\toprule
			\multicolumn{1}{c}{Fixed $g=0.6$} & $a$ & 0.02 & 0.03 & 0.04 & 0.05 & 0.06 & 0.07 & 0.08 \\
			\midrule
			$r_{sh}$                          &     & 4.57926 & 4.66687 & 4.75611 & 4.84744 & 4.94095 & 5.03674 & 5.13489 \\
			\midrule
			\multicolumn{1}{c}{Fixed $a=0.05$} & $g$ & 0.45 & 0.48 & 0.50 & 0.52 & 0.55 & 0.58 & 0.60 \\
			\midrule
			$r_{sh}$                          &     & 5.22535 & 5.16497 & 5.12116 & 5.07421 & 4.99720 & 4.91099 & 4.84744 \\
			\bottomrule
		\end{tabular}
	}
\end{table}
\section{The physical properties and image of the thin accretion disk}
\label{section4}
\subsection{The physical properties of thin accretion disk}
\begin{figure*}[htbp]
	\centering
	\begin{subfigure}{0.45\textwidth}
		\includegraphics[width=3in, height=5.5in, keepaspectratio]{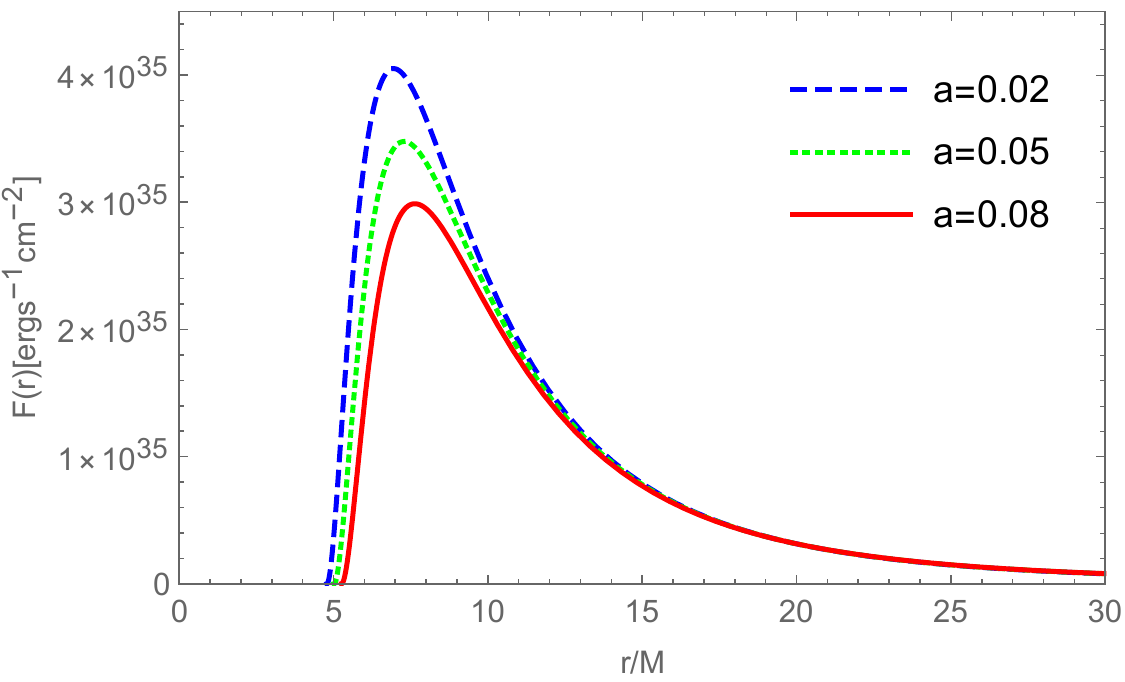}
	\end{subfigure}
	\hfill
	\begin{subfigure}{0.45\textwidth}
		\includegraphics[width=3in, height=5.5in, keepaspectratio]{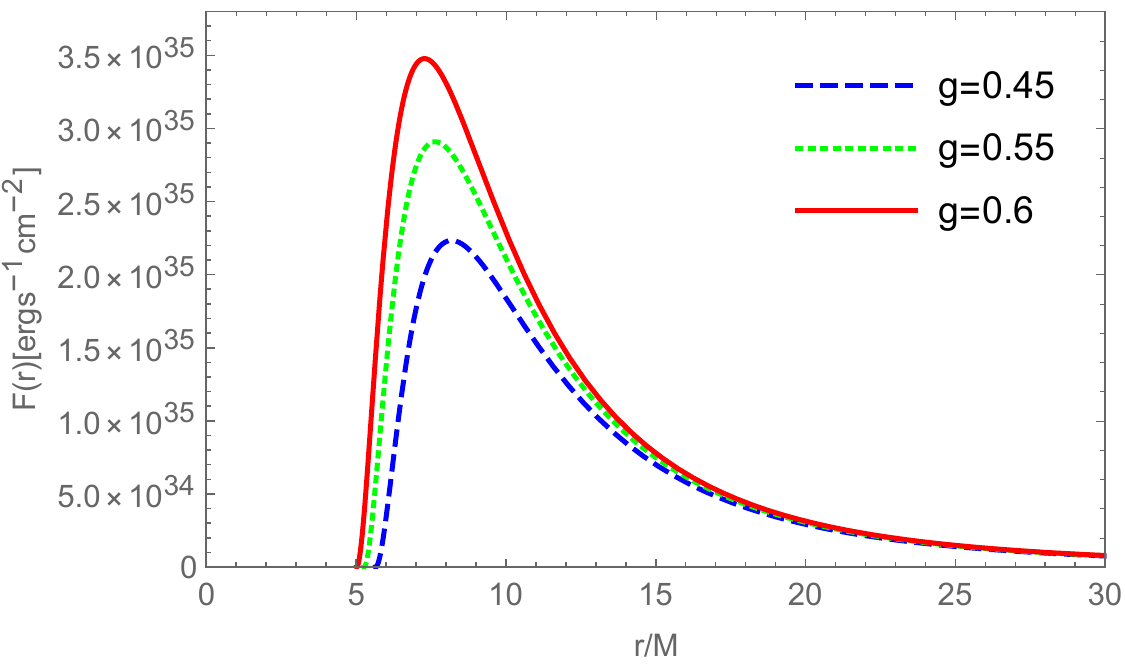}
	\end{subfigure}
	
	\caption{The radial profile of the energy flux $F(r)$ from a thin accretion disk surrounding an ABG BH coupled with a CS, shown for various values of the parameters $g$ and $a$. Left panel: $g=0.6$ is fixed; Right panel: $a=0.05$ is fixed.}
	\label{F}
\end{figure*}

\begin{figure*}[htbp]
	\centering
	\begin{subfigure}{0.45\textwidth}
		\includegraphics[width=3in, height=5.5in, keepaspectratio]{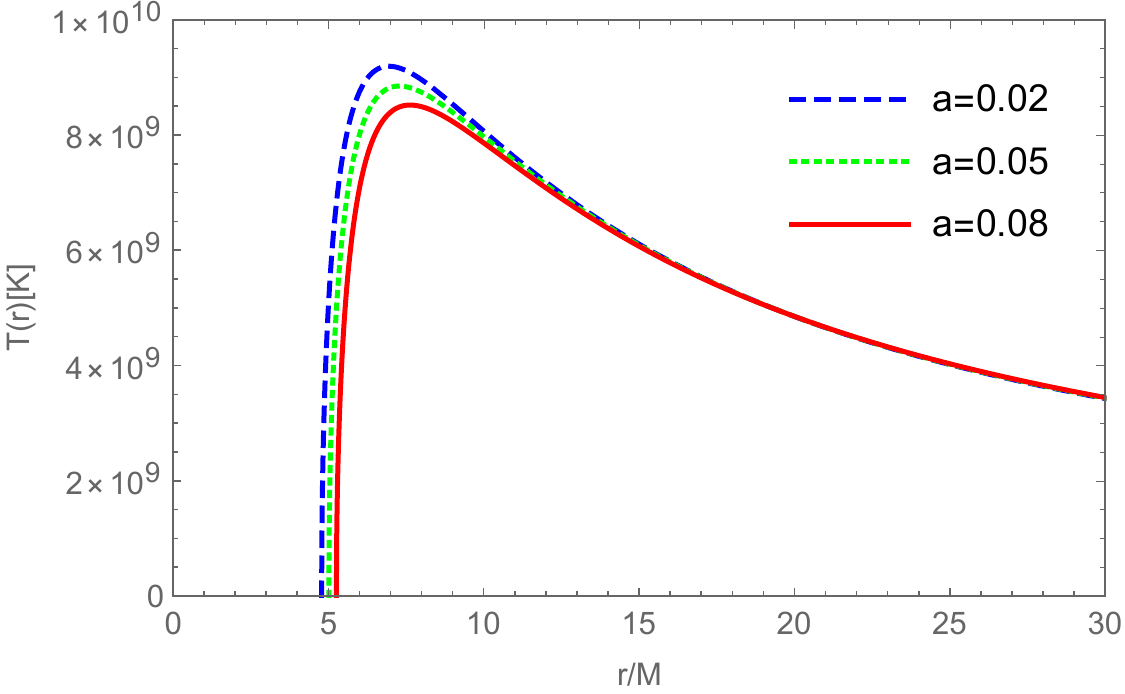}
	\end{subfigure}
	\hfill
	\begin{subfigure}{0.45\textwidth}
		\includegraphics[width=3in, height=5.5in, keepaspectratio]{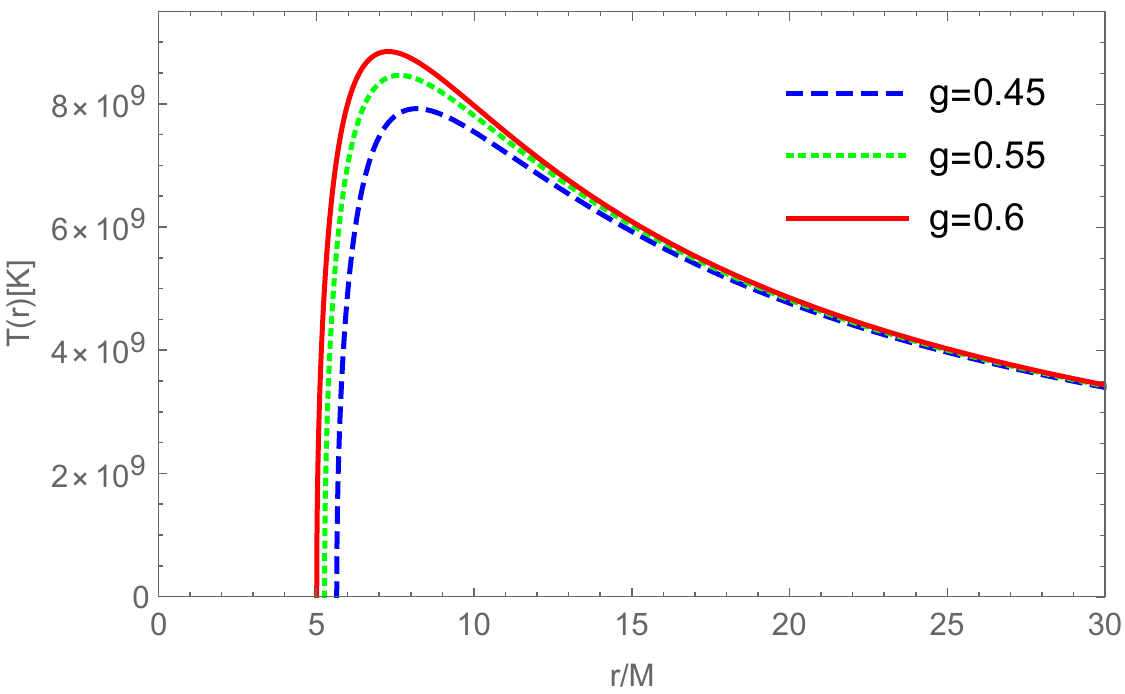}
	\end{subfigure}
	
	\caption{Variation of the disk radiation temperature $T(r)$ as a function of radius, for different values of the parameters $g$ and $a$, for a thin accretion disk around an ABG BH coupled with a CS. Left panel: $g=0.6$ is fixed; Right panel: $a=0.05$ is fixed.}
	\label{T}
\end{figure*}

In this subsection, we investigate the accretion mechanisms in thin disks around an ABG BH coupled with CS. A comprehensive analysis is presented on the influence of the parameters 
$g$ and $\alpha$ on the emitted energy flux, radiation temperature, and observable luminosity. The following physical constants and disk properties are employed in our calculations: $c=2.997\times10^{10}\text{cms}^{-1}$, $\dot{M}_{0}=2\times10^{-6}M_{\odot}\text{yr}^{-1}$, $1\text{yr}=3.156\times10^{7}\text{s}$, $\sigma_{\text{SB}}=5.67\times10^{-5}\text{ergs}^{-1}\text{cm}^{-2}\text{K}^{-4}$, $h=6.625\times10^{-27}\text{ergs}$, $k_{\text{B}}=1.38\times10^{-16}\text{erg}\text{K}^{-1}$, $M_{\odot}=1.989\times10^{33}\text{g}$, and the mass of BH $M=2\times10^{6}M_{\odot}$. The surface energy flux is calculated using the formalism developed by Page and Thorne \cite{Page:1974he}
\begin{equation}
	\label{21}
	F(r)=-\frac{\dot{M}_{0}\Omega_{,r}}{4\pi\sqrt{-\tilde{g}}(E-\Omega L)^{2}}\int^{r}_{r_{isco}}(E-\Omega L)L_{,r}dr.
\end{equation}
In the above expression, $\dot{M}_{0}$ represents the mass accretion rate, and $\tilde{g}$ denotes the determinant of the metric tensor. The quantities $E$, $L$, and $\Omega$ correspond to the specific energy, angular momentum, and angular velocity of a particle in a circular orbit, respectively. Figure.~\ref{F} displays the radial profile of the emitted energy flux $F(r)$ from a thin disk around an ABG BH coupled with a CS, for varying values of the parameters $a$ and $g$. It is observed that, for a fixed value of $g$, the energy flux $F(r)$ decreases as $a$ increases. Conversely, for a fixed $a$, the energy flux increases with increasing $g$. For a general static and spherically symmetric spacetime described by the metric $ds^{2}=g_{tt}dt^{2}+g_{rr}dr^{2}+g_{\theta\theta}d\theta^{2}+g_{\phi\phi}d\phi^{2}$, the energy $E$, angular momentum $L$, and angular velocity $\Omega$ are given by
\begin{equation}
	\label{22}
	E=-\frac{g_{tt}}{\sqrt{-g_{tt}-g_{\phi\phi}\Omega^{2}}},
\end{equation}
\begin{equation}
	\label{23}
	L=\frac{g_{\phi\phi}\Omega}{\sqrt{-g_{tt}-g_{\phi\phi}\Omega^{2}}},
\end{equation}
\begin{equation}
	\label{24}
	\Omega=\frac{d\phi}{dt}=\sqrt{-\frac{g_{tt,r}}{g_{\phi\phi,r}}}.
\end{equation}
Within the framework of the Novikov–Thorne model, the accreted matter is assumed to be in thermodynamic equilibrium. As a result, the radiation emitted by the disk closely approximates that of a black body. The local radiation temperature $T(r)$ of the disk is related to the emitted energy flux $F(r)$ through the Stefan–Boltzmann law: $F(r)=\sigma_{\text{SB}}T^{4}(r)$, where $\sigma_{\text{SB}}$ denotes the Stefan–Boltzmann constant. Figure.~\ref{T} presents the radial profile of the radiation temperature $T(r)$ for a thin accretion disk around an ABG BH coupled with a CS. It can be observed that, for a fixed value of $g$, the temperature $T(r)$ decreases as $a$ increases. Conversely, for a fixed $a$, the temperature increases with increasing $g$.

The red-shifted blackbody spectrum of the observed luminosity $L(\nu)$ emitted by a thin accretion disk around a BH is derived in \cite{Torres:2002td}
\begin{equation}
	\label{25}
	L(\nu)=4\pi d^{2}I(\nu)=\frac{8\pi h\cos\theta}{c^{2}}\int_{r_{i}}^{r_{f}}\int_{0}^{2\pi}\frac{\nu_{e}^{3}r}{e^{\frac{h\nu_{e}}{k_{B}T}}-1}drd\phi.
\end{equation}
In the above expression, $d$ denotes the distance to the center of the disk, and $I(\nu)$ represents the thermal energy flux radiated by the disk surface. Here, $h$ is the Planck constant, $k_{B}$ is the Boltzmann constant, and $\theta$ is the inclination angle of the disk with respect to the observer. For simplicity, we assume a face-on orientation and set $\theta=0$. The parameters $r_{f}$ and $r_{i}$ correspond to the outer and inner radii of the disk, respectively.

\begin{figure*}[htbp]
	\centering
	\begin{subfigure}{0.45\textwidth}
		\includegraphics[width=3in, height=5.5in, keepaspectratio]{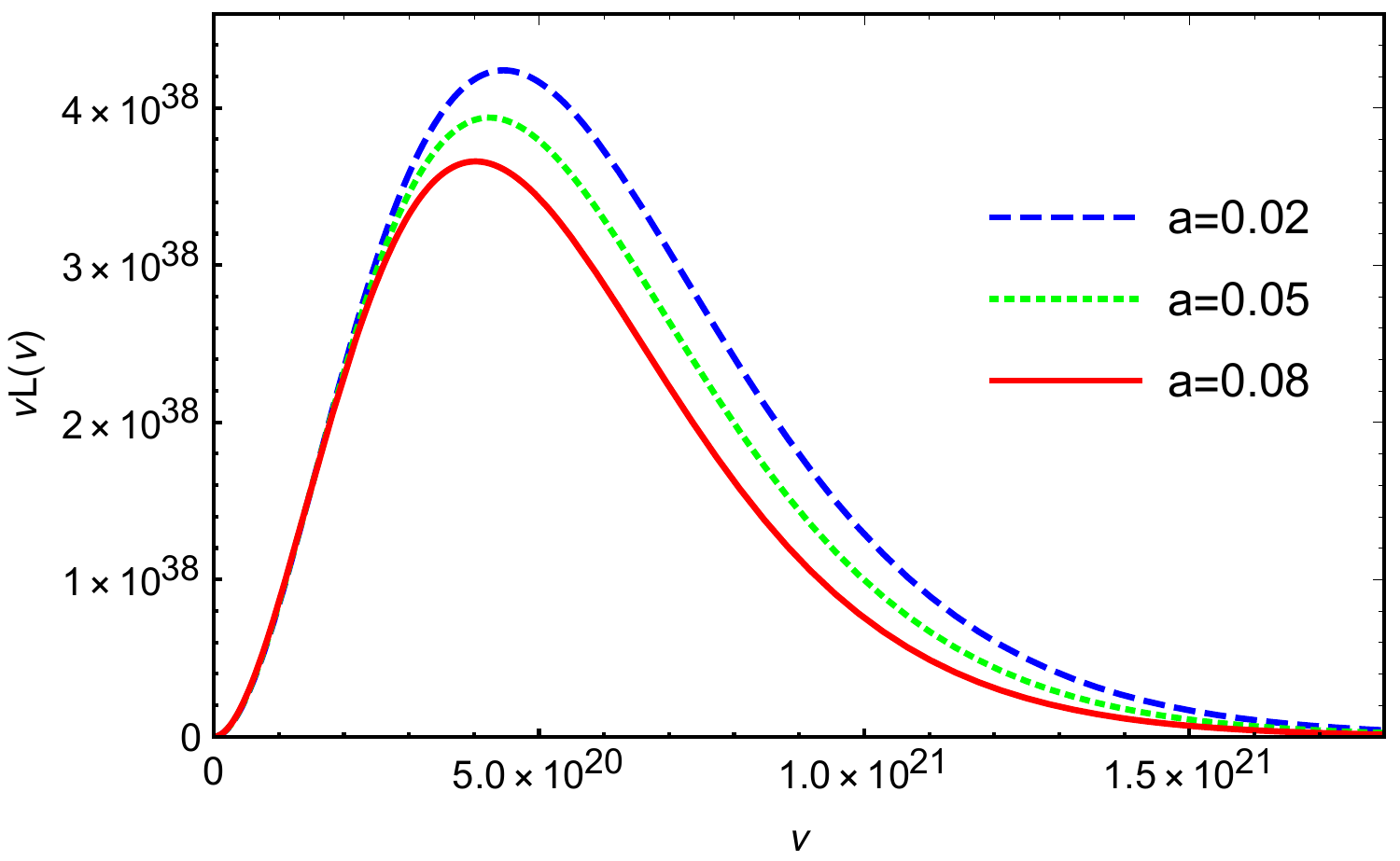}
	\end{subfigure}
	\hfill
	\begin{subfigure}{0.45\textwidth}
		\includegraphics[width=3in, height=5.5in, keepaspectratio]{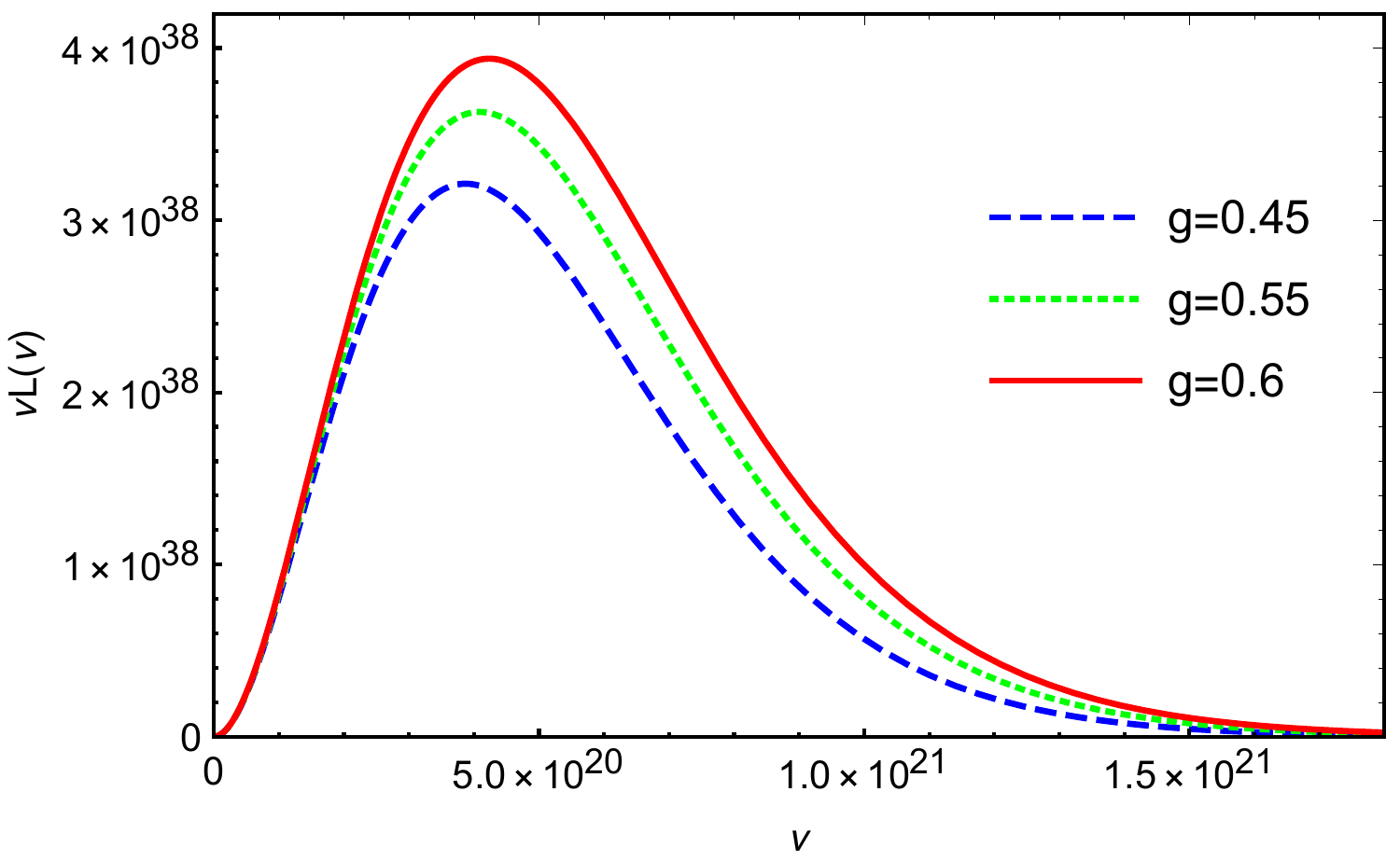}
	\end{subfigure}
	
	\caption{The emission spectrum $\nu L(\nu)$ for a thin accretion disk surrounding an ABG BH coupled with a CS, shown for different values of the parameters $g$ and $a$. Left panel: $g=0.6$ is fixed; Right panel: $a=0.05$ is fixed.}
	\label{vLv}
\end{figure*}

Under the assumption that the radiation flux vanishes at the disk edges, we adopt $r_{i}=r_{isco}$ and take the limit $r_{f}\rightarrow\infty$ when computing the total observed luminosity $L(\nu)$. The emitted frequency is related to the observed frequency via $\nu_{e}=\nu(1+z)$, where $z$ is the gravitational redshift factor. This redshift factor can be expressed as follows \cite{Luminet:1979nyg}:
\begin{equation}
	\label{26}
	1+z=\frac{1+\Omega b\sin\theta\cos\alpha}{\sqrt{-g_{tt}-g_{\phi\phi}\Omega^{2}}},
\end{equation}
we denote the angle $\alpha$ appearing in Eq. (17) of Ref. \cite{Luminet:1979nyg} as $\alpha^{\prime}$. The relationship between $\alpha$ and $\alpha^{\prime}$ is defined by $\alpha=\frac{\pi}{2}-\alpha^{\prime}$. For simplicity and clarity in the present analysis, we neglect the effects of light bending, following the approach adopted in \cite{Bhattacharyya:2000kt}. Consequently, the expression for the impact parameter becomes $b\cos\alpha = r\sin\phi$ and the redshift factor $z$ can be reformulated as follows:
\begin{equation}
	\label{27}
	1+z=\frac{1+\Omega r\sin\theta\sin\phi}{\sqrt{-g_{tt}-g_{\phi\phi}\Omega^{2}}}.
\end{equation}
Figure.~\ref{vLv} illustrates the variations in the spectral energy distribution, which follow a similar trend to those observed in the energy flux and disk temperature.
\subsection{Observation coordinate system}
\begin{figure}[htbp]
	\centering
	\begin{subfigure}{0.45\textwidth}
		\includegraphics[width=3.2in, height=5.5in, keepaspectratio]{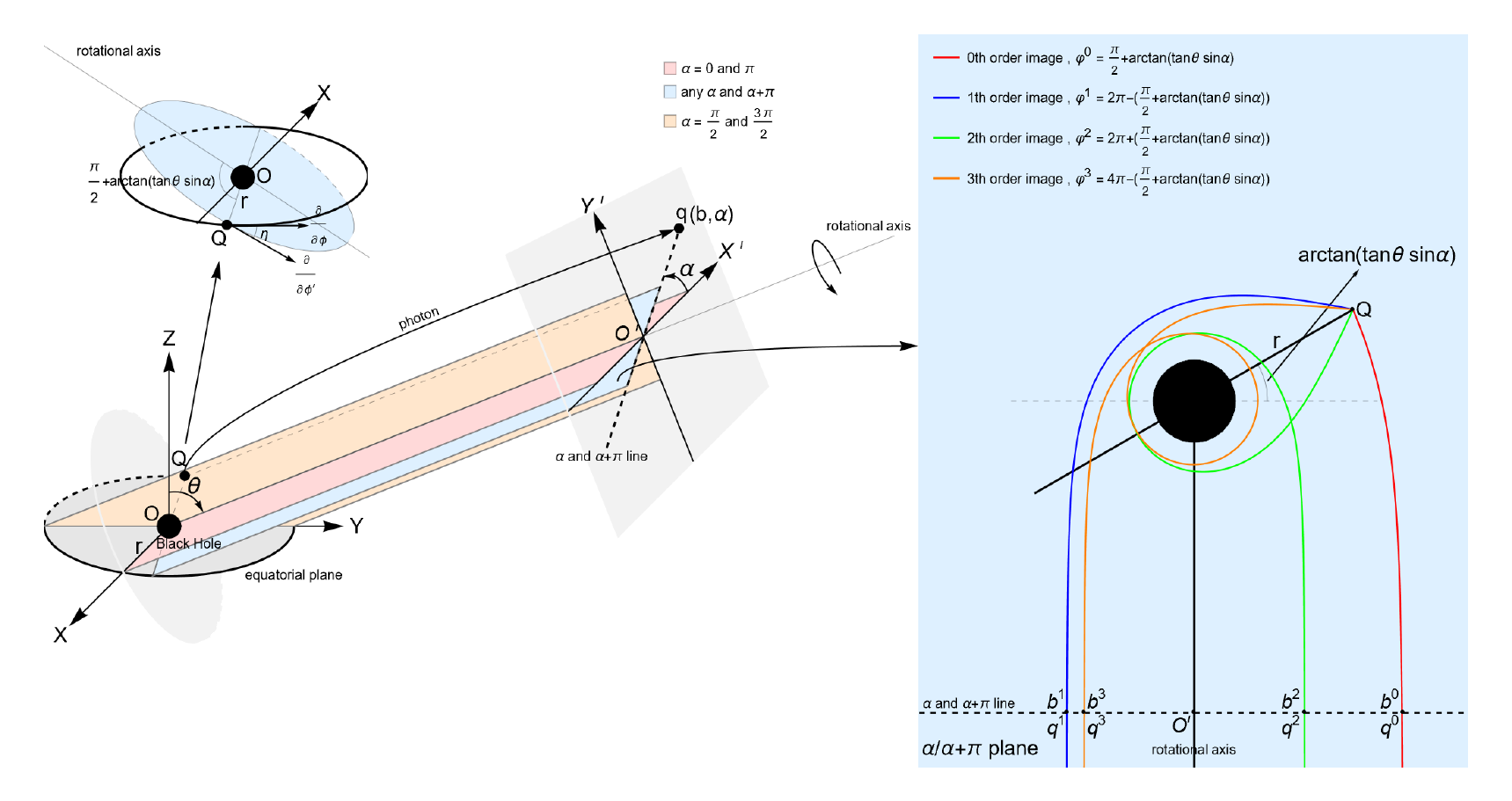}
	\end{subfigure}
	\caption{Coordinate system is indicated in Ref. \cite{You:2024uql}.}
	\label{coordinate}
\end{figure}

To investigate the imaging characteristics of a thin accretion disk, we adopt an observational coordinate system, as depicted in Fig.~\ref{coordinate}. The observer is located at $(\infty,\theta,0)$ in the BH's spherical coordinate system $(r,\theta,\phi)$, where the origin is set at the center of the BH $(r=0)$.

Within the observer-based coordinate system $O^{\prime}X^{\prime}Y^{\prime}$, consider a photon originating from point $q(b,\alpha)$ and propagating vertically. Here, $b$ denotes the photon's impact parameter. This photon intersects the accretion disk at point $Q(r,\frac{\pi}{2},\phi)$. By invoking the principle of optical path reversibility, a photon emitted from point $Q(r,\frac{\pi}{2},\phi)$ on the accretion disk will trace a trajectory that ultimately reaches the image point $q(b,\alpha)$ in the observer's field of view.

When the radial distance $r$ is held constant, the corresponding image represents a circular orbit of fixed radius. As illustrated on the left side of Fig.~\ref{coordinate}, each $\alpha/\alpha+\pi$ plane intersects the equatorial, constant-$r$ orbit at two distinct points, which differ in their azimuthal angle $\phi$ by exactly $\pi$.

In the adopted coordinate system, the $X^{\prime}$-axis is defined by setting $\alpha=0$, while the $X$-axis aligns with $\phi=0$. Geometrically, this configuration enables the determination of the angle $\varphi$ formed between the rotational axis and the line segment $OQ$
\begin{equation}
	\label{28}
	\varphi=\frac{\pi}{2}+\arctan(\tan\theta\sin\alpha).
\end{equation}
As the impact parameter $b$ approaches the critical value $b_{c}$, the degree of light bending increases significantly. This can lead to a single source point $Q$ on the accretion disk producing multiple corresponding image points $q$ in the observer's sky. These image points are indexed according to the increasing order of their azimuthal angles $\varphi$ as $q^{n}$, where $(n\in\mathbb{N})$ and $n$ represents the image order. As shown in the right panel of Fig.~\ref{coordinate}, all even-order images of $Q$ appear on the same side $(\alpha)$ as the source point itself, whereas all odd-order images are located on the opposite side $(\alpha+\pi)$. The angular deflections associated with the formation of the $n^{th}$-order image are denoted by $\varphi^{n}$
\begin{equation} 
	\label{29}
	\varphi^{n}= 
	\begin{cases} 
		\frac{n}{2}2\pi+(-1)^{n}[\frac{\pi}{2}+\arctan(\tan\theta\sin\alpha)], & \text{when $n$ is even,} \\
		\frac{n+1}{2}2\pi+(-1)^{n}[\frac{\pi}{2}+\arctan(\tan\theta\sin\alpha)], & \text{when $n$ is odd.}
	\end{cases}
\end{equation}
The parameter $n$ denotes the image order observed in the lensed accretion disk. Specifically, the case $n=0$ corresponds to the primary (or direct) image, which represents the unobstructed view of the disk as seen by the observer. Higher-order images, with $n=1$, $2$, $3$..., correspond to successive lensed replicas of the source, referred to as secondary, tertiary, and higher-order images, respectively.
\subsection{Direct and secondary images of BH}
\begin{figure*}[htbp]
	\centering
	\begin{subfigure}{0.3\textwidth}
		\includegraphics[width=\textwidth, keepaspectratio]{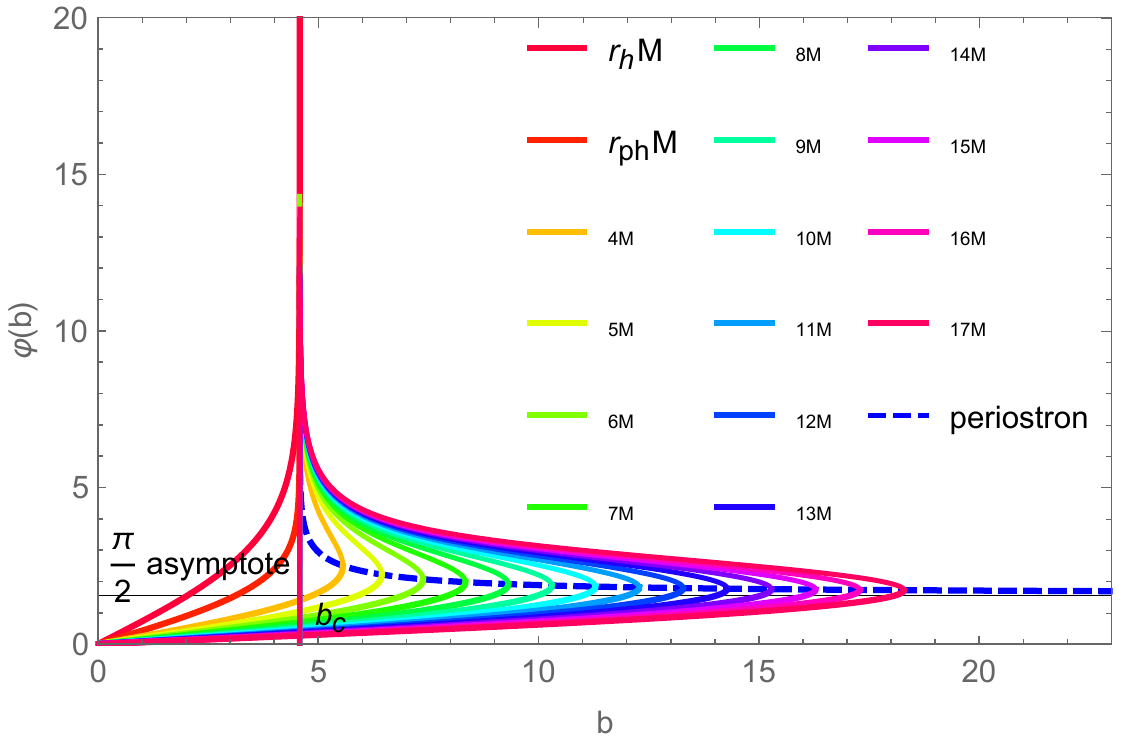}
	\end{subfigure}
	\hfill
	\begin{subfigure}{0.3\textwidth}
		\includegraphics[width=\textwidth, keepaspectratio]{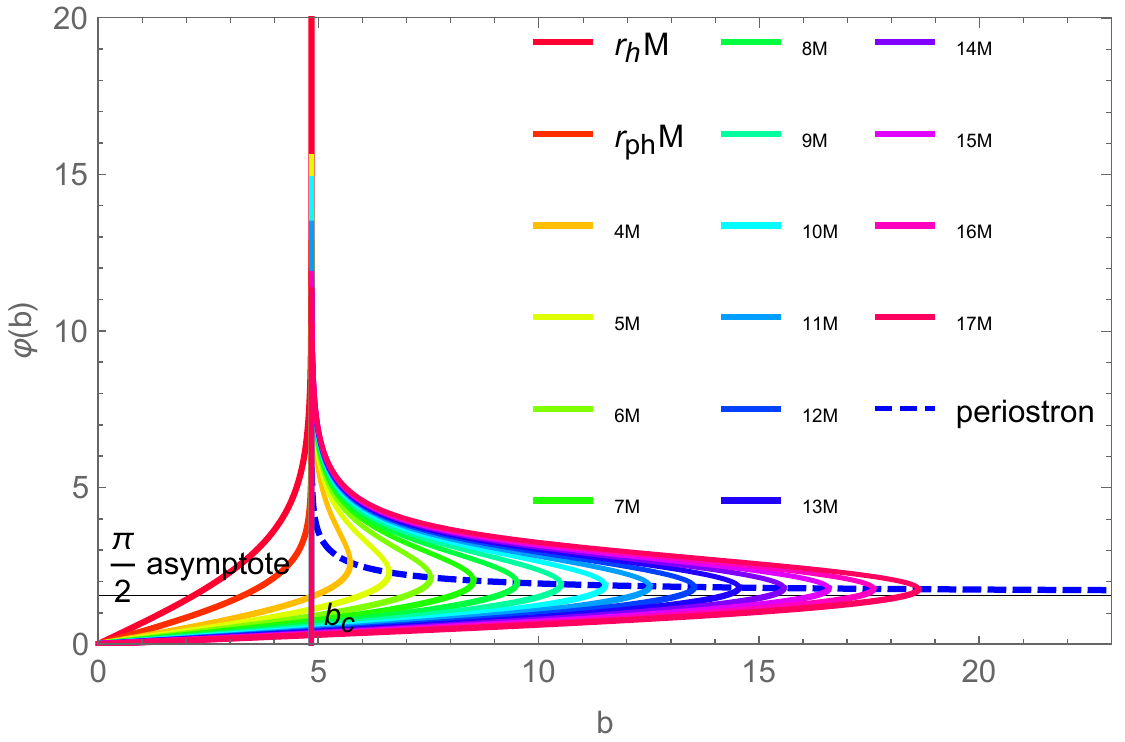}
	\end{subfigure}
	\hfill
	\begin{subfigure}{0.3\textwidth}
		\includegraphics[width=\textwidth, keepaspectratio]{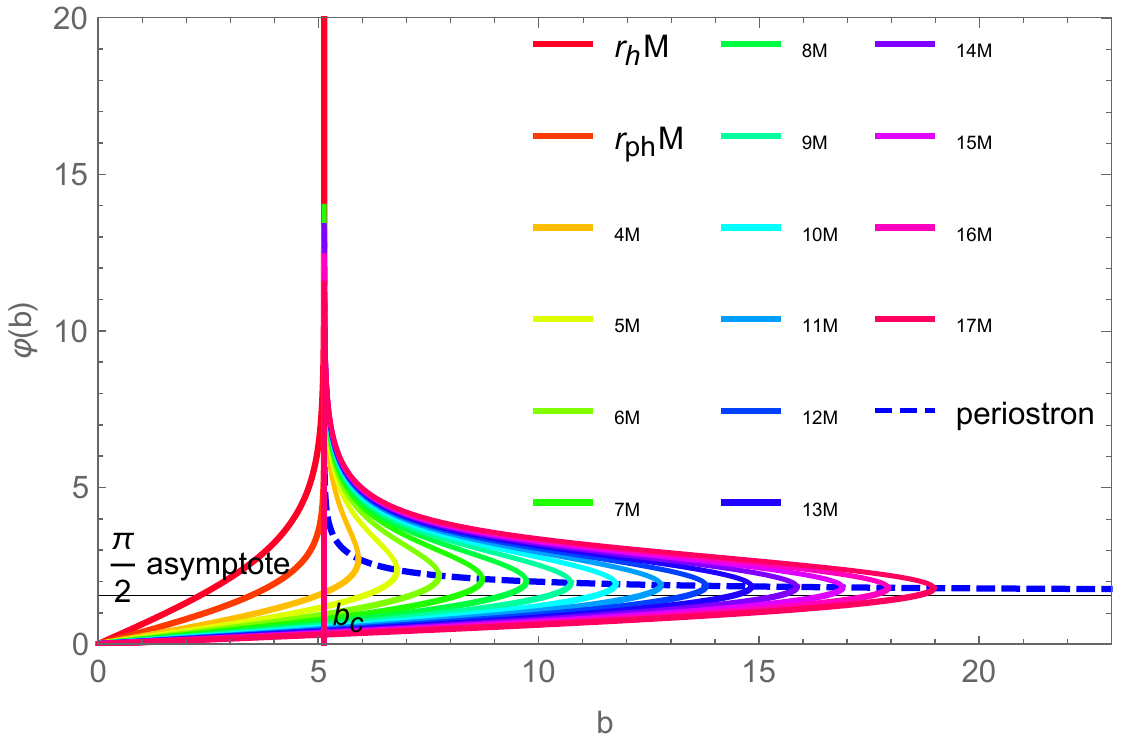}
	\end{subfigure}
	
	\caption{Deflection angle $\varphi$ as a function of the impact parameter $b$ for different radii $r$. For the left panel, we set $g = 0.6, a = 0.02$; for the middle panel, $g = 0.6, a = 0.05$; and for the right panel, $g = 0.6, a = 0.08$.}
	\label{faib}
\end{figure*}

Photons arriving from spatial infinity with different impact parameters $b$ intersect the equatorial circular orbit at distinct points. Figure.~\ref{faib} illustrates the function $\varphi(b)$, which characterizes the angular deflection as a function of $b$. As shown in the figure, for a fixed value of $g=0.6$, the curve $\varphi(b)$ shifts to the right as $a$ increases. The blue dashed line in the plot represents the reference curve $\varphi_{1}(b)$. Using this line as a boundary, the curves lying below it are labeled as $\varphi_{2}(b)$, while those above it are denoted as $\varphi_{3}(b)$. Accordingly, we may define:
\begin{equation}
	\label{30}
	\varphi_{1}(b)=\int_{0}^{u_{min}}\frac{1}{\sqrt{G(u)}}du,
\end{equation}
\begin{equation}
	\label{31}
	\varphi_{2}(b)=\int_{0}^{u_{r}}\frac{1}{\sqrt{G(u)}}du,
\end{equation}
\begin{equation}
	\label{32}
	\varphi_{3}(b)=2\int_{0}^{u_{min}}\frac{1}{\sqrt{G(u)}}du-\int_{0}^{u_{r}}\frac{1}{\sqrt{G(u)}}du.
\end{equation}

Each colored curve in the figures corresponds to an orbit of constant radius $r$, where each point $(b,\varphi)$ represents the angular deflection $\varphi$ experienced by a photon with impact parameter $b$ as it intersects the equatorial circular orbit at radius $r$. The blue dashed line intersects these curves at their respective maxima, marking the deflection angle at the point of closest approach—referred to as the perihelion distance $r_{pe}$. It is evident that the blue dashed line asymptotically approaches $\varphi=\frac{\pi}{2}$, which corresponds to the limiting case of photons with infinitely large impact parameters $(b\rightarrow\infty)$ traveling along nearly straight paths. These trajectories are tangent to the circular orbit at infinity $(r\rightarrow\infty)$, resulting in a deflection angle of $\varphi=\frac{\pi}{2}$.

By simultaneously solving the system of Eqs. (\ref{28}), (\ref{30}), and (\ref{31}) and employing numerical integration techniques to determine all corresponding $(b,\alpha)$ pairs, one can reconstruct the projected image of the accretion disk in the observer's plane.
\begin{figure*}[htbp]
	\centering
	
	\begin{subfigure}{0.31\textwidth}
		\includegraphics[width=\linewidth]{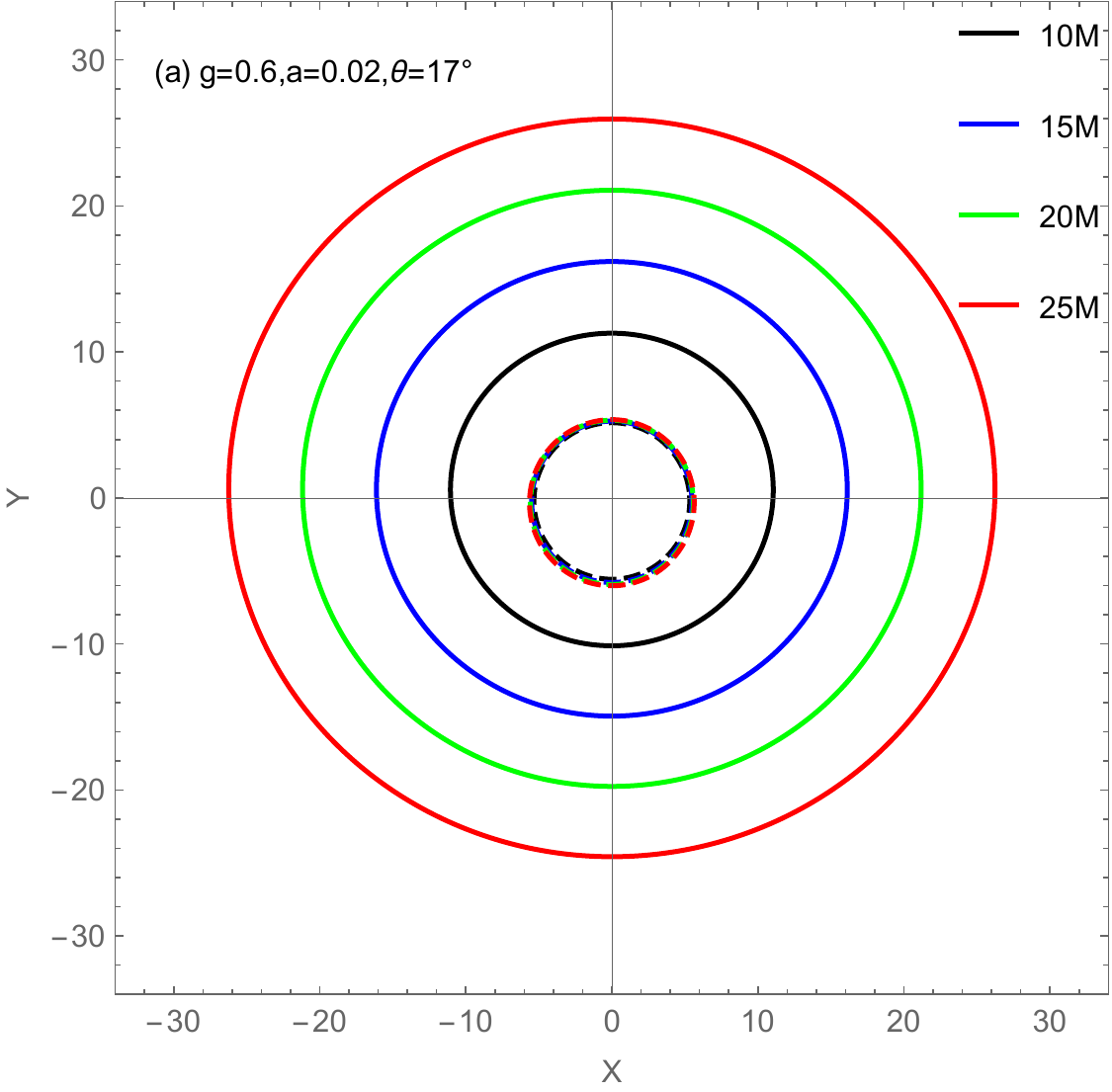}
	\end{subfigure}
	\hspace{0.02\textwidth}
	\begin{subfigure}{0.31\textwidth}
		\includegraphics[width=\linewidth]{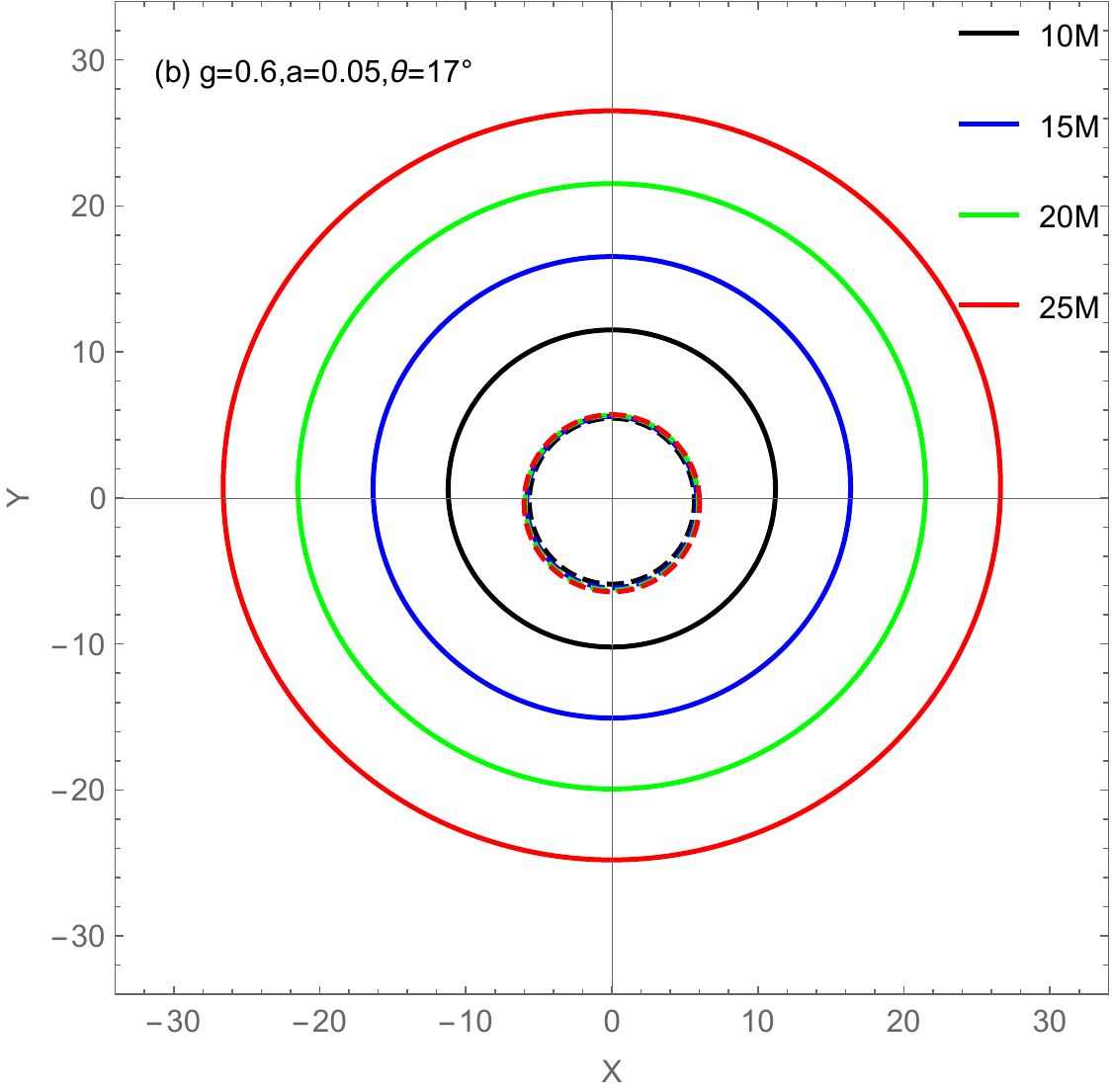}
	\end{subfigure}
	\hspace{0.02\textwidth}
	\begin{subfigure}{0.31\textwidth}
		\includegraphics[width=\linewidth]{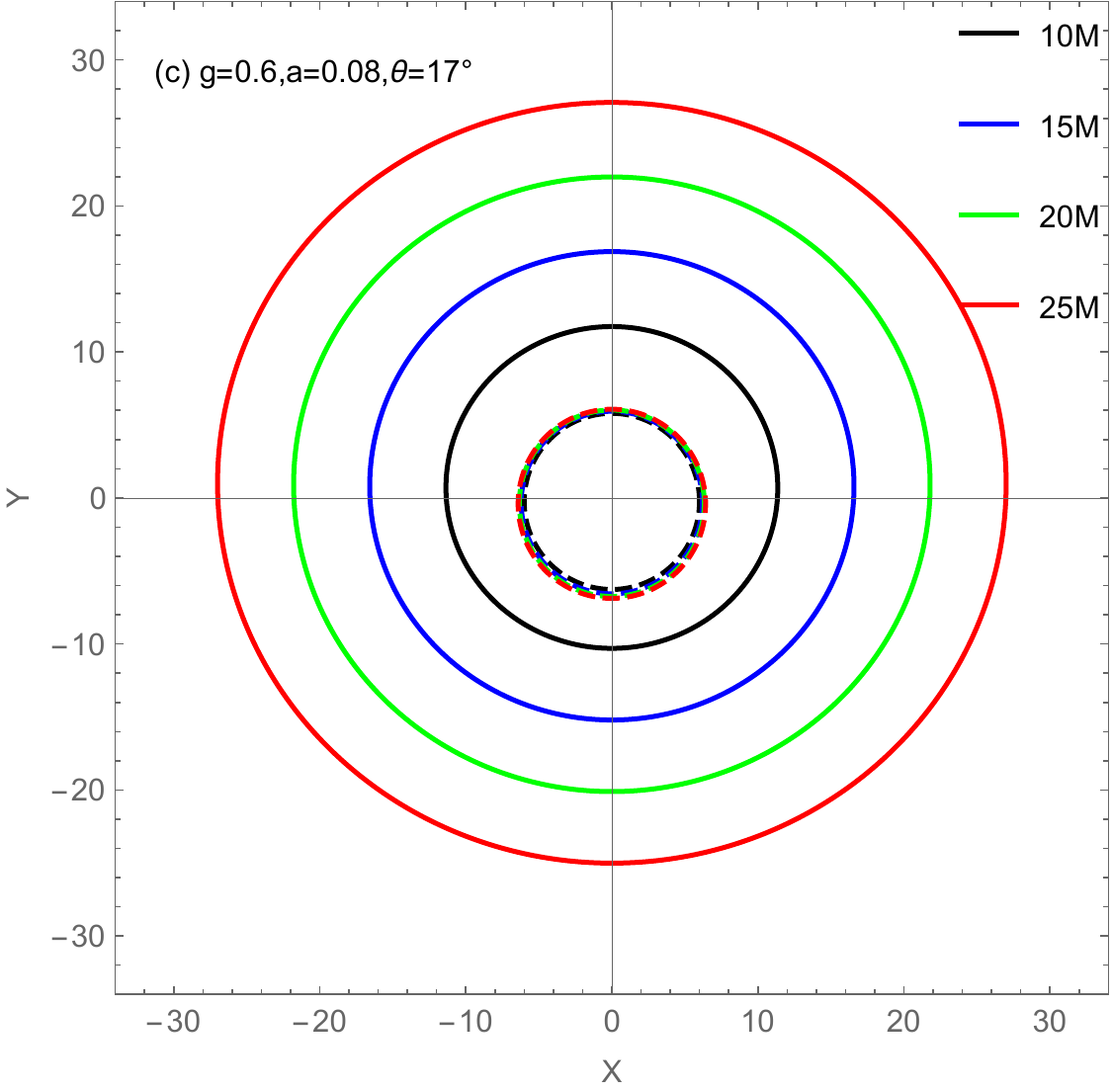}
	\end{subfigure}
	
	\vspace{0.3cm}
	
	\begin{subfigure}{0.31\textwidth}
		\includegraphics[width=\linewidth]{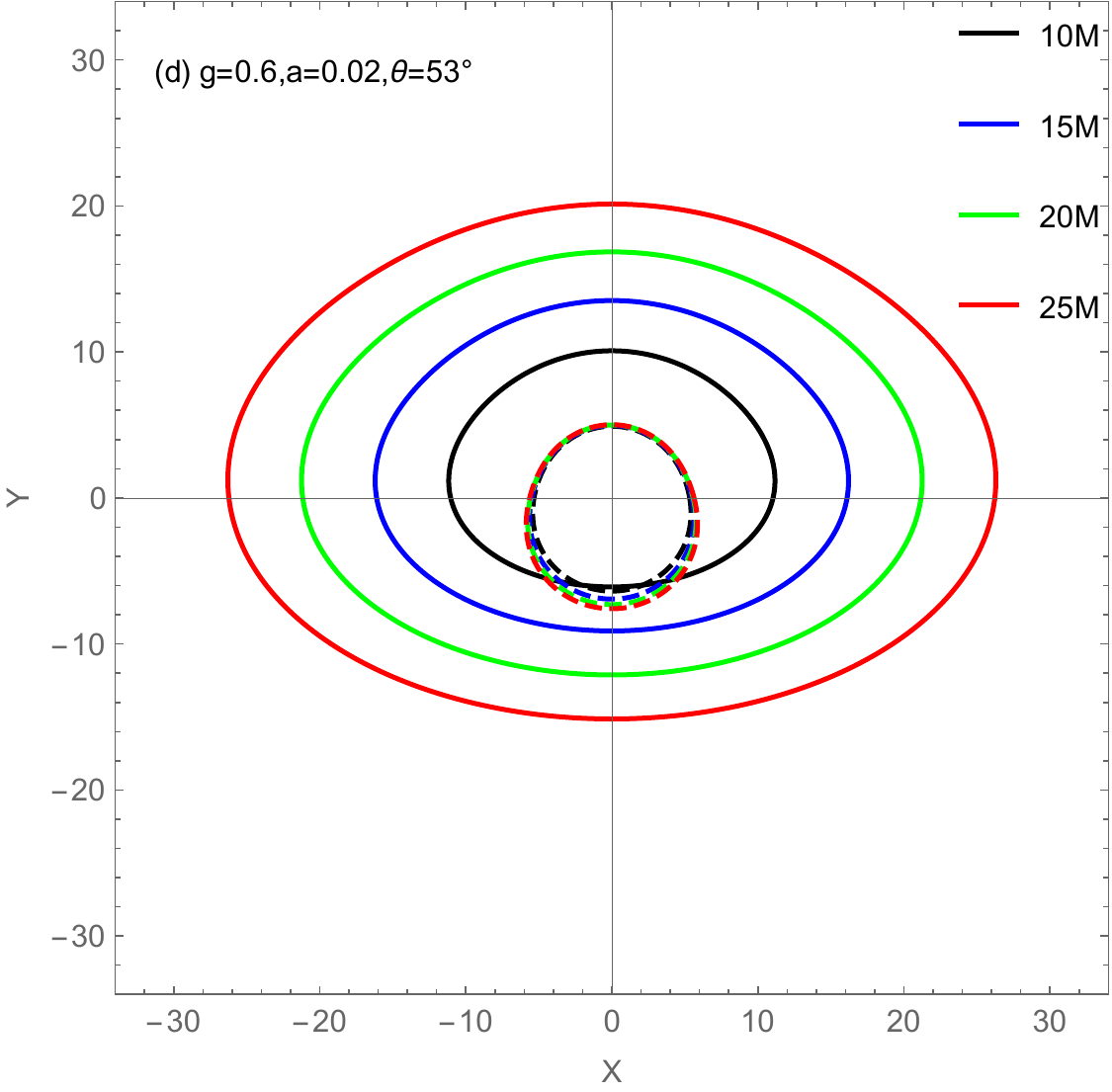}
	\end{subfigure}
	\hspace{0.02\textwidth}
	\begin{subfigure}{0.31\textwidth}
		\includegraphics[width=\linewidth]{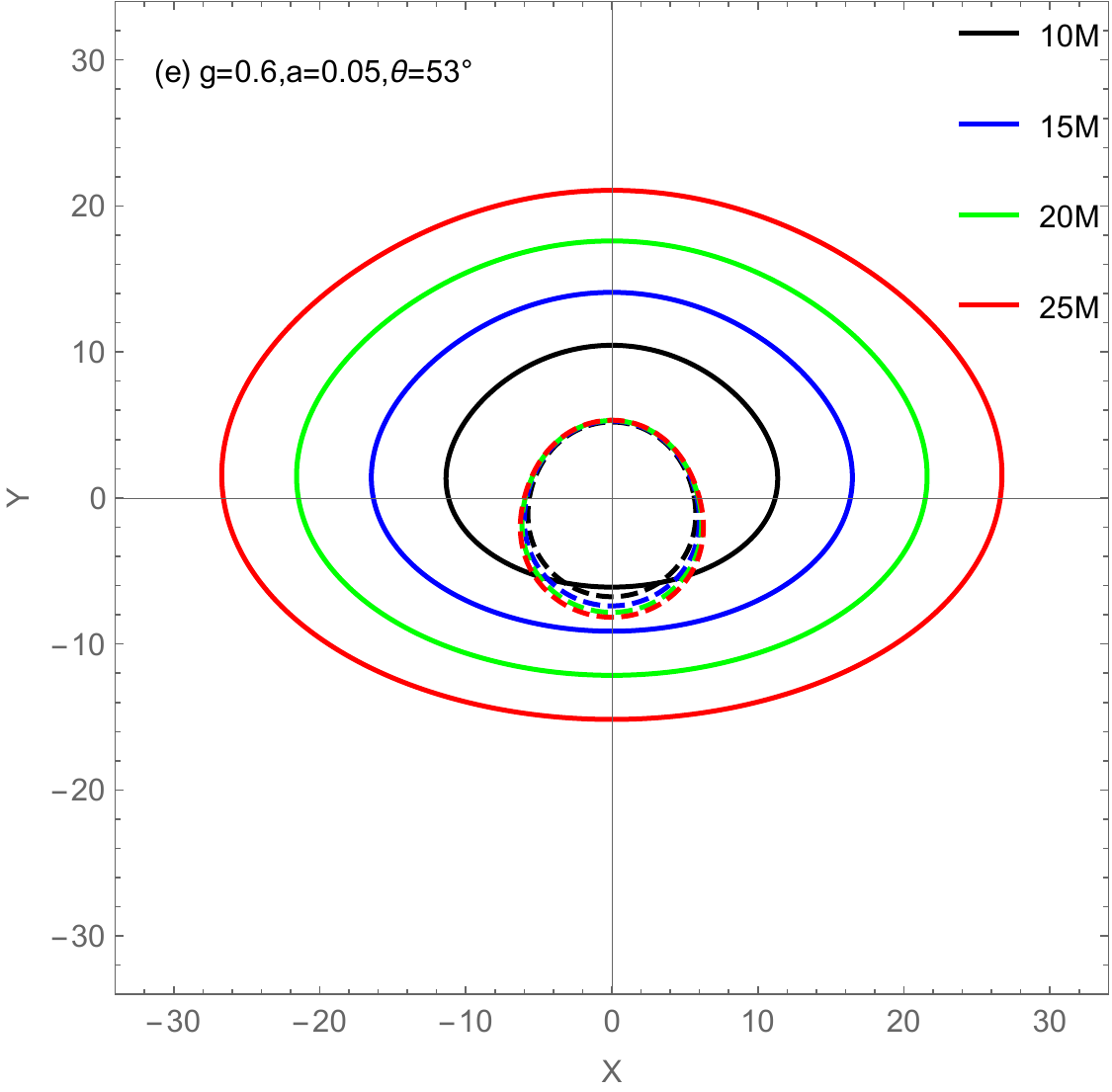}
	\end{subfigure}
	\hspace{0.02\textwidth}
	\begin{subfigure}{0.31\textwidth}
		\includegraphics[width=\linewidth]{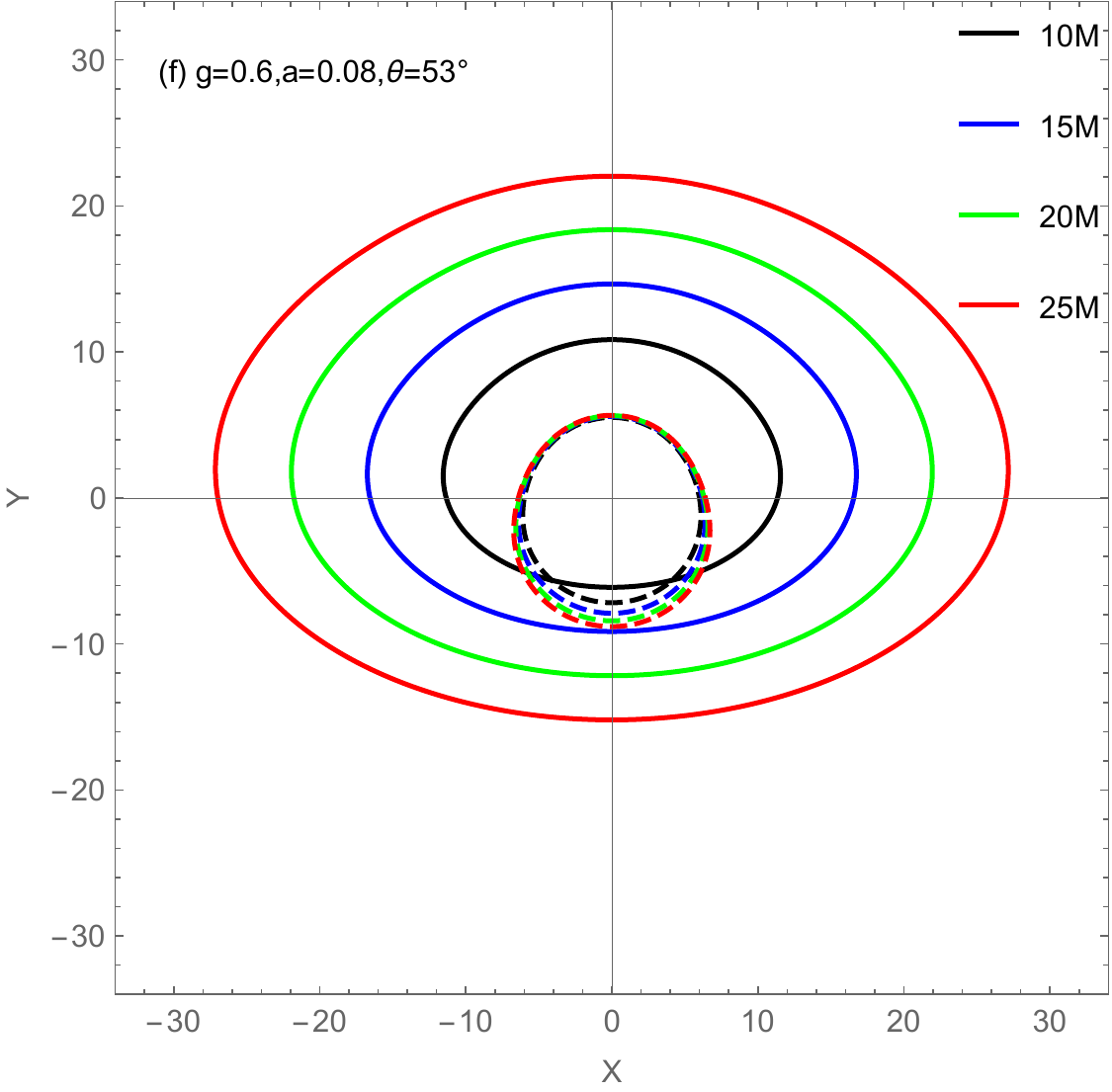}
	\end{subfigure}
	
	\vspace{0.3cm}
	
	\begin{subfigure}{0.31\textwidth}
		\includegraphics[width=\linewidth]{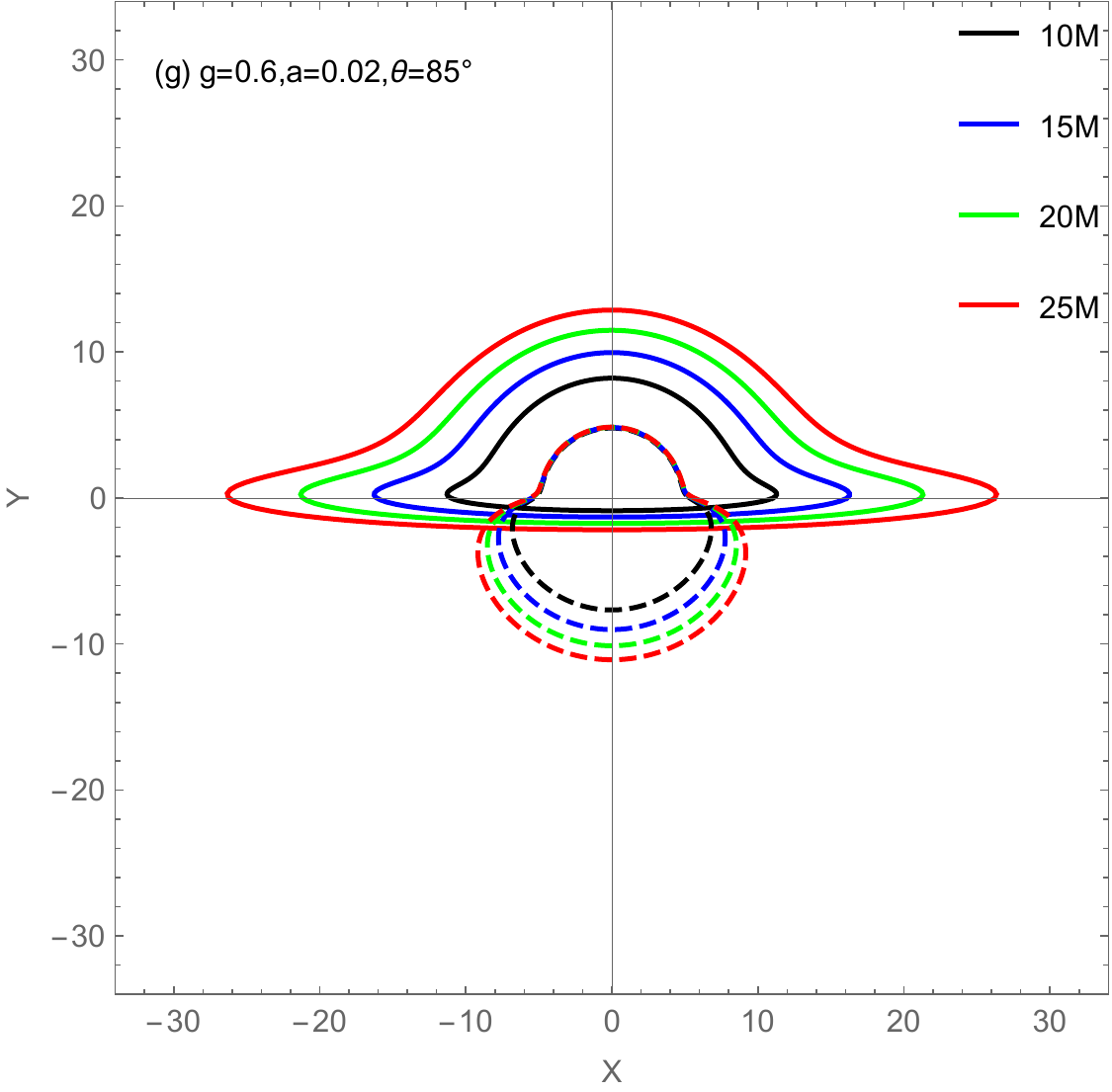}
	\end{subfigure}
	\hspace{0.02\textwidth}
	\begin{subfigure}{0.31\textwidth}
		\includegraphics[width=\linewidth]{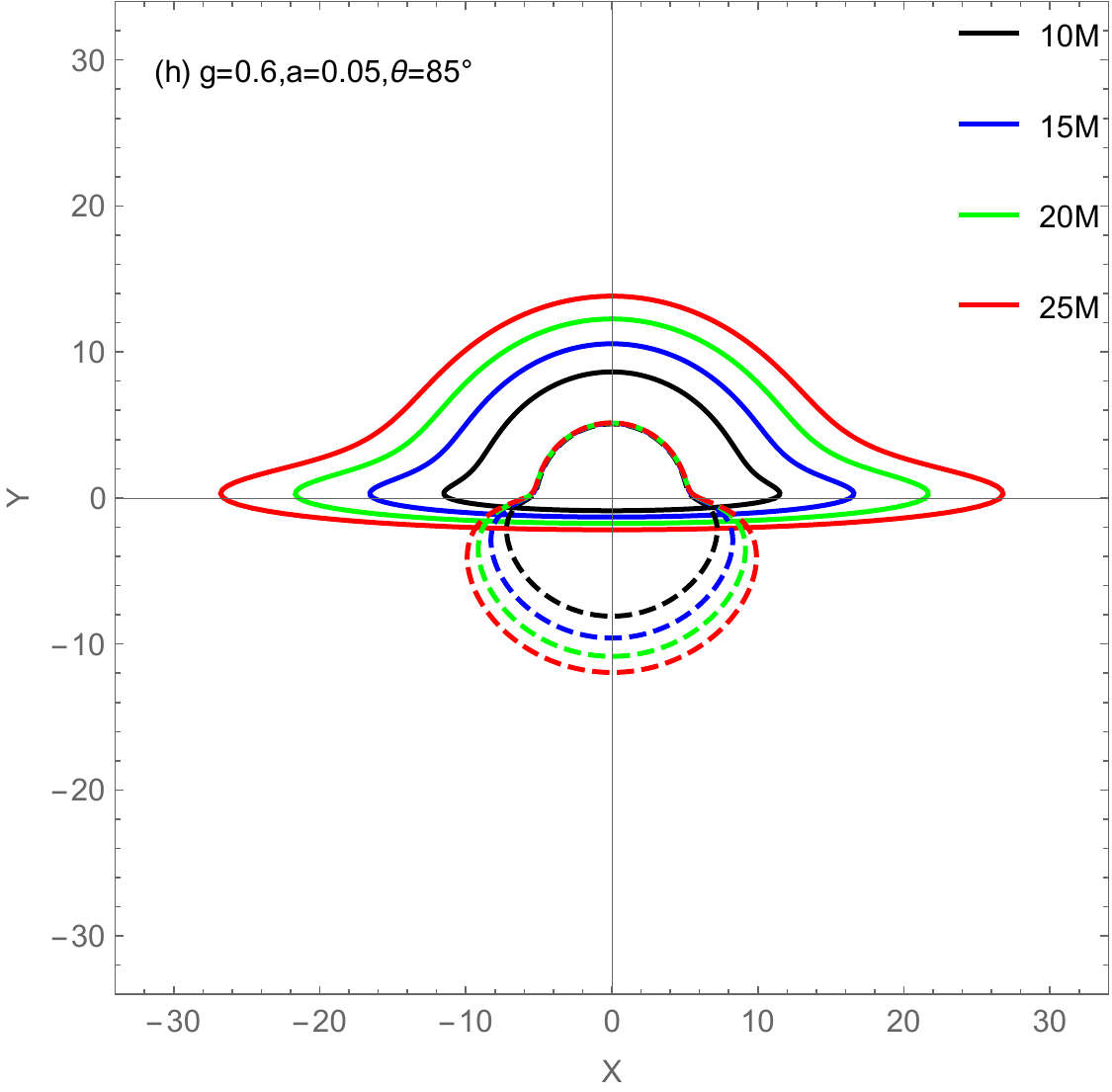}
	\end{subfigure}
	\hspace{0.02\textwidth}
	\begin{subfigure}{0.31\textwidth}
		\includegraphics[width=\linewidth]{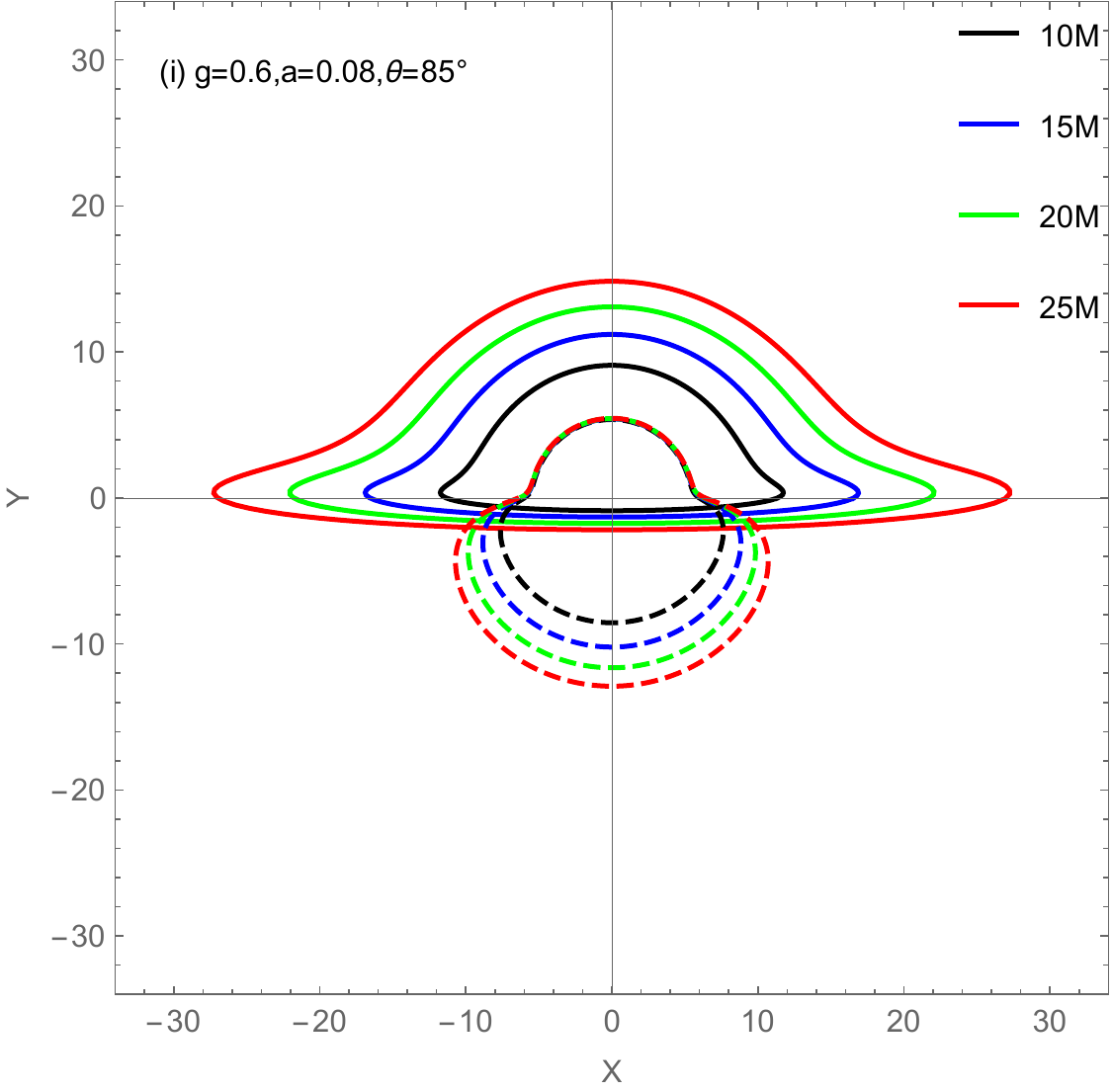}
	\end{subfigure}
	
	\caption{The direct (solid line) and secondary (dashed line) image of the thin accretion disk.}
	\label{dengr}
\end{figure*}

Figure.~\ref{dengr} presents both the direct (primary) and secondary images of selected stable circular orbits around an ABG BH coupled with a CS, as viewed by a distant observer at different inclination angles. Each column, from top to bottom, corresponds to inclination angles of $17^{\circ}$, $53^{\circ}$, and $85^{\circ}$, respectively. Each row, from left to right, is computed for a fixed coupling parameter $g=0.6$ and represents increasing values of $a=0.02$, $0.05$, and $0.08$. These images correspond to stable circular orbits at radii $r=10$, $15$, $20$, and $25$, ordered from the innermost to the outermost. As the inclination angle decreases, the secondary image becomes increasingly embedded within the direct image, forming a structure that resembles a photon ring. This behavior suggests an alternative interpretation of what may be observed as a photon ring in gravitational lensing studies. In contrast, as the inclination angle increases, the separation between the direct and secondary images becomes more distinct. Furthermore, with increasing $a$, both the direct and secondary images exhibit a noticeable increase in size.
\subsection{Observed Flux}
\begin{figure*}[htbp]
	\centering
	\begin{tabular}{ccc}
		\begin{minipage}[t]{0.3\textwidth}
			\centering
			\begin{overpic}[width=0.75\textwidth]{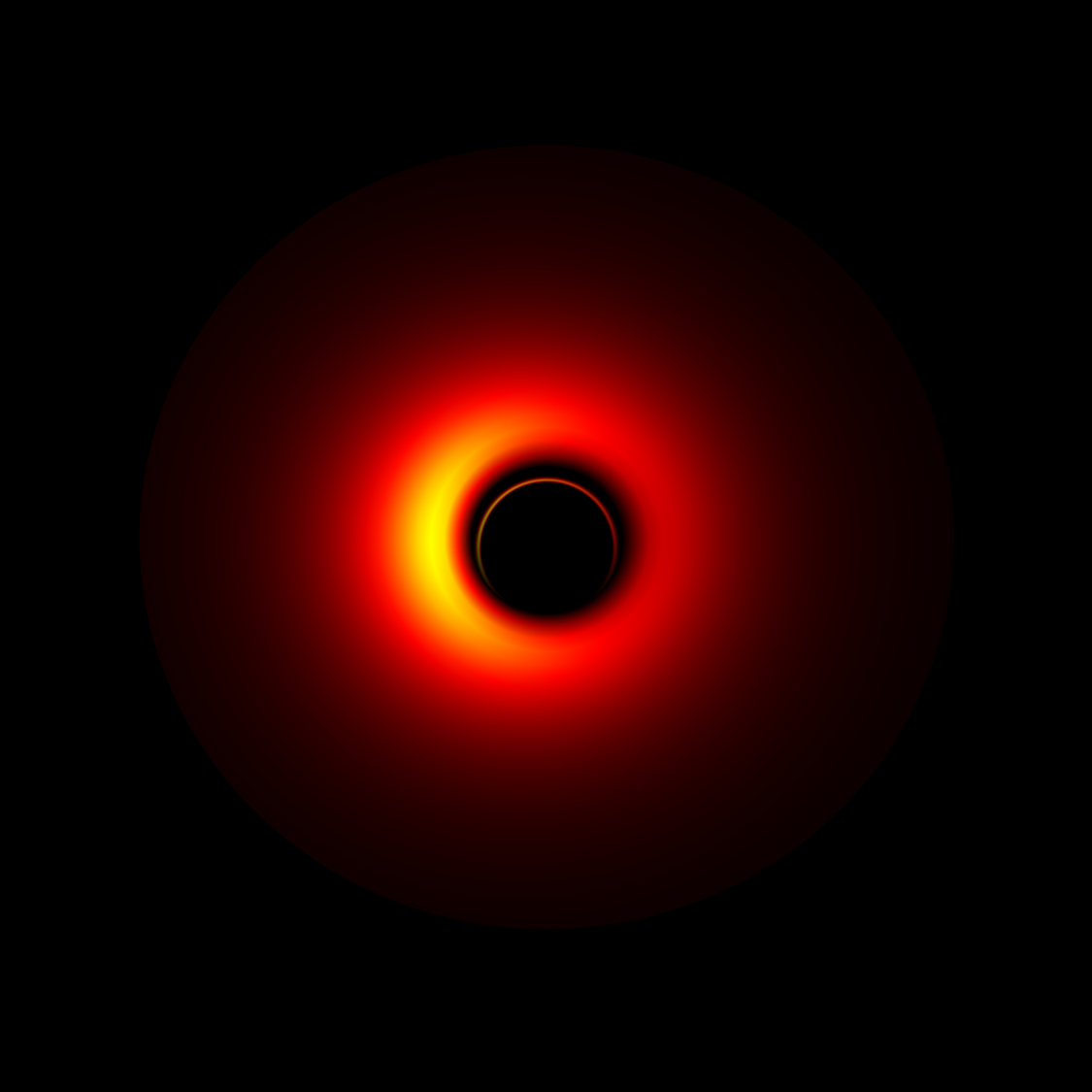} 
				\put(15,103){\color{black}\large $a=0.02, \theta=17^{\circ}$} 
				\put(-10,48){\color{black} Y}
				\put(48,-10){\color{black} X}
			\end{overpic}
		\end{minipage}
		&
		\begin{minipage}[t]{0.3\textwidth}
			\centering
			\begin{overpic}[width=0.75\textwidth]{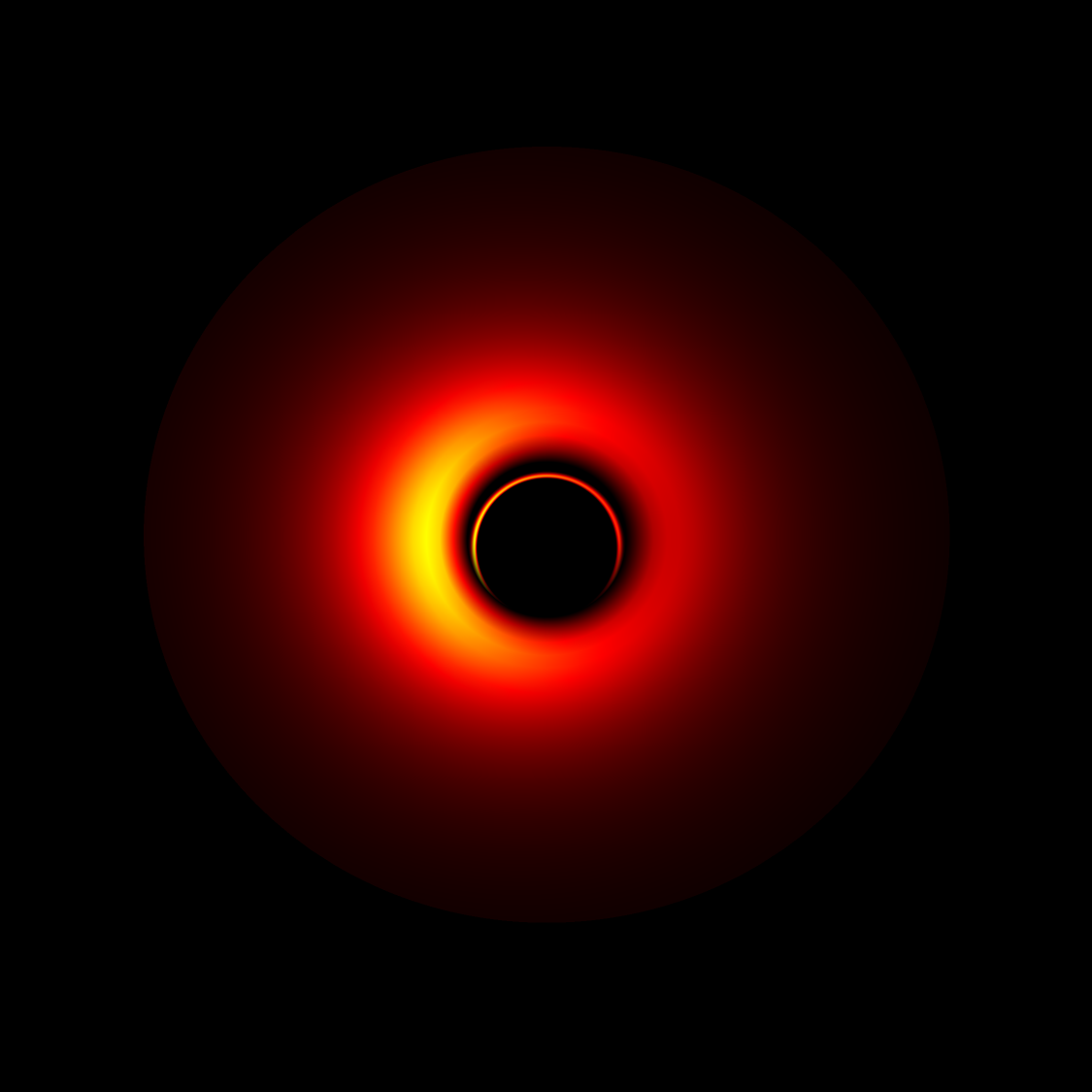} 
				\put(15,103){\color{black}\large $a=0.05, \theta=17^{\circ}$} 
				\put(-10,48){\color{black} Y}
				\put(48,-10){\color{black} X}
			\end{overpic}
		\end{minipage}
		&
		\begin{minipage}[t]{0.3\textwidth}
			\centering
			\begin{overpic}[width=0.75\textwidth]{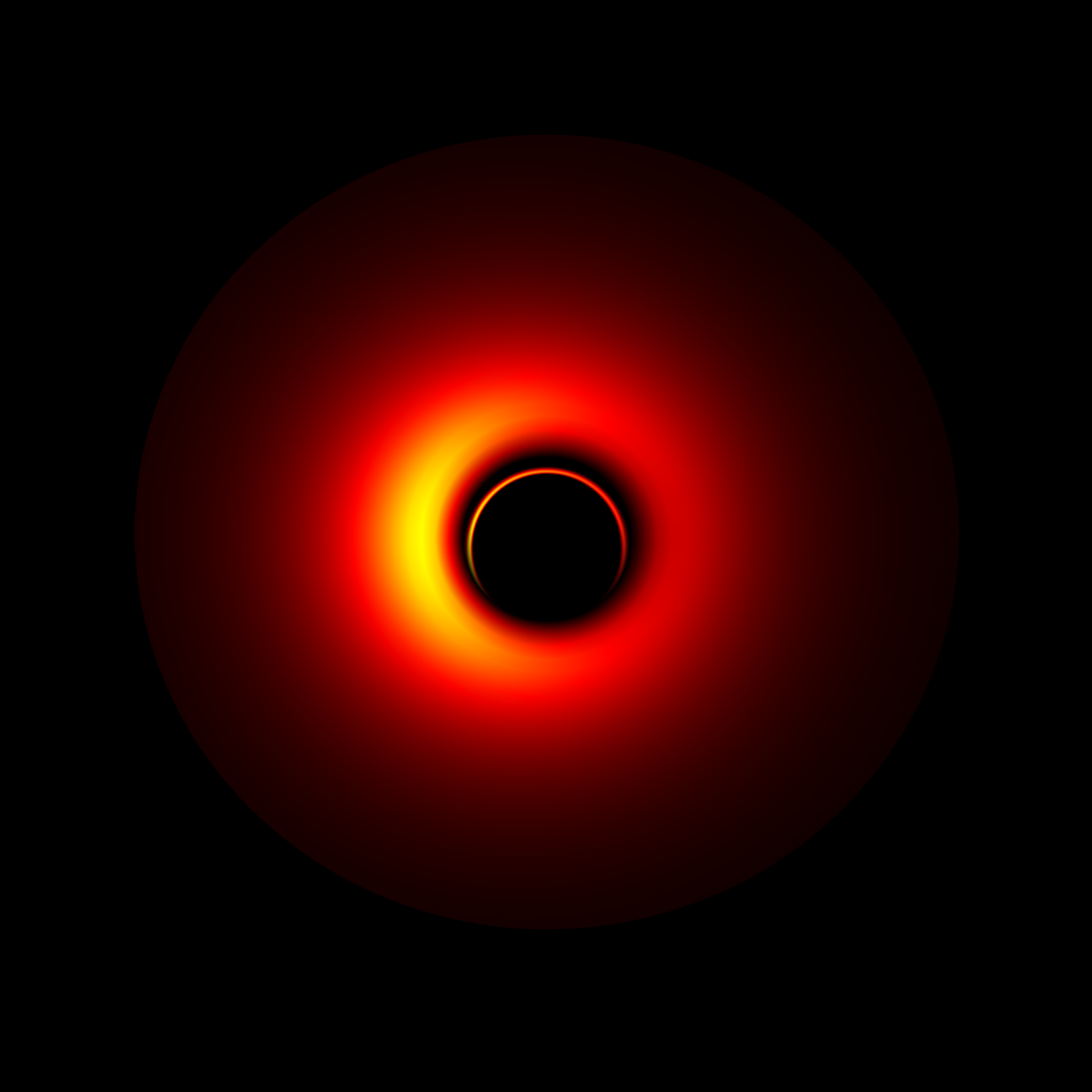}
				\put(14,103){\color{black}\large $a=0.08, \theta=17^{\circ}$} 
				\put(-10,48){\color{black} Y}
				\put(48,-10){\color{black} X}
			\end{overpic}
		\end{minipage}
		\vspace{40pt} 
		\\ 
		\begin{minipage}[t]{0.3\textwidth}
			\centering
			\begin{overpic}[width=0.75\textwidth]{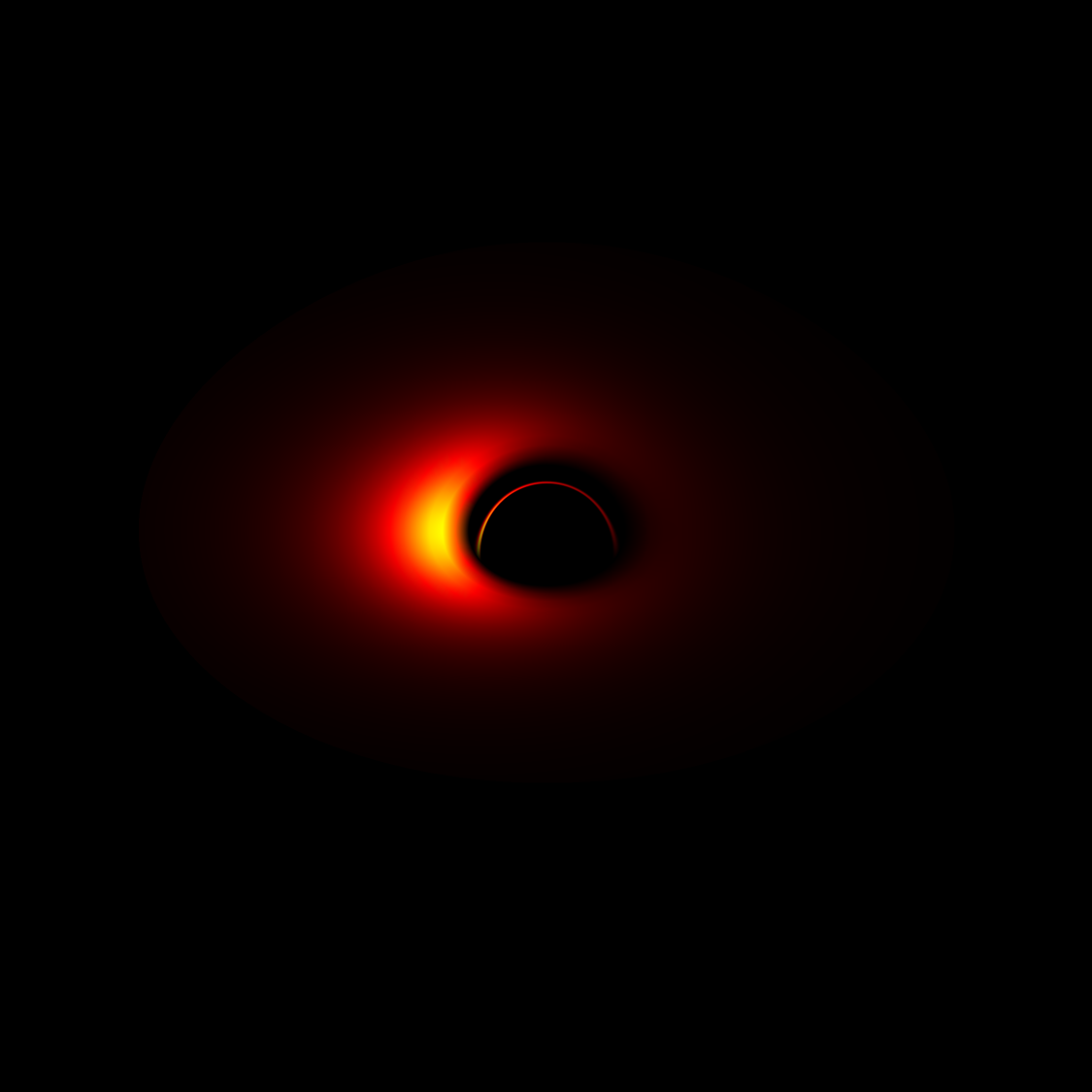} 
				\put(15,103){\color{black}\large $a=0.02, \theta=53^{\circ}$} 
				\put(-10,48){\color{black} Y}
				\put(48,-10){\color{black} X}
			\end{overpic}
		\end{minipage}
		&
		\begin{minipage}[t]{0.3\textwidth}
			\centering
			\begin{overpic}[width=0.75\textwidth]{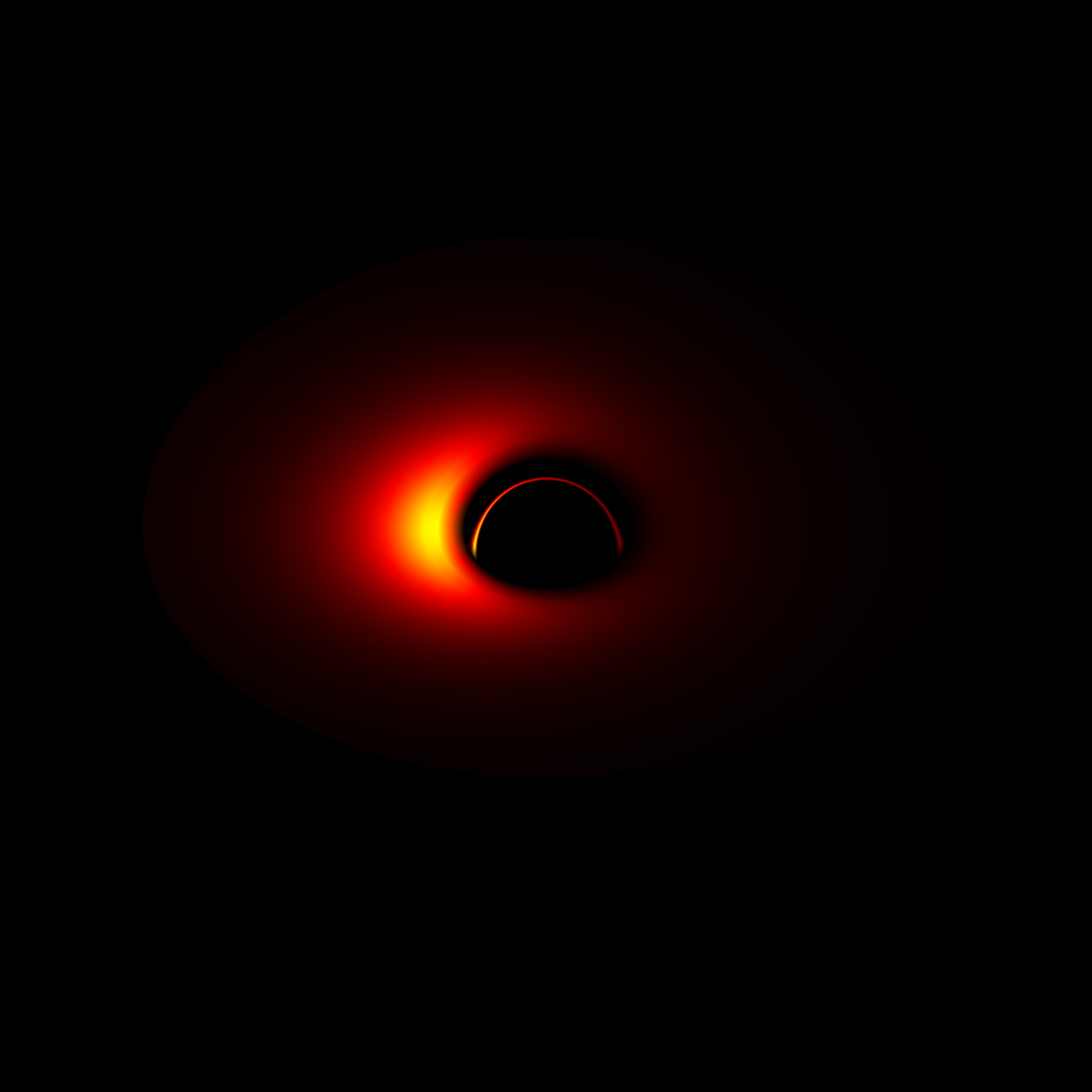} 
				\put(15,103){\color{black}\large $a=0.05, \theta=53^{\circ}$} 
				\put(-10,48){\color{black} Y}
				\put(48,-10){\color{black} X}
			\end{overpic}
		\end{minipage}
		&
		\begin{minipage}[t]{0.3\textwidth}
			\centering
			\begin{overpic}[width=0.75\textwidth]{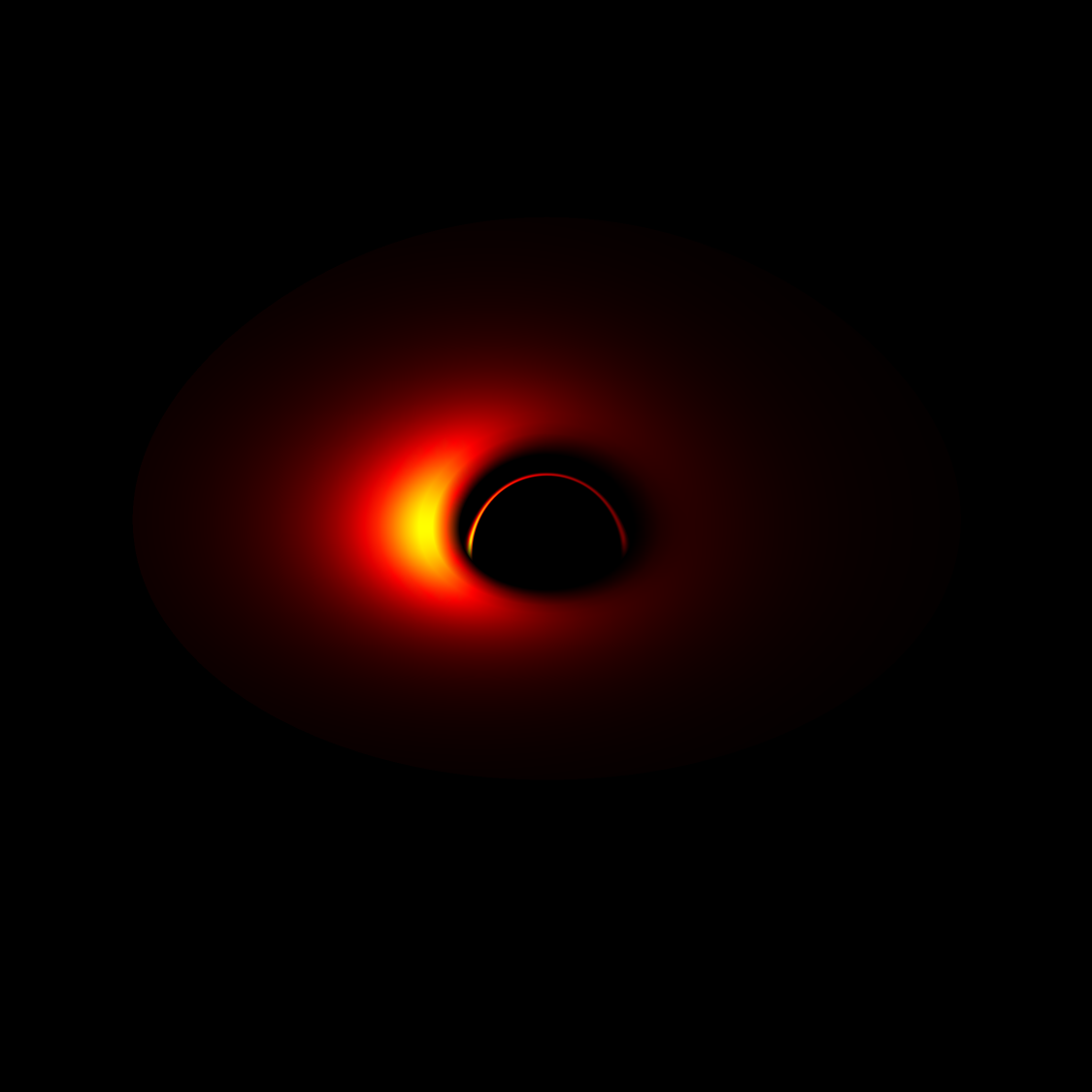} 
				\put(14,103){\color{black}\large $a=0.08, \theta=53^{\circ}$} 
				\put(-10,48){\color{black} Y}
				\put(48,-10){\color{black} X}
			\end{overpic}
		\end{minipage}
		\vspace{40pt} 
		\\
		\begin{minipage}[t]{0.3\textwidth}
			\centering
			\begin{overpic}[width=0.75\textwidth]{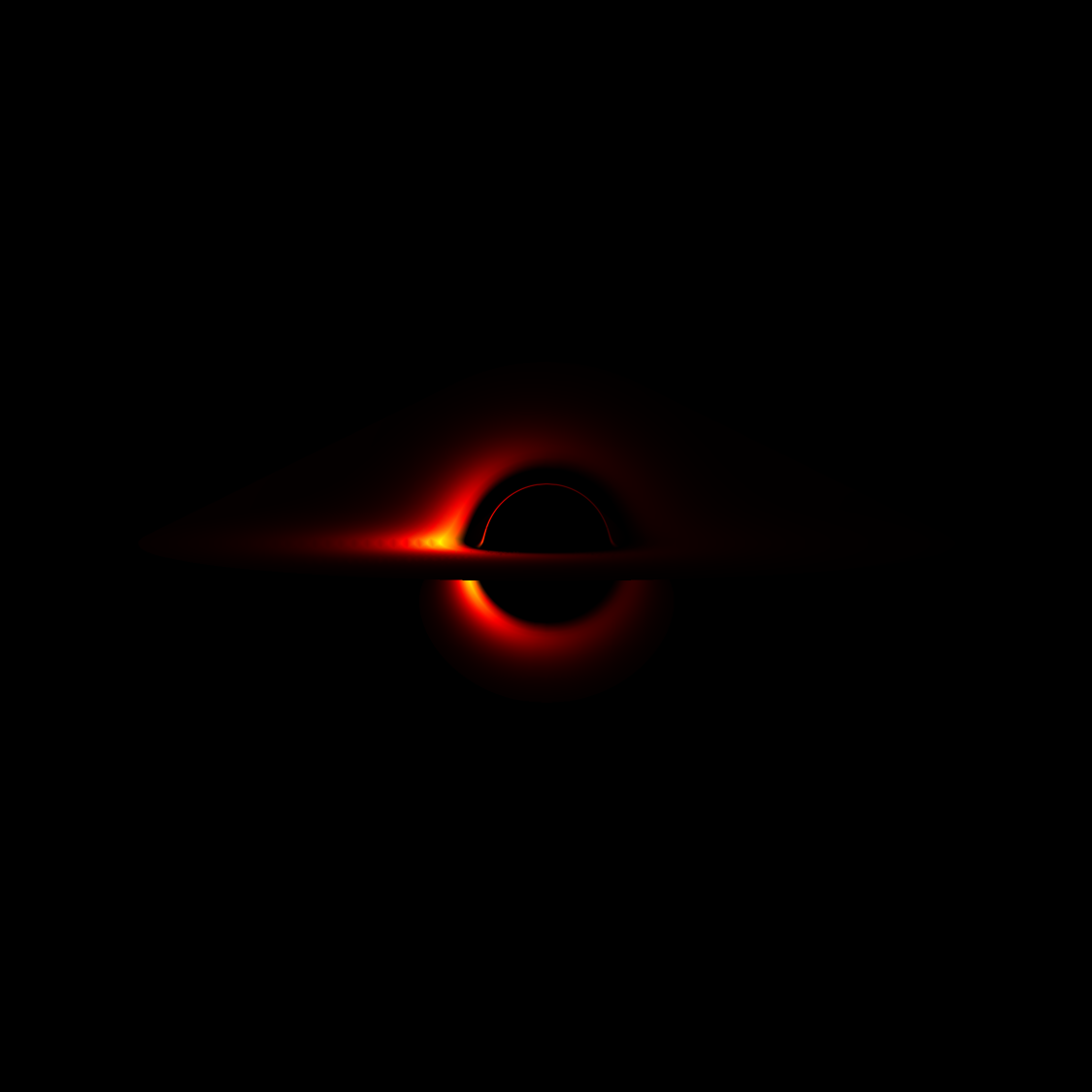}
				\put(15,103){\color{black}\large $a=0.02, \theta=85^{\circ}$}
				\put(-10,48){\color{black} Y}
				\put(48,-10){\color{black} X}
			\end{overpic}
		\end{minipage}
		&
		\begin{minipage}[t]{0.3\textwidth}
			\centering
			\begin{overpic}[width=0.75\textwidth]{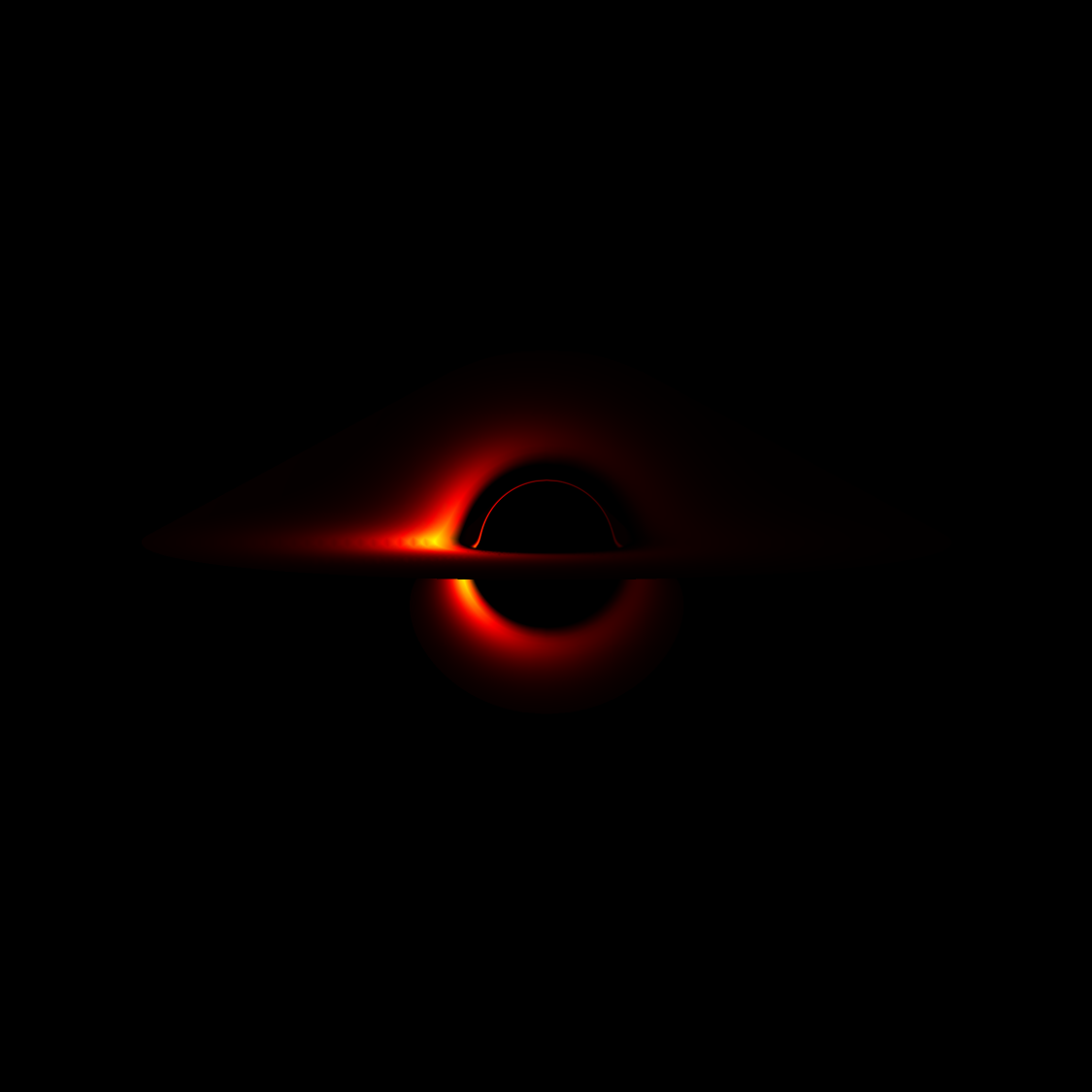}
				\put(15,103){\color{black}\large $a=0.05, \theta=85^{\circ}$} 
				\put(-10,48){\color{black} Y}
				\put(48,-10){\color{black} X}
			\end{overpic}
		\end{minipage}
		&
		\begin{minipage}[t]{0.3\textwidth}
			\centering
			\begin{overpic}[width=0.75\textwidth]{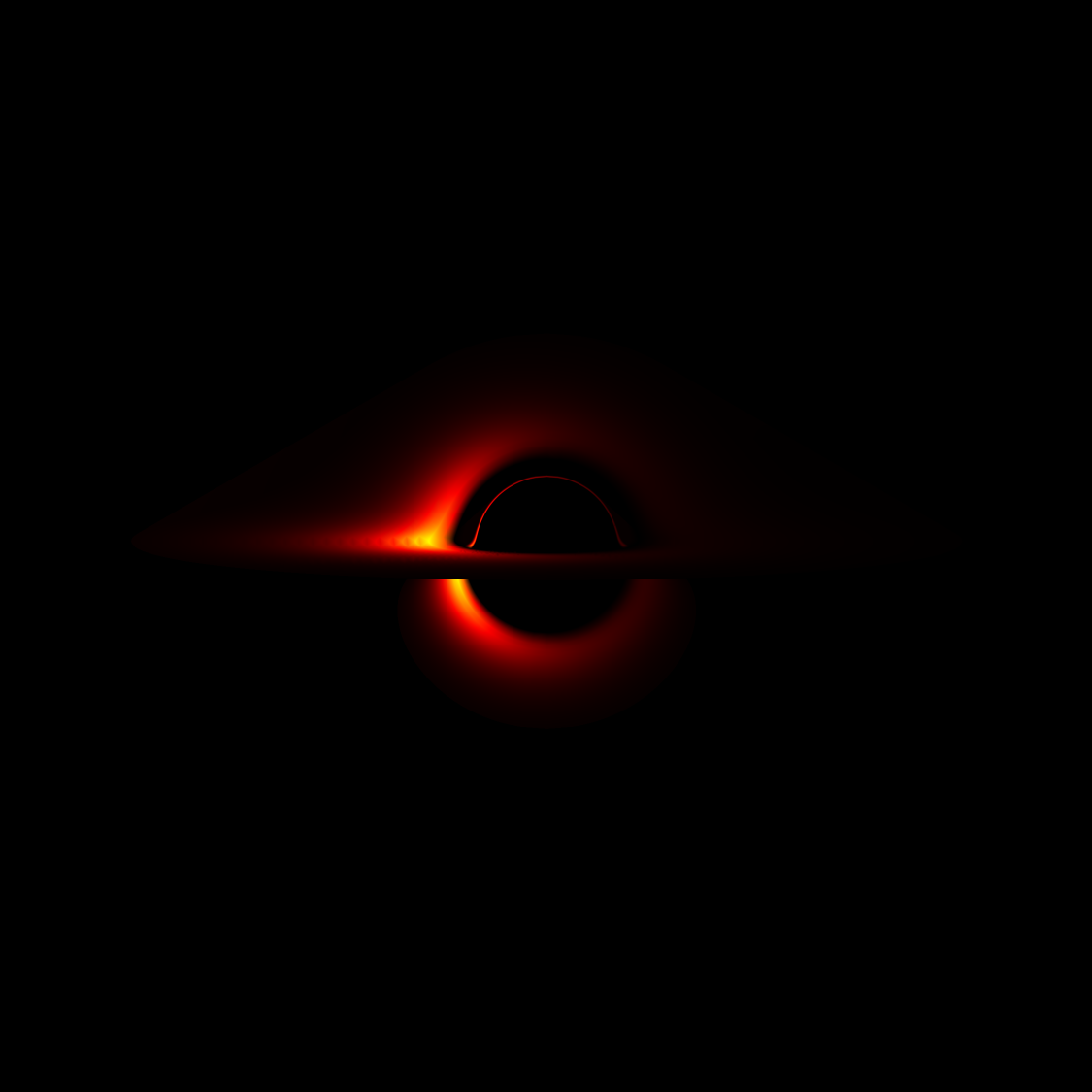}
				\put(14,103){\color{black}\large $a=0.08, \theta=85^{\circ}$}
				\put(-10,48){\color{black} Y}
				\put(48,-10){\color{black} X}
			\end{overpic}
		\end{minipage}
	\end{tabular}
	\caption{Complete apparent images of a thin accretion disk for different values of the parameter $a$ and inclination angle. We set $g=0.6$.}
	\label{xuanran}
\end{figure*}
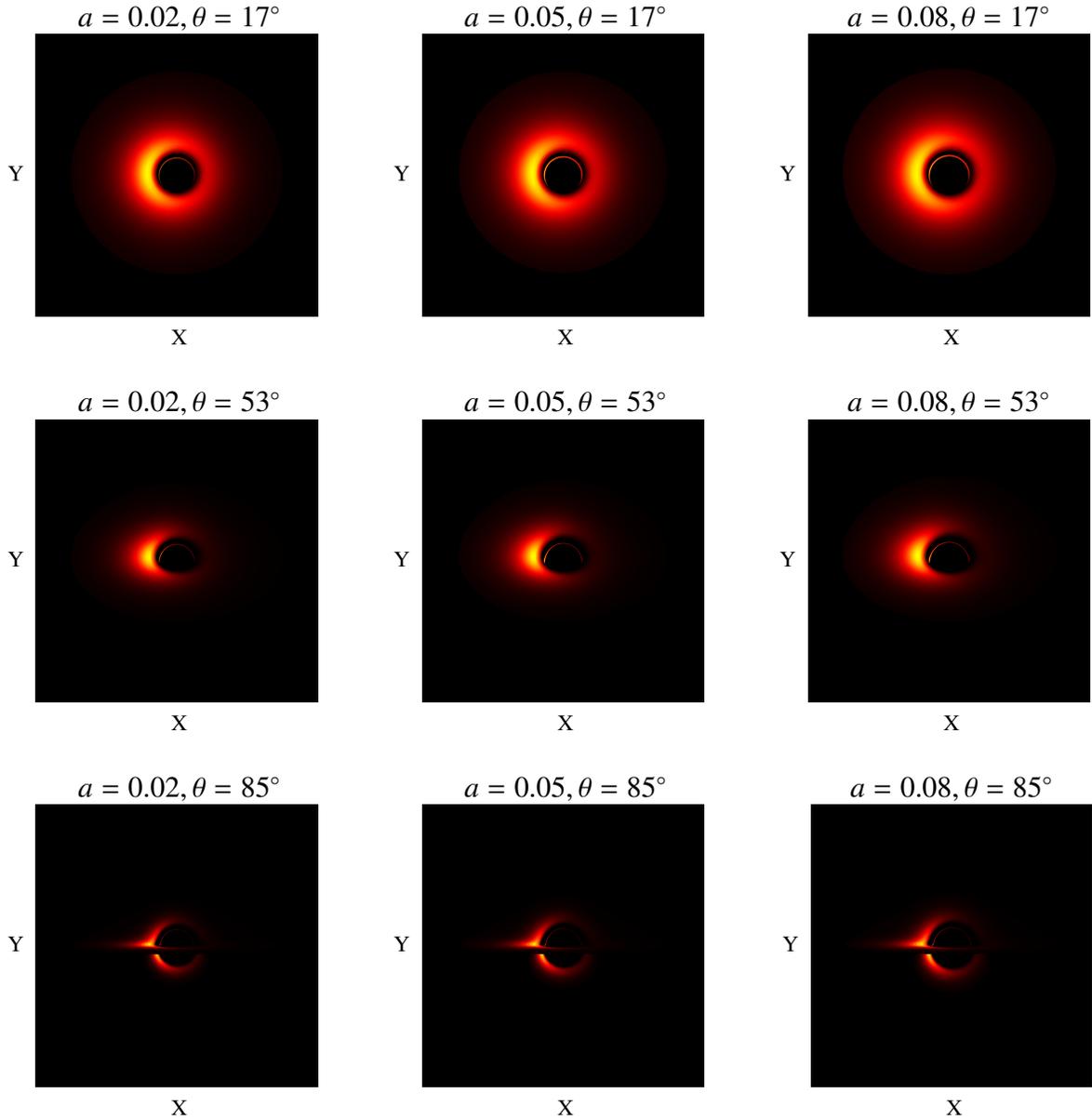

To determine the observed flux $F_{obs}$ at a given point on the celestial sphere, it is essential to account for the effects of gravitational redshift, denoted by $z$. Consequently, the following relation is obtained:
\begin{equation}
	\label{33}
	F_{obs}=\frac{F(r)}{(1+z)^{4}}.
\end{equation}
From Eqs. (\ref{21}), (\ref{26}), and (\ref{33}), the observed flux $F_{obs}$ is given by
\begin{equation}
	\label{34}
	F_{obs}=\frac{-\frac{\dot{M}\Omega_{,r}}{4\pi\sqrt{-g}(E-\Omega L)^{2}}\int^{r}_{r_{isco}}(E-\Omega L)L_{,r}dr}{(\frac{1+\Omega b\sin\theta\cos\alpha}{\sqrt{-g_{tt}-g_{\phi\phi}\Omega^{2}}})^{4}}.
\end{equation}
Based on the preceding analysis, we present the observed flux distribution of the accretion disk in Fig.~\ref{xuanran}. This figure illustrates the results for an ABG BH coupled with a CS under varying model parameters. It is evident that the inclination angle of observation has a significant impact on the resulting image morphology. As the inclination angle increases, the observed image of the accretion disk transitions into a distinctive ``hat-like" shape. Each column in the figure, from top to bottom, corresponds to inclination angles of $17^{\circ}$, $53^{\circ}$, and $85^{\circ}$, respectively. Each row, from left to right, is computed for a fixed coupling parameter $g=0.6$ and represents increasing values of $a=0.02$, $0.05$, and $0.08$.

We examine the redshift distributions—presented as contour maps of the redshift parameter $z$—in the direct images, considering various inclination angles and model parameters $g$ and $a$, as shown in Fig.~\ref{hongyi1}. The influence of $g$ and $a$ on the redshift structure is also apparent in the secondary images, which are displayed in Fig.~\ref{hongyi2}.

For observers with high inclination angles, a significant blueshift $(z<0)$ appears in the left half of the image plane, surpassing the gravitational redshift $(z>0)$ induced by the presence of the BH. In contrast, no blueshift features are observed at low inclination angles. For a fixed value of $g=0.6$, increasing $\alpha$ results in an expansion of the redshifted region in the case of an ABG BH coupled with a CS.
\begin{figure*}[htbp]
	\centering
	\begin{tabular}{ccc}
		\begin{minipage}[t]{0.3\textwidth}
			\centering
			\begin{overpic}[width=0.75\textwidth]{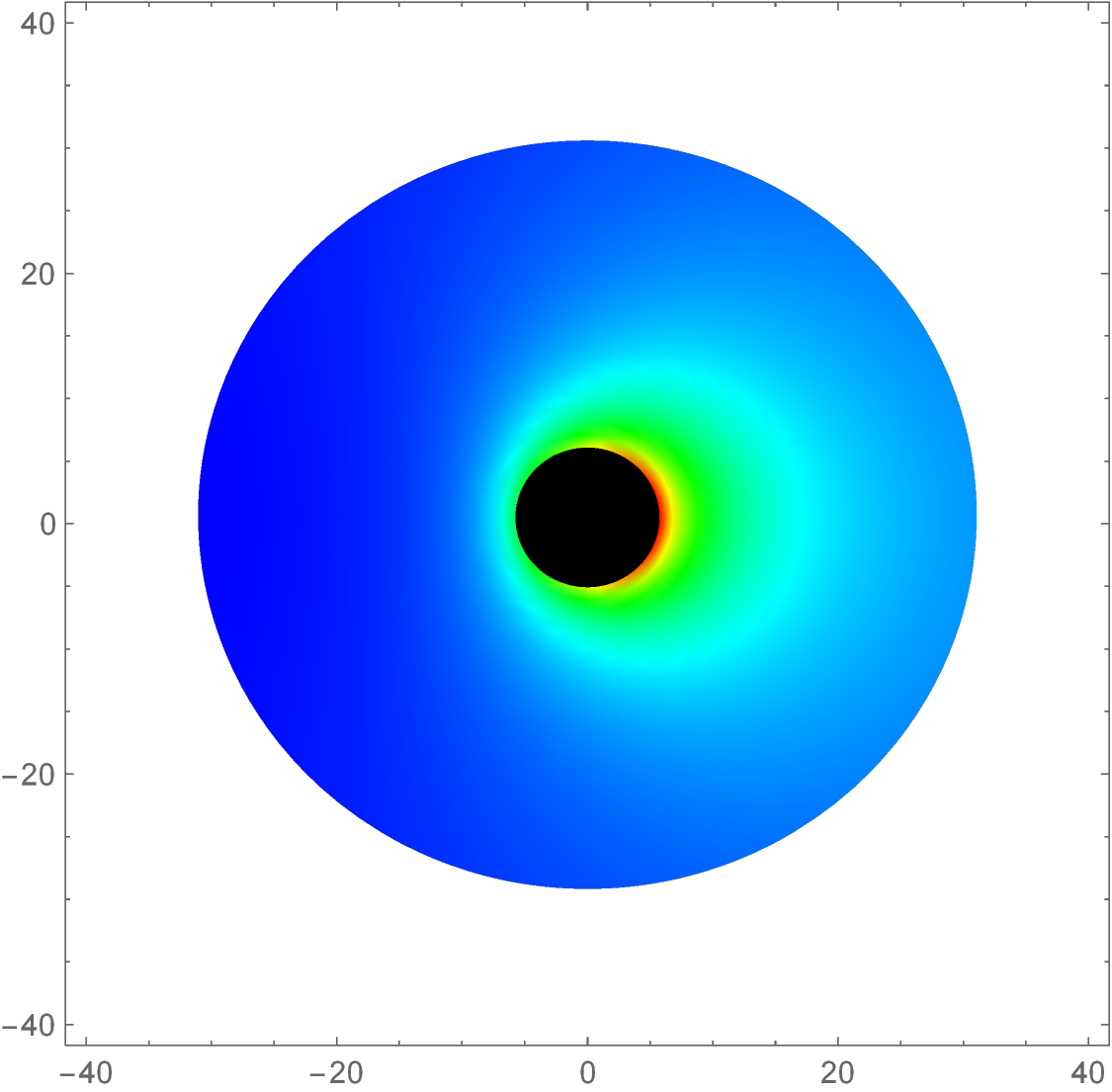}
				\put(18,100){\color{black}\large $a=0.02, \theta=17^{\circ}$}
				\put(-8,48){\color{black} Y}
				\put(48,-10){\color{black} X}
			\end{overpic}
			\raisebox{0.05\height}{ 
				\begin{overpic}[width=0.07\textwidth]{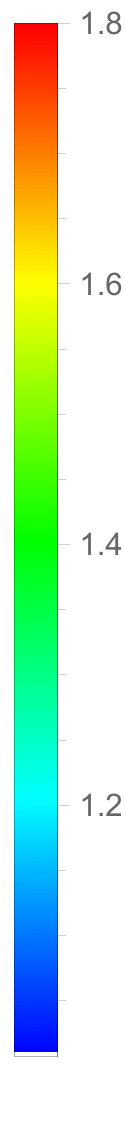}
					\put(0,103){\color{black}\large $z$} 
				\end{overpic}
			}
		\end{minipage}
		&
		\begin{minipage}[t]{0.3\textwidth}
			\centering
			\begin{overpic}[width=0.75\textwidth]{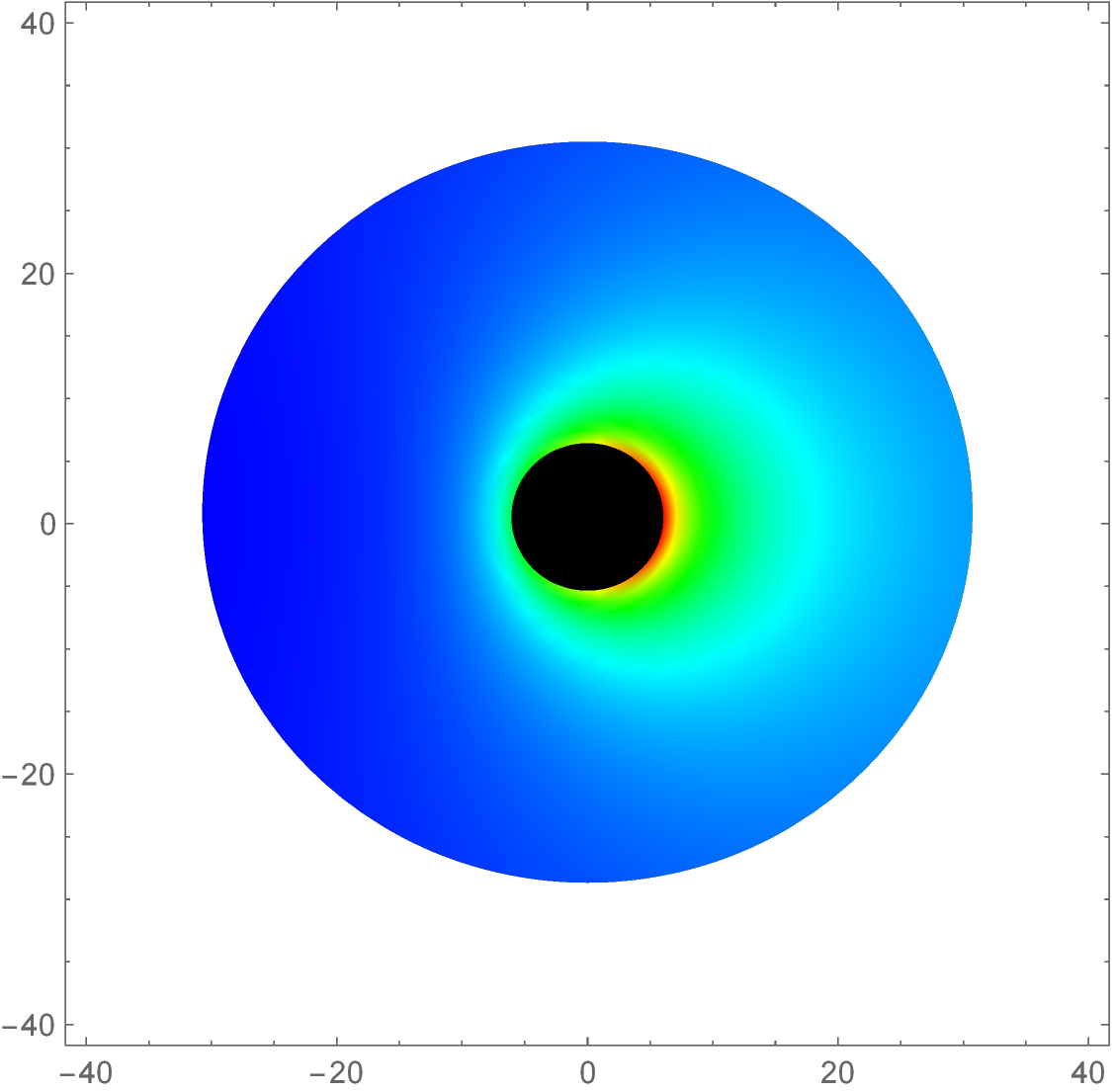}
				\put(18,100){\color{black}\large $a=0.05, \theta=17^{\circ}$} 
				\put(-8,48){\color{black} Y}
				\put(48,-10){\color{black} X}
			\end{overpic}
			\raisebox{0.05\height}{ 
				\begin{overpic}[width=0.07\textwidth]{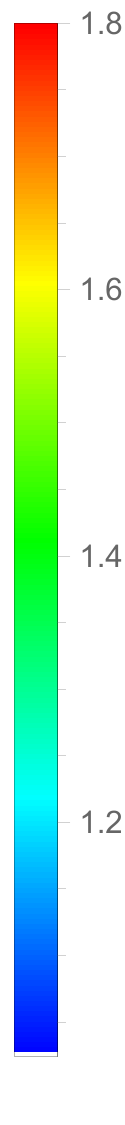} 
					\put(0,103){\color{black}\large $z$} 
				\end{overpic}
			}			
		\end{minipage}
		&
		\begin{minipage}[t]{0.3\textwidth}
			\centering
			\begin{overpic}[width=0.75\textwidth]{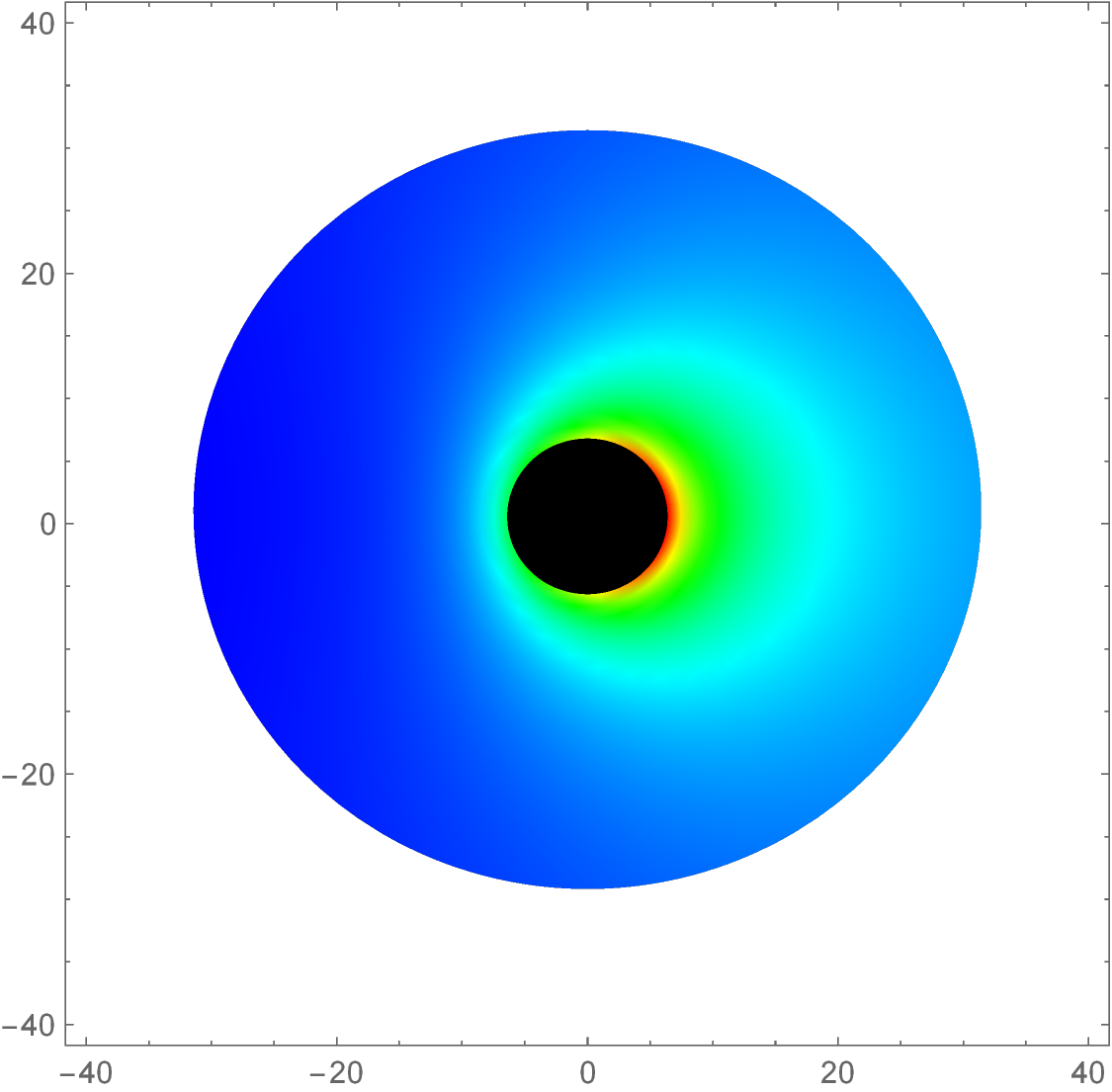}
				\put(18,100){\color{black}\large $a=0.08, \theta=17^{\circ}$} 
				\put(-8,48){\color{black} Y}
				\put(48,-10){\color{black} X}
			\end{overpic}
			\raisebox{0.05\height}{
				\begin{overpic}[width=0.07\textwidth]{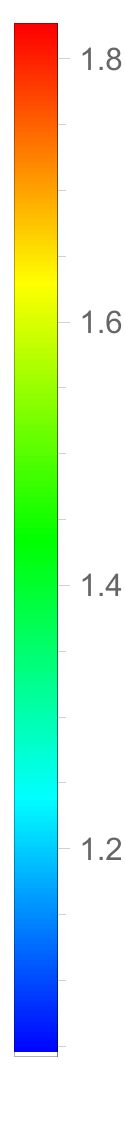}
					\put(0,103){\color{black}\large $z$}
				\end{overpic}
			}
		\end{minipage}
		\vspace{40pt} 
		\\ 
		\begin{minipage}[t]{0.3\textwidth}
			\centering
			\begin{overpic}[width=0.75\textwidth]{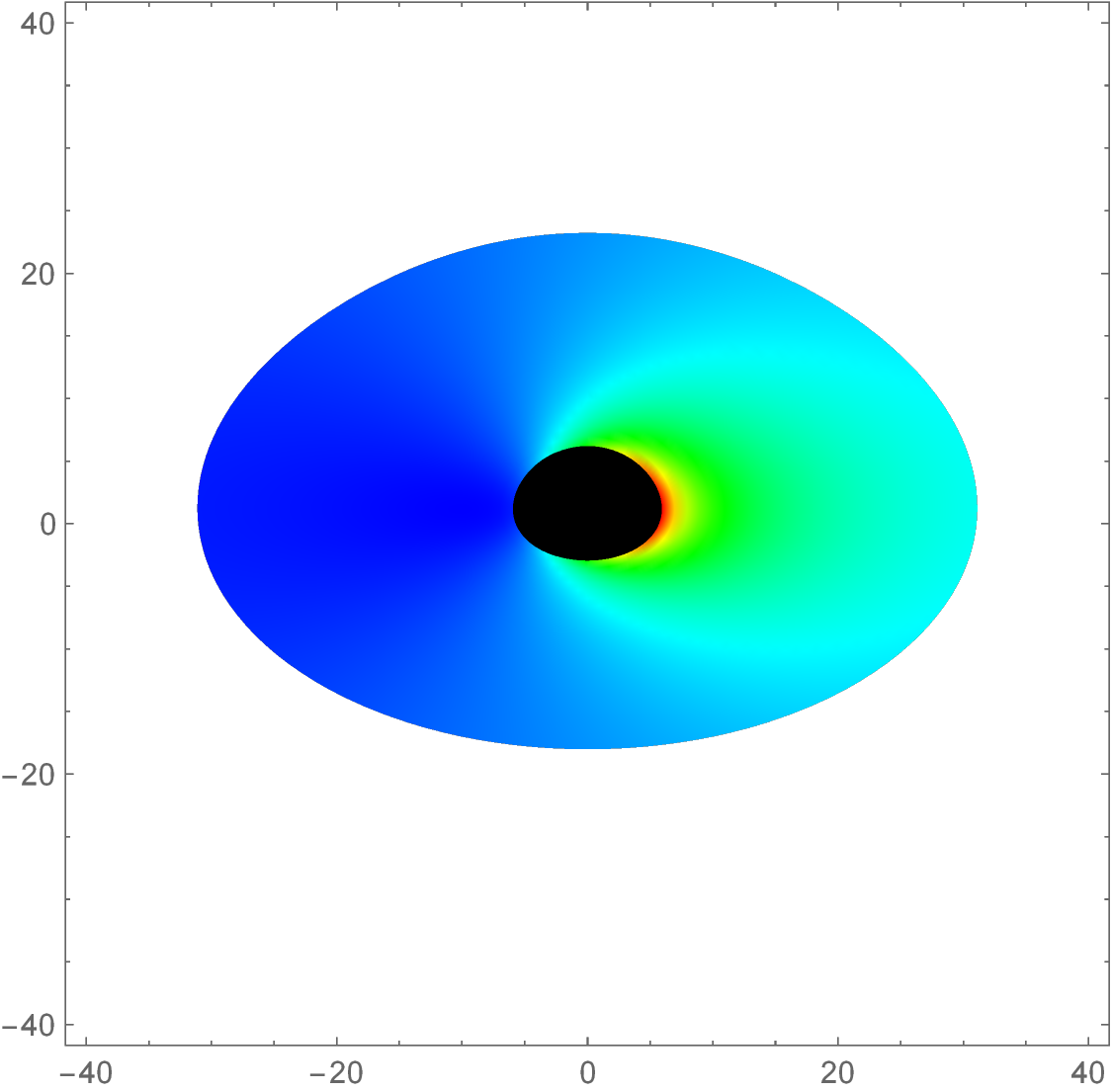}
				\put(18,100){\color{black}\large $a=0.02, \theta=53^{\circ}$} 
				\put(-8,48){\color{black} Y}
				\put(48,-10){\color{black} X}
			\end{overpic}
			\raisebox{0.05\height}{
				\begin{overpic}[width=0.07\textwidth]{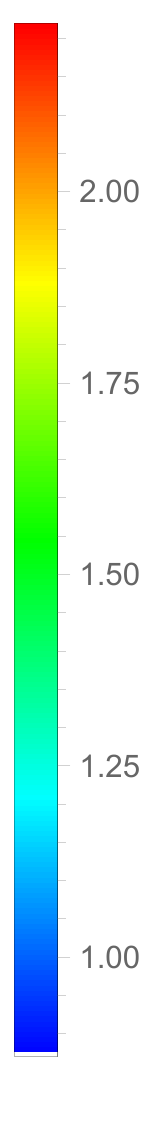} 
					\put(0,103){\color{black}\large $z$} 
				\end{overpic}
			}
		\end{minipage}
		&
		\begin{minipage}[t]{0.3\textwidth}
			\centering
			\begin{overpic}[width=0.75\textwidth]{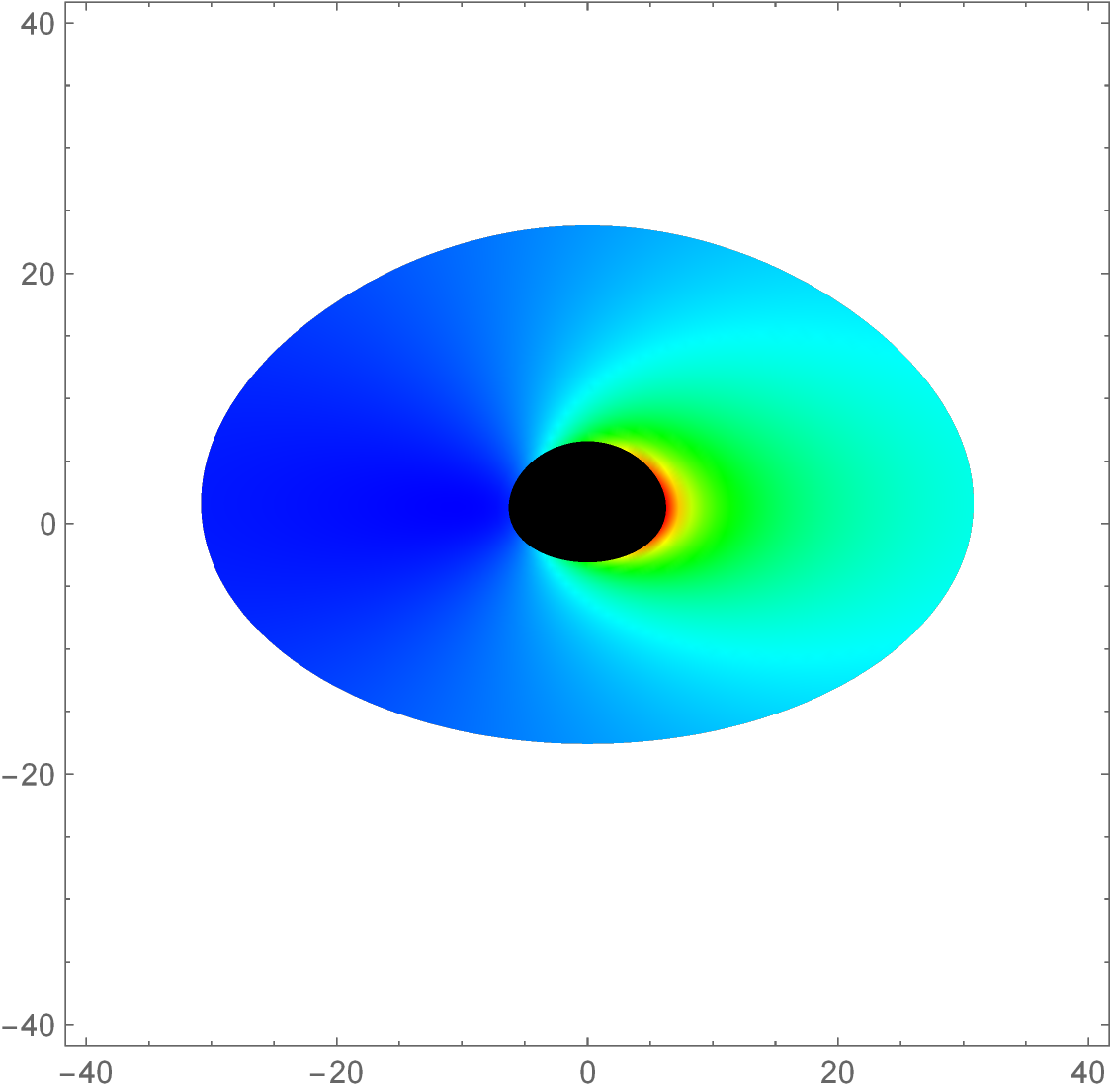} 
				\put(18,100){\color{black}\large $a=0.05, \theta=53^{\circ}$} 
				\put(-8,48){\color{black} Y}
				\put(48,-10){\color{black} X}
			\end{overpic}
			\raisebox{0.05\height}{ 
				\begin{overpic}[width=0.07\textwidth]{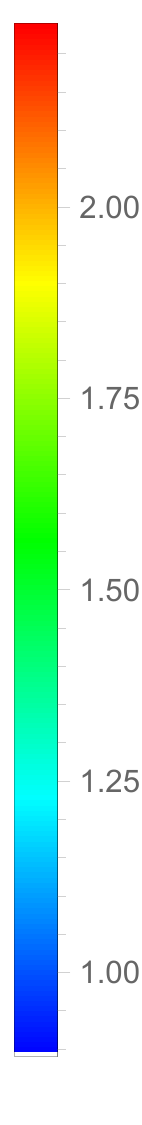} 
					\put(0,103){\color{black}\large $z$} 
				\end{overpic}
			}
		\end{minipage}
		&
		\begin{minipage}[t]{0.3\textwidth}
			\centering
			\begin{overpic}[width=0.75\textwidth]{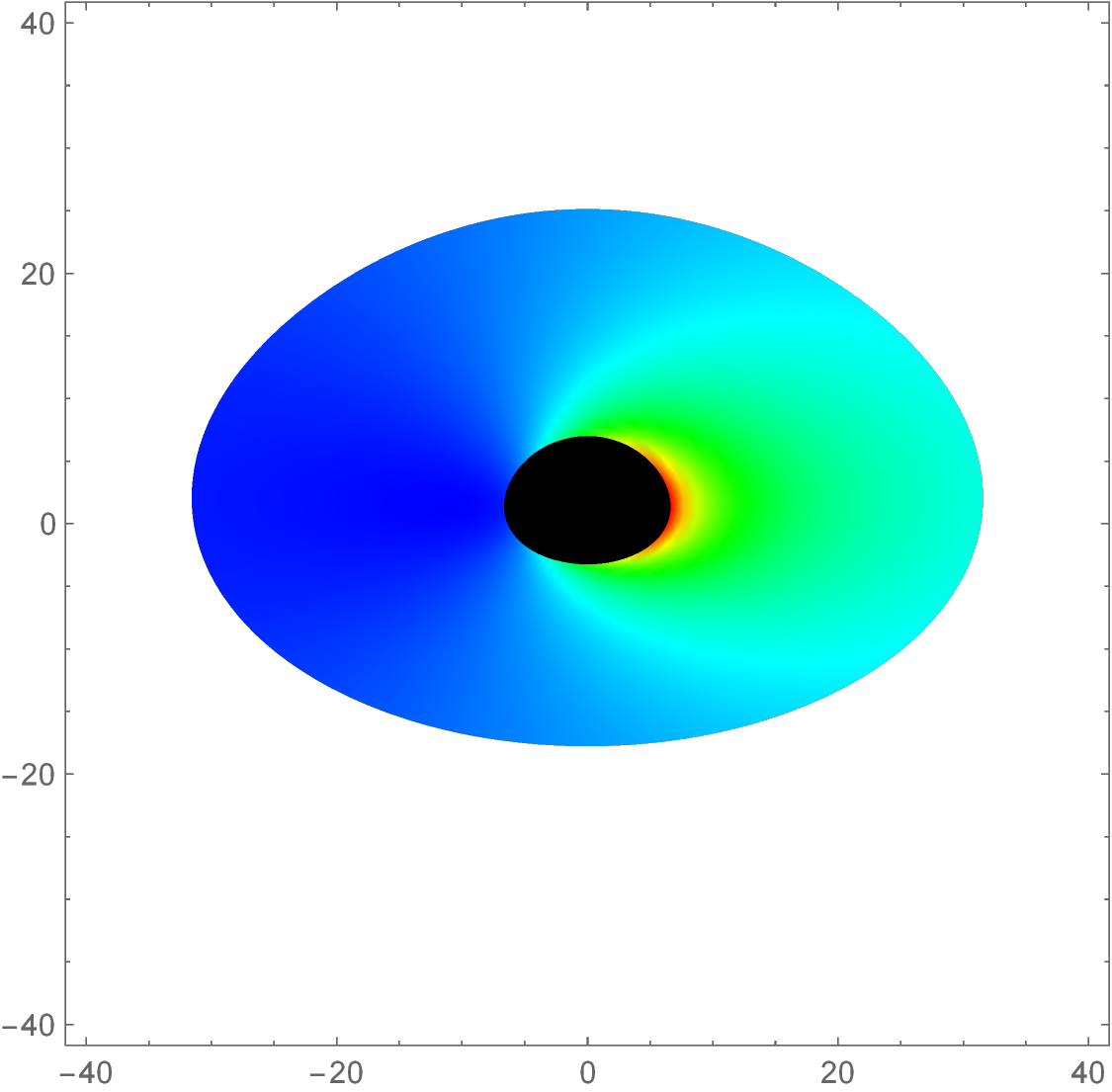} 
				\put(18,100){\color{black}\large $a=0.08, \theta=53^{\circ}$} 
				\put(-8,48){\color{black} Y}
				\put(48,-10){\color{black} X}
			\end{overpic}
			\raisebox{0.05\height}{
				\begin{overpic}[width=0.07\textwidth]{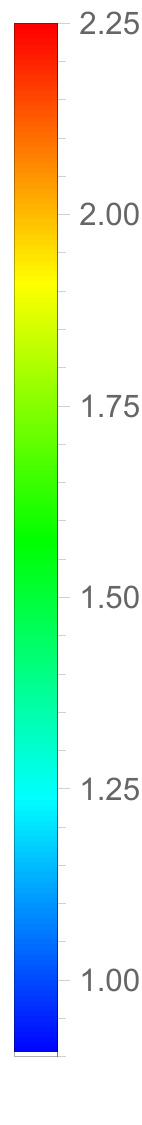}
					\put(0,103){\color{black}\large $z$} 
				\end{overpic}
			}
		\end{minipage}
		\vspace{40pt} 
		\\ 
		\begin{minipage}[t]{0.3\textwidth}
			\centering
			\begin{overpic}[width=0.75\textwidth]{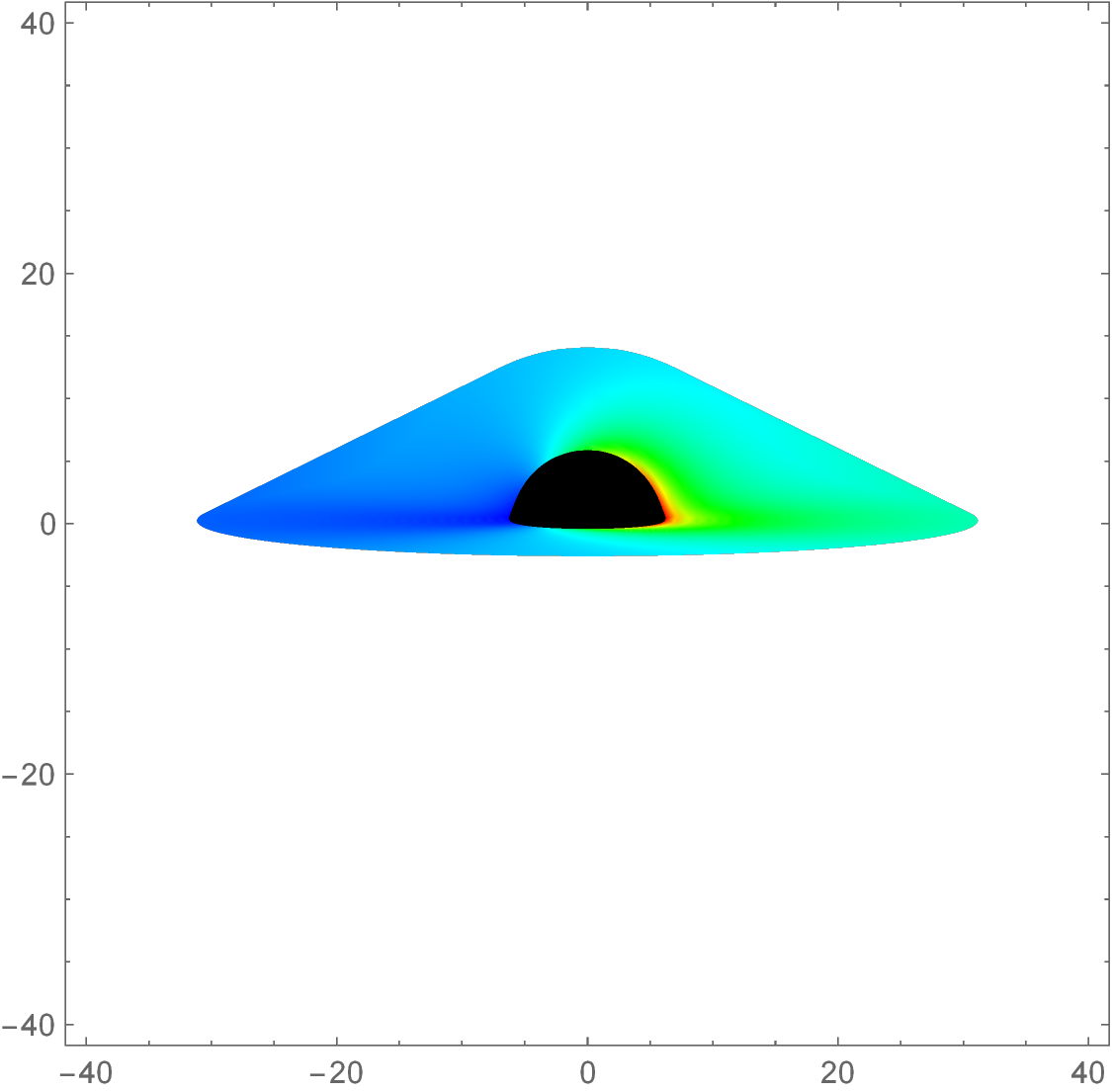} 
				\put(18,100){\color{black}\large $a=0.02, \theta=85^{\circ}$}
				\put(-8,48){\color{black} Y}
				\put(48,-10){\color{black} X}
			\end{overpic}
			\raisebox{0.13\height}{ 
				\begin{overpic}[width=0.07\textwidth]{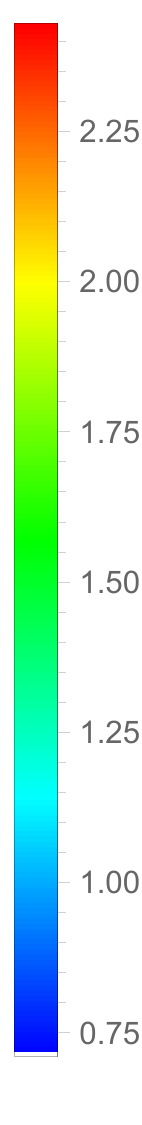}
					\put(0,103){\color{black}\large $z$} 
				\end{overpic}
			}
		\end{minipage}
		&
		\begin{minipage}[t]{0.3\textwidth}
			\centering
			\begin{overpic}[width=0.75\textwidth]{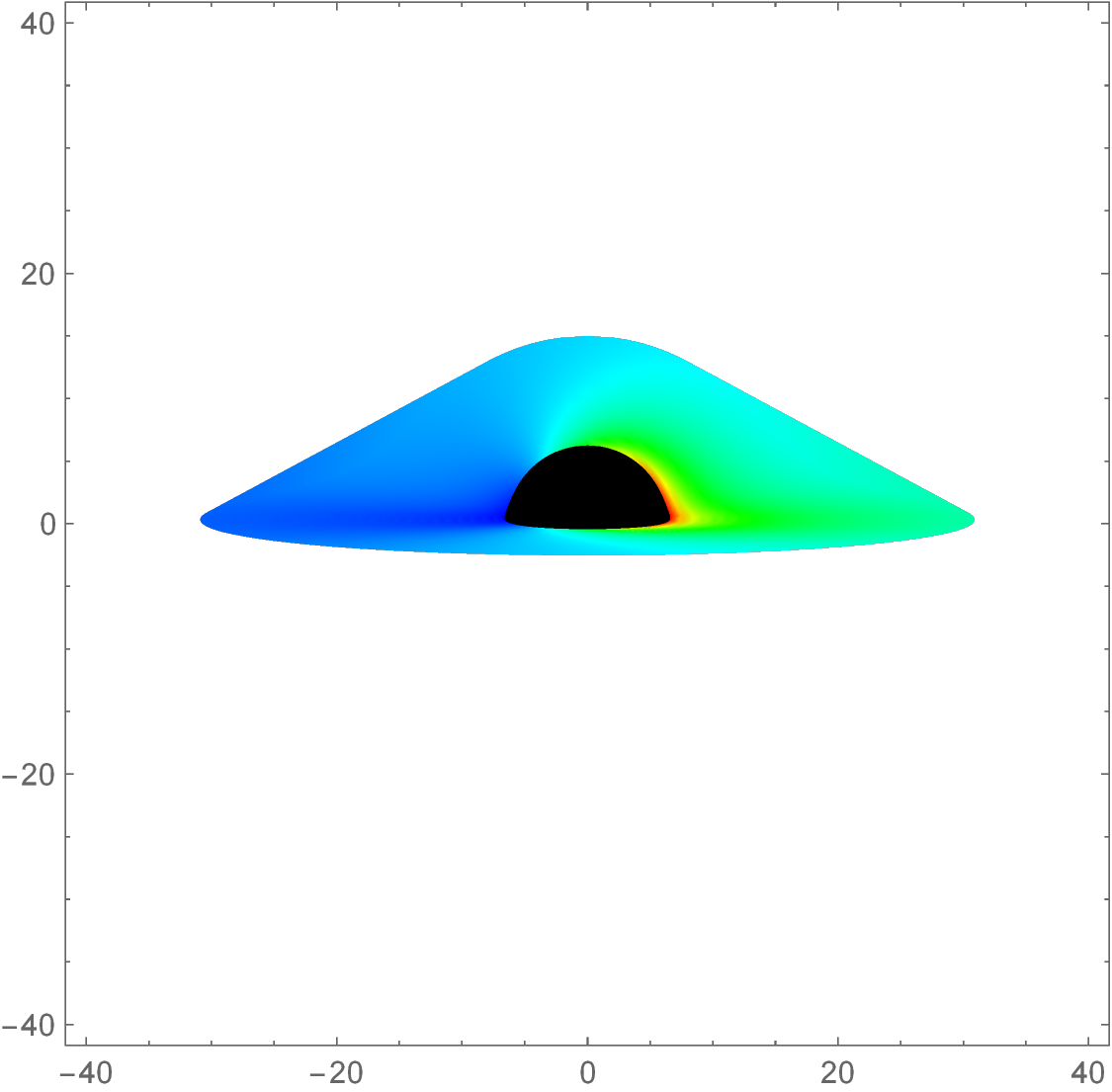} 
				\put(18,100){\color{black}\large $a=0.05, \theta=85^{\circ}$} 
				\put(-8,48){\color{black} Y}
				\put(48,-10){\color{black} X}
			\end{overpic}
			\raisebox{0.13\height}{ 
				\begin{overpic}[width=0.07\textwidth]{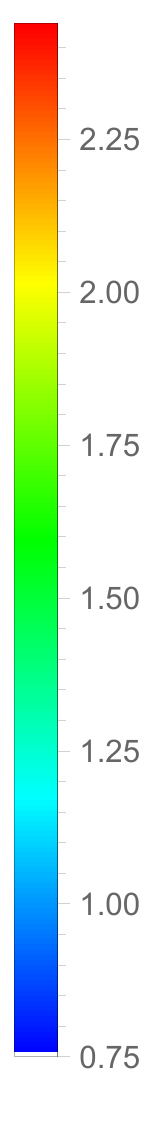} 
					\put(0,103){\color{black}\large $z$} 
				\end{overpic}
			}
		\end{minipage}
		&
		\begin{minipage}[t]{0.3\textwidth}
			\centering
			\begin{overpic}[width=0.75\textwidth]{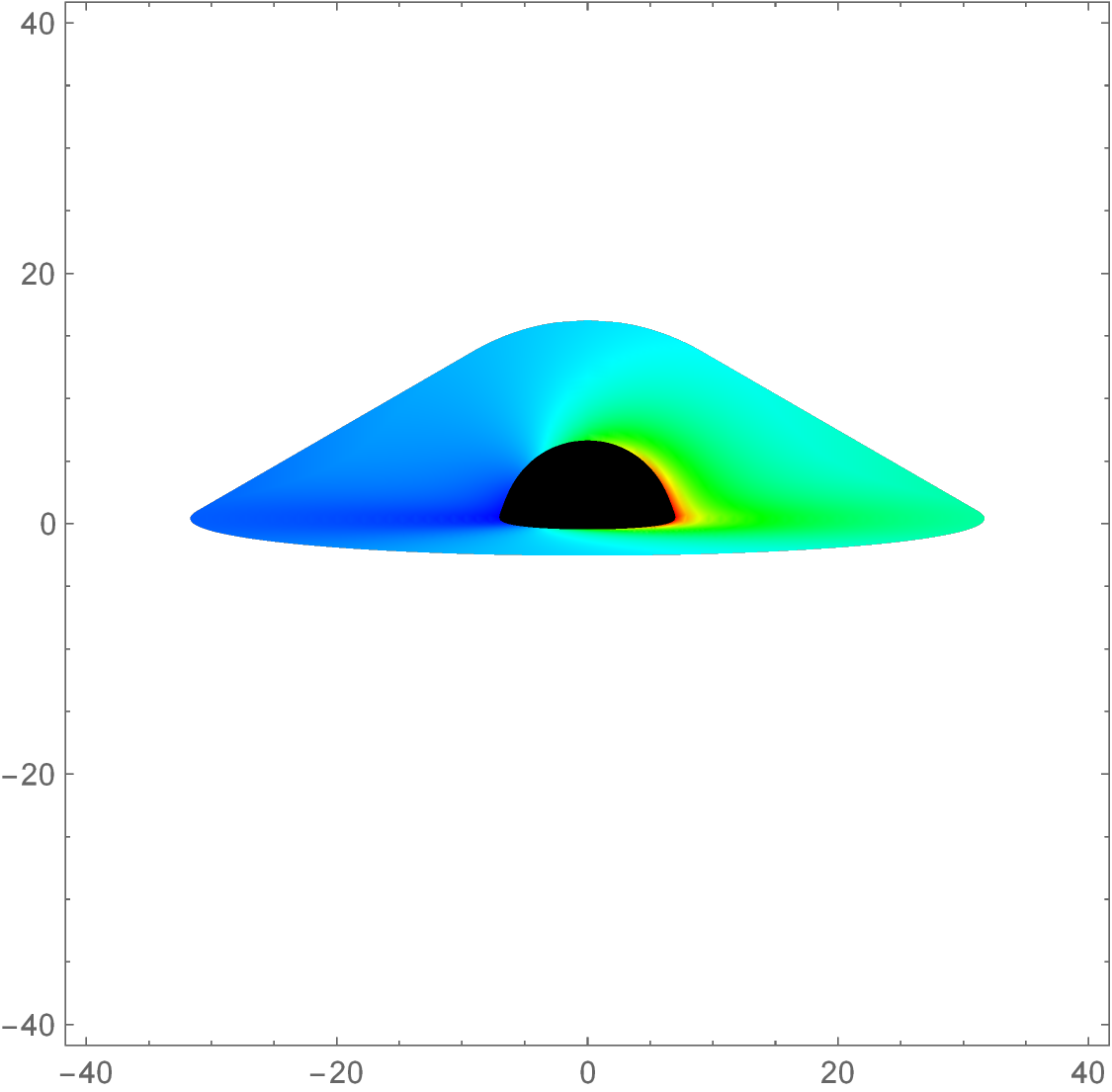}
				\put(18,100){\color{black}\large $a=0.08, \theta=85^{\circ}$}
				\put(-8,48){\color{black} Y}
				\put(48,-10){\color{black} X}
			\end{overpic}
			\raisebox{0.13\height}{ 
				\begin{overpic}[width=0.07\textwidth]{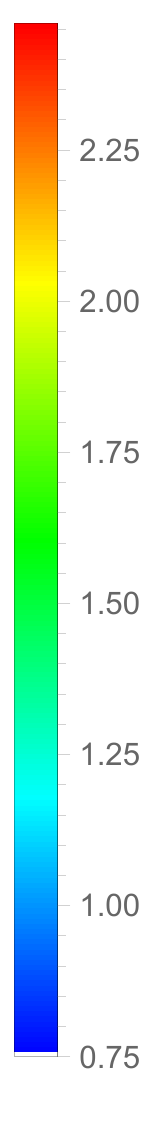} 
					\put(0,103){\color{black}\large $z$} 
				\end{overpic}
			}
		\end{minipage}
	\end{tabular}
	\caption{Continuous distributions of redshift $z$ in the direct images cast by the accretion disk around an ABG BH coupled with a CS, under different values of parameter $\alpha$ and inclination angles. Each column, from top to bottom, corresponds to inclination angles of $17^{\circ}$, $53^{\circ}$, and $85^{\circ}$, while each row, from left to right, represents $a$ values of $0.02$, $0.05$, and $0.08$. We set $g=0.6$.}
	\label{hongyi1}
\end{figure*}
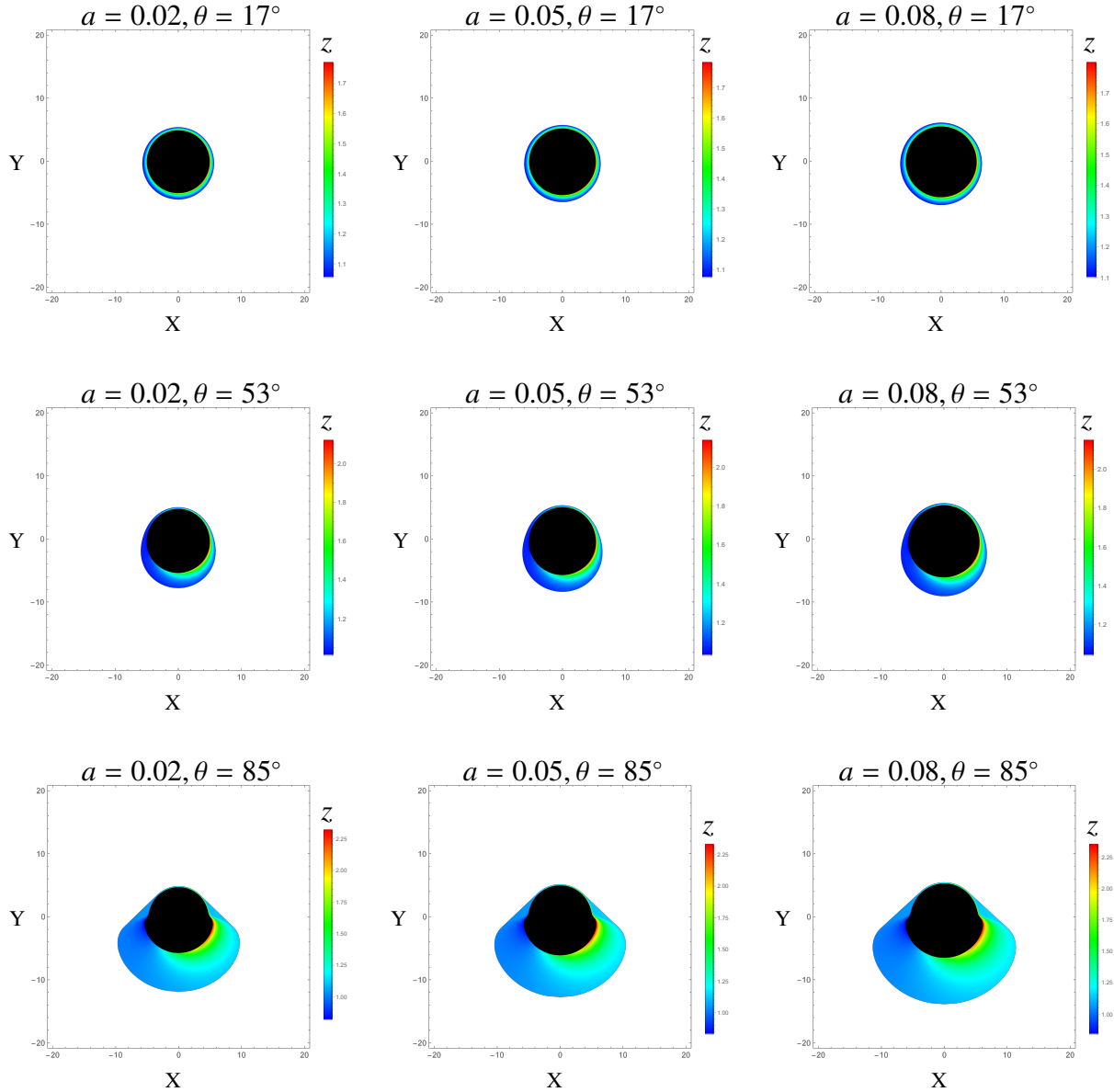
\begin{figure*}[htbp]
	\centering
	\begin{tabular}{ccc}
		\begin{minipage}[t]{0.3\textwidth}
			\centering
			\begin{overpic}[width=0.75\textwidth]{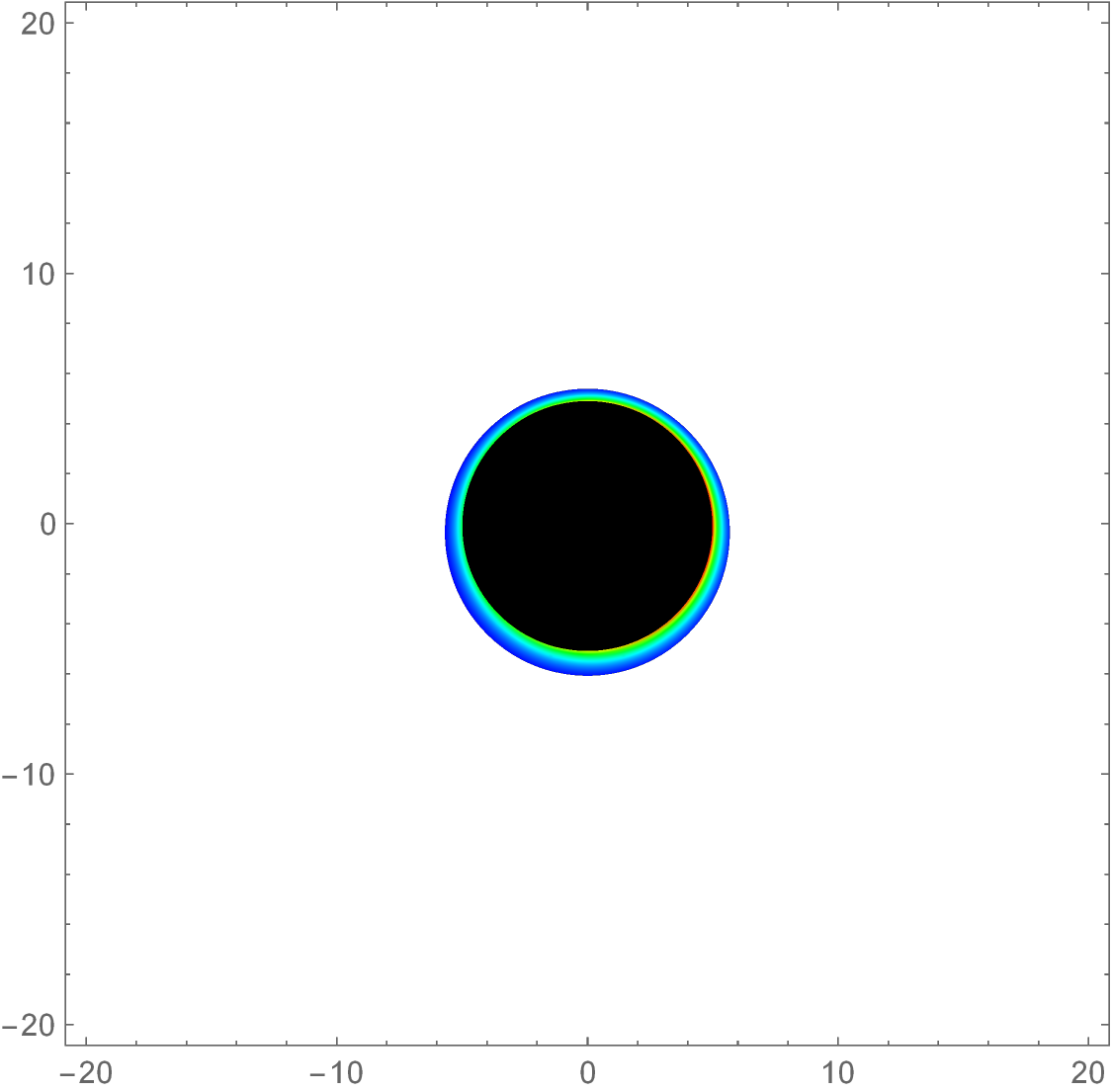}
				\put(18,100){\color{black}\large $a=0.02, \theta=17^{\circ}$} 
				\put(-8,48){\color{black} Y}
				\put(48,-10){\color{black} X}
			\end{overpic}
			\raisebox{0.05\height}{ 
				\begin{overpic}[width=0.07\textwidth]{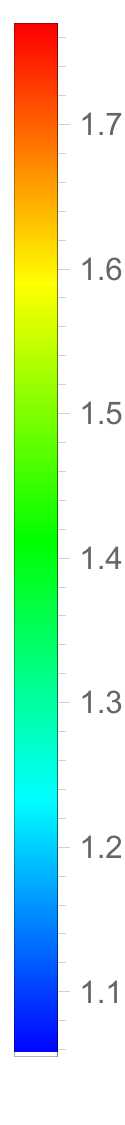}
					\put(0,103){\color{black}\large $z$}
				\end{overpic}
			}
		\end{minipage}
		&
		\begin{minipage}[t]{0.3\textwidth}
			\centering
			\begin{overpic}[width=0.75\textwidth]{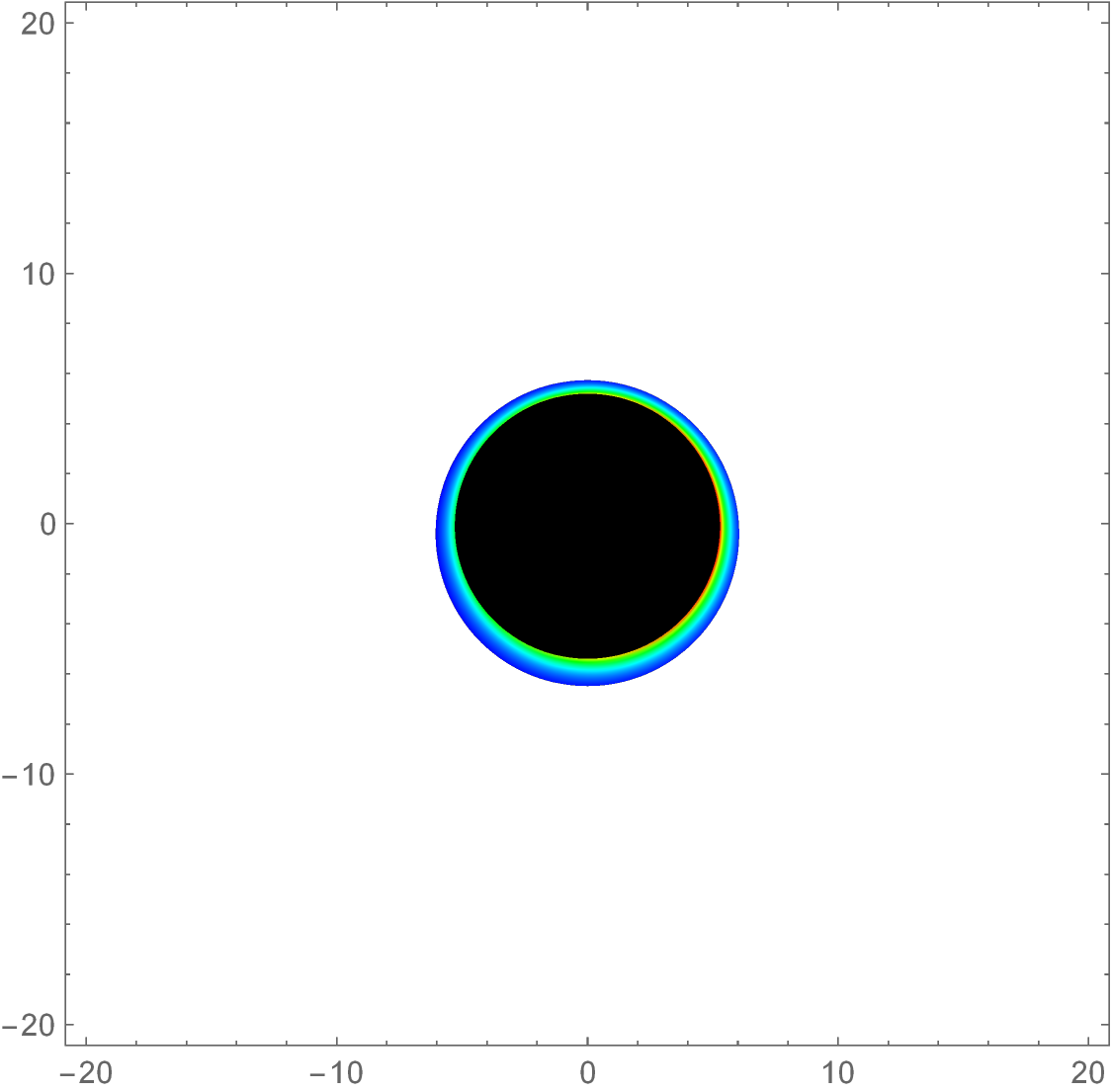} 
				\put(18,100){\color{black}\large $a=0.05, \theta=17^{\circ}$} 
				\put(-8,48){\color{black} Y}
				\put(48,-10){\color{black} X}
			\end{overpic}
			\raisebox{0.05\height}{ 
				\begin{overpic}[width=0.07\textwidth]{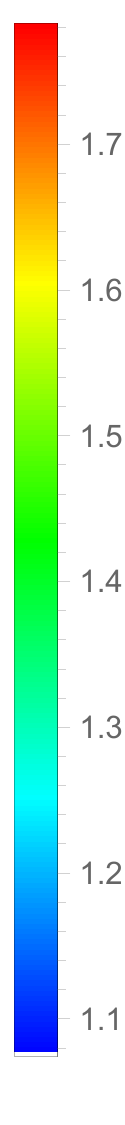}
					\put(0,103){\color{black}\large $z$}
				\end{overpic}
			}
		\end{minipage}
		&
		\begin{minipage}[t]{0.3\textwidth}
			\centering
			\begin{overpic}[width=0.75\textwidth]{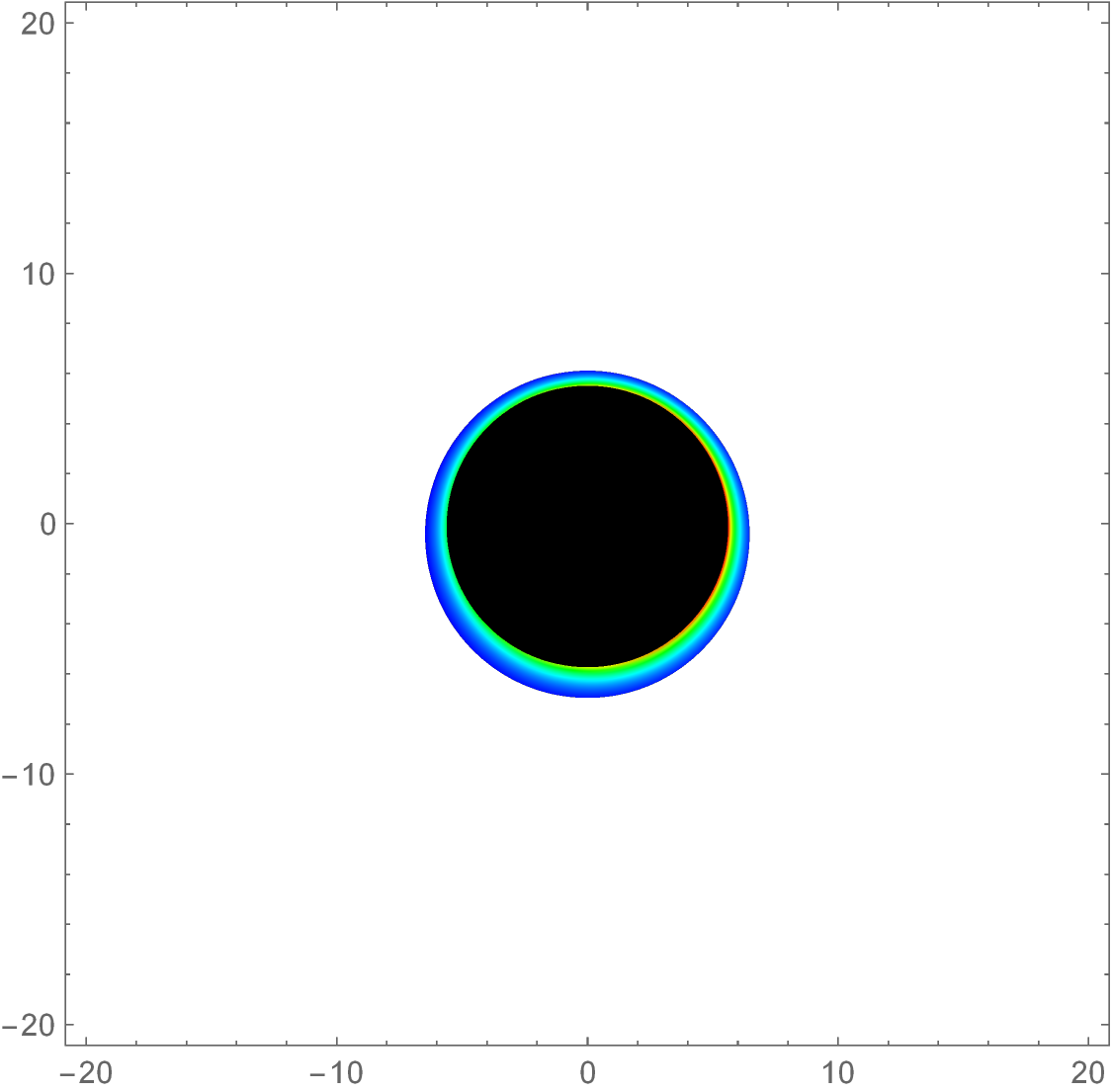} 
				\put(15,100){\color{black}\large $a=0.08, \theta=17^{\circ}$}
				\put(-8,48){\color{black} Y}
				\put(48,-10){\color{black} X}
			\end{overpic}
			\raisebox{0.05\height}{ 
				\begin{overpic}[width=0.07\textwidth]{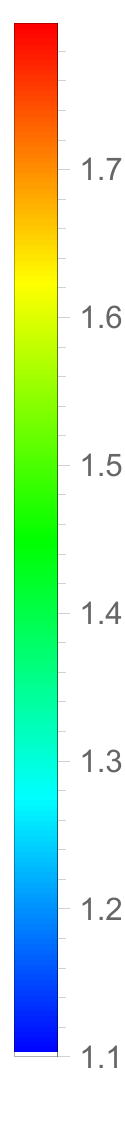}
					\put(0,103){\color{black}\large $z$} 
				\end{overpic}
			}
		\end{minipage}
		\vspace{40pt} 
		\\ 
		\begin{minipage}[t]{0.3\textwidth}
			\centering
			\begin{overpic}[width=0.75\textwidth]{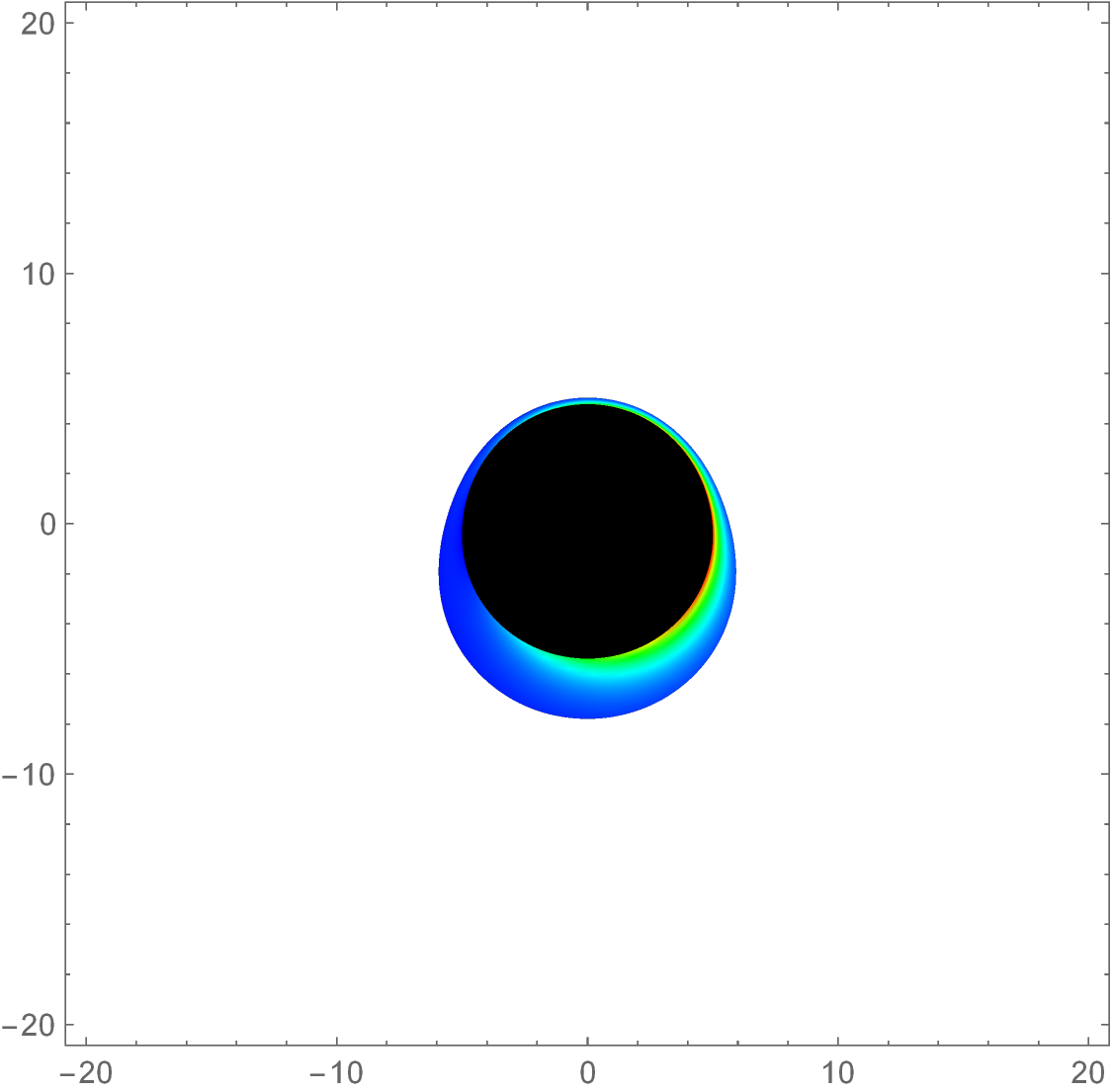}
				\put(18,100){\color{black}\large $a=0.02, \theta=53^{\circ}$}
				\put(-8,48){\color{black} Y}
				\put(48,-10){\color{black} X}
			\end{overpic}
			\raisebox{0.05\height}{
				\begin{overpic}[width=0.07\textwidth]{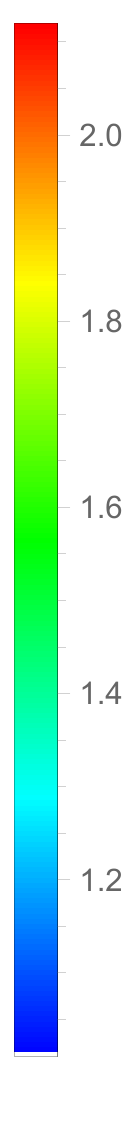}
					\put(0,103){\color{black}\large $z$} 
				\end{overpic}
			}
		\end{minipage}
		&
		\begin{minipage}[t]{0.3\textwidth}
			\centering
			\begin{overpic}[width=0.75\textwidth]{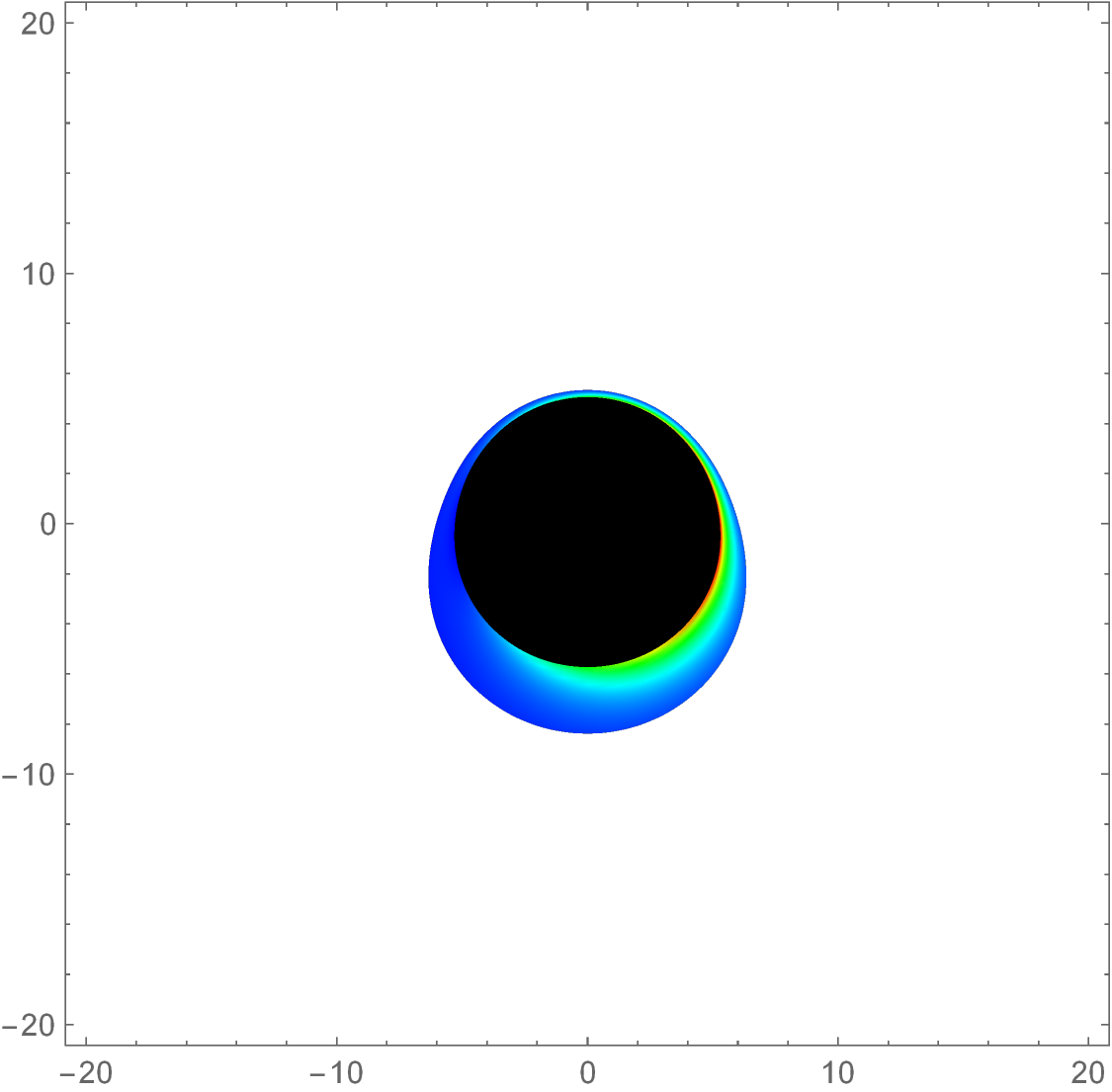}
				\put(18,100){\color{black}\large $a=0.05, \theta=53^{\circ}$} 
				\put(-8,48){\color{black} Y}
				\put(48,-10){\color{black} X}
			\end{overpic}
			\raisebox{0.05\height}{ 
				\begin{overpic}[width=0.07\textwidth]{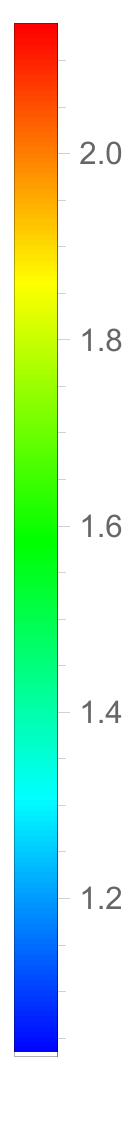} 
					\put(0,103){\color{black}\large $z$}
				\end{overpic}
			}
		\end{minipage}
		&
		\begin{minipage}[t]{0.3\textwidth}
			\centering
			\begin{overpic}[width=0.75\textwidth]{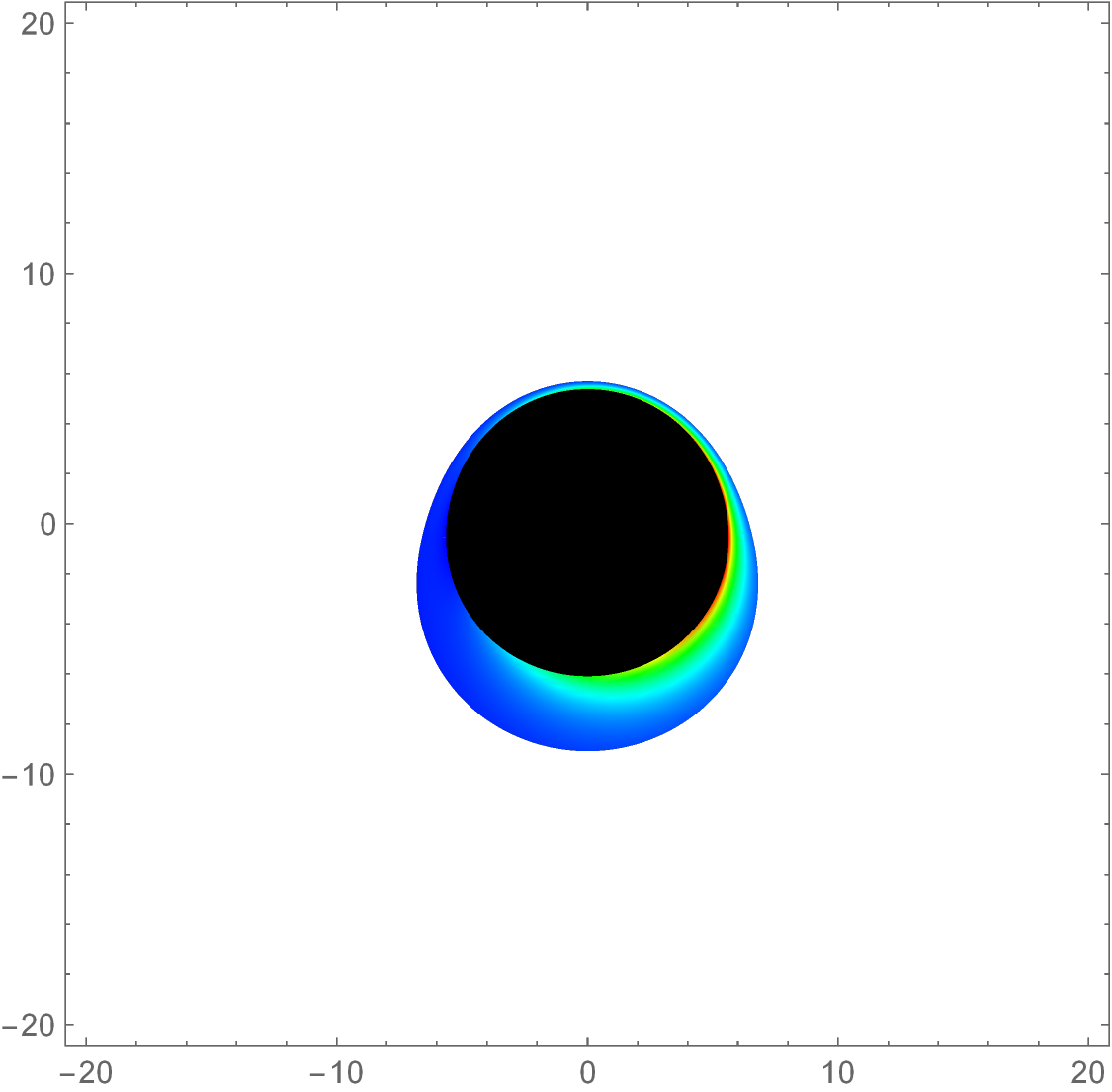} 
				\put(15,100){\color{black}\large $a=0.08, \theta=53^{\circ}$} 
				\put(-8,48){\color{black} Y}
				\put(48,-10){\color{black} X}
			\end{overpic}
			\raisebox{0.05\height}{ 
				\begin{overpic}[width=0.07\textwidth]{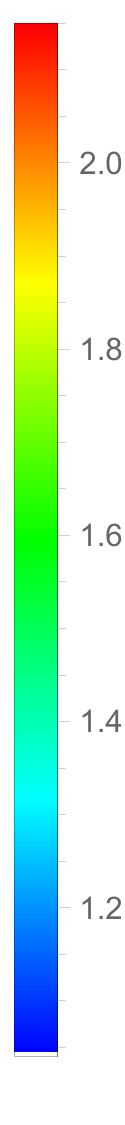} 
					\put(0,103){\color{black}\large $z$} 
				\end{overpic}
			}
		\end{minipage}
		\vspace{40pt} 
		\\ 
		\begin{minipage}[t]{0.3\textwidth}
			\centering
			\begin{overpic}[width=0.75\textwidth]{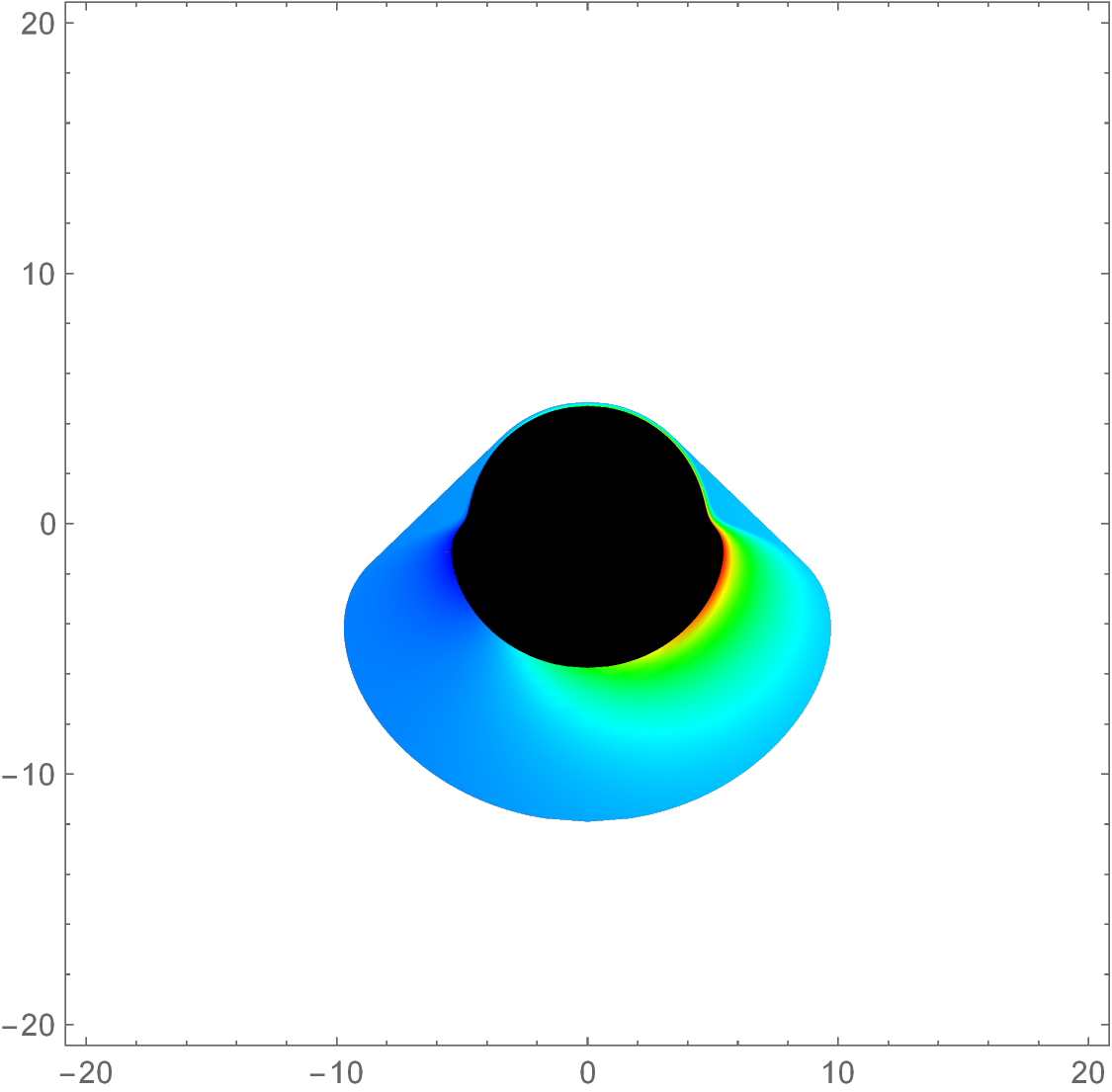}
				\put(18,100){\color{black}\large $a=0.02, \theta=85^{\circ}$}
				\put(-8,48){\color{black} Y}
				\put(48,-10){\color{black} X}
			\end{overpic}
			\raisebox{0.13\height}{ 
				\begin{overpic}[width=0.07\textwidth]{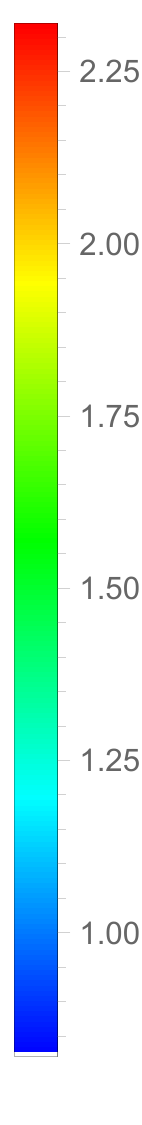}
					\put(0,103){\color{black}\large $z$} 
				\end{overpic}
			}
		\end{minipage}
		&
		\begin{minipage}[t]{0.3\textwidth}
			\centering
			\begin{overpic}[width=0.75\textwidth]{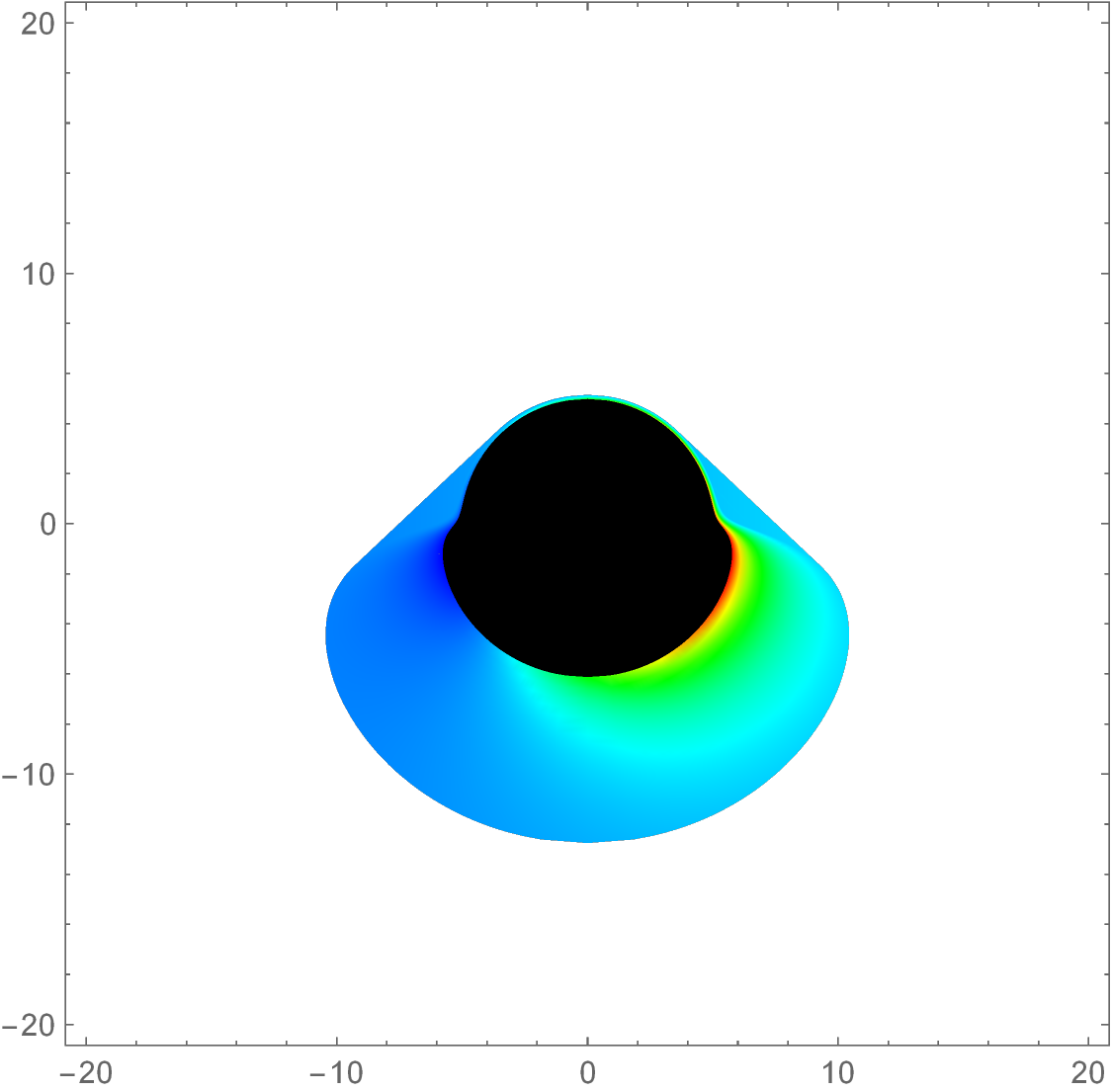}
				\put(18,100){\color{black}\large $a=0.05, \theta=85^{\circ}$}
				\put(-8,48){\color{black} Y}
				\put(48,-10){\color{black} X}
			\end{overpic}
			\raisebox{0.06\height}{ 
				\begin{overpic}[width=0.07\textwidth]{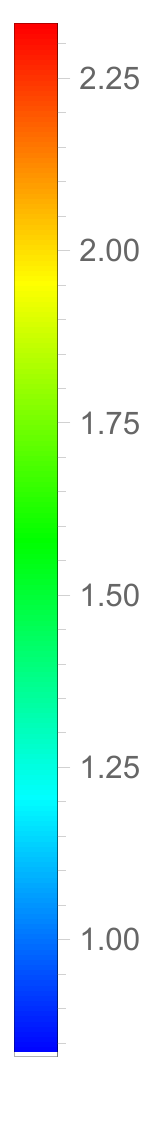} 
					\put(0,103){\color{black}\large $z$} 
				\end{overpic}
			}
		\end{minipage}
		&
		\begin{minipage}[t]{0.3\textwidth}
			\centering
			\begin{overpic}[width=0.75\textwidth]{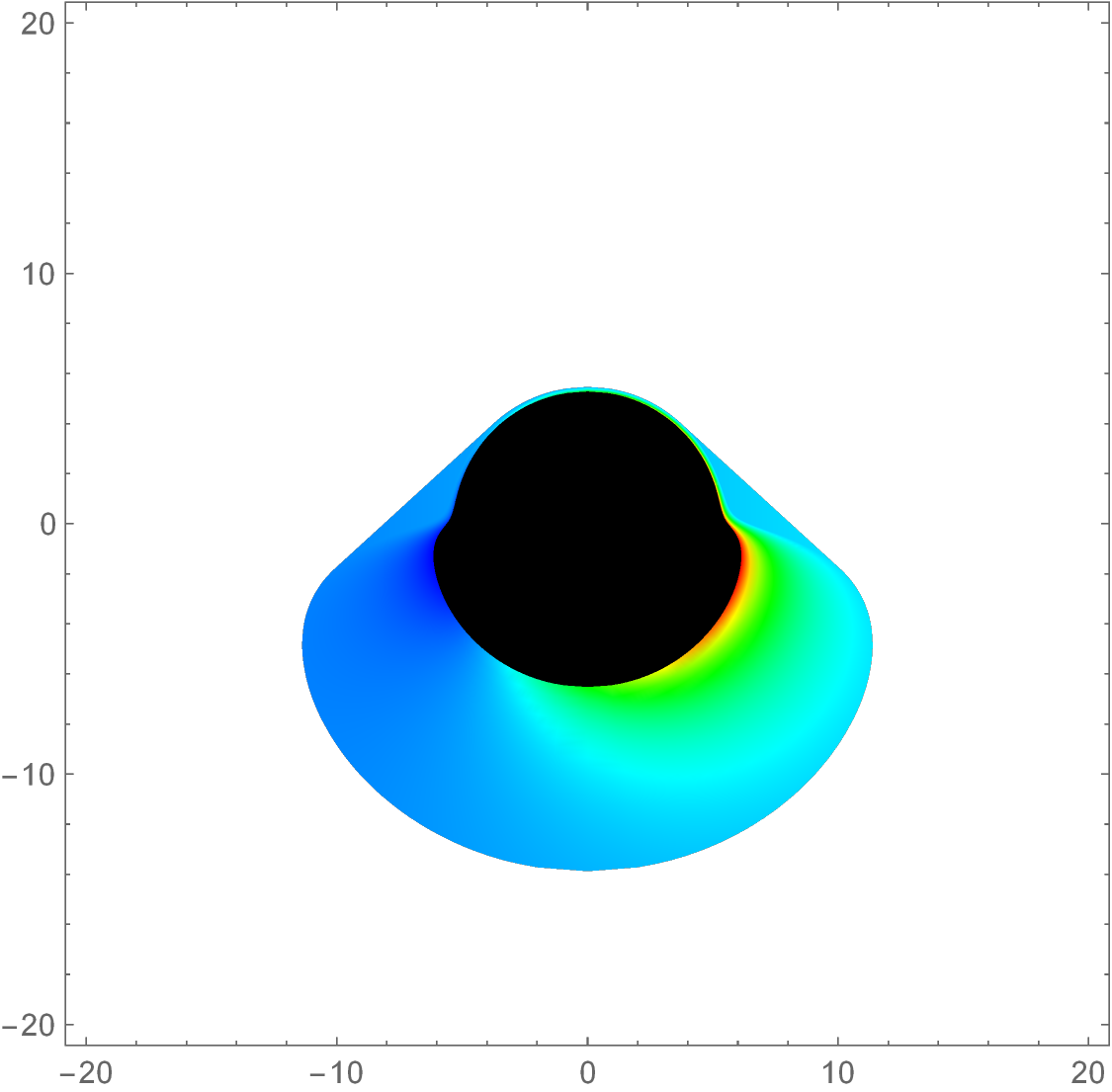}
				\put(15,100){\color{black}\large $a=0.08, \theta=85^{\circ}$}
				\put(-8,48){\color{black} Y}
				\put(48,-10){\color{black} X}
			\end{overpic}
			\raisebox{0.06\height}{ 
				\begin{overpic}[width=0.07\textwidth]{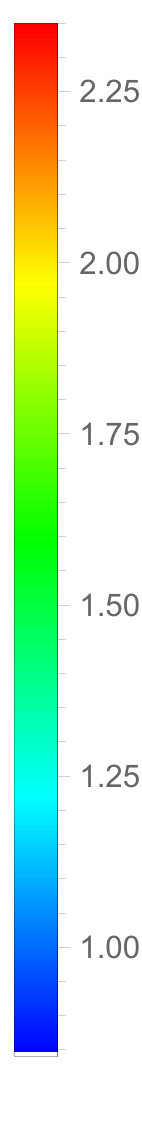} 
					\put(0,103){\color{black}\large $z$} 
				\end{overpic}
			}
		\end{minipage}
	\end{tabular}
	\caption{Similar to Fig.~\ref{hongyi1}, but for the secondary images.}
	\label{hongyi2}
\end{figure*}
\section{Conclusion} 
\label{section5}
Our study delved into the physical properties and optical appearance of a thin accretion disk around an ABG BH coupled with a CS. This study utilizes Event Horizon Telescope (EHT) observational data on the shadow diameters of the supermassive black holes M87$^*$ and Sgr A$^*$ to place the first joint observational constraints on two key parameters, $a$ and $g$, in the Ay\'{o}n-Beato-Garc\'{i}a black hole solution coupled with nonlinear electrodynamics and a cloud of strings. Compared to previous theoretical studies—such as the original ABG paper \cite{Ayon-Beato:1998hmi}, which derived a theoretical bound on the NLED charge parameter $(|g|\leq0.6M)$ based on the existence of a black hole horizon, and Letelier's pioneering work on clouds of strings \cite{Letelier:1979ej}, which established a theoretical range for the string cloud parameter $(0<a<1)$—our analysis provides independent and complementary constraints on the interdependent relationship between these two parameters, based on direct imaging at event horizon scales. For instance, when $g=0.5M$, observations of M87$^*$ constrain $a$ within the range of $0.006$ to $0.158$, significantly narrower than the maximum theoretically allowed value, demonstrating the power of EHT observations in testing theoretical parameters. To date, no study in the literature has performed a joint analysis of these two parameters using the same high-precision astrophysical data set. Our results fill this gap and lay the foundation for further constraining these fundamental parameters with next-generation EHT observations. Further investigation into the shadows and observational features of the ABG BH coupled with a CS showed that higher values of the parameter $a$ result in a larger BH shadow, whereas an increase in the parameter $g$ leads to a smaller shadow. Lastly, we analyzed the physical properties and imaging of the accretion disk, finding that the impact of these parameters on BH properties is negligible.

\begin{acknowledgments}
	This research was supported by the National Natural Science Foundation of China (Grant No. 12265007).
\end{acknowledgments}

\end{document}